\documentclass[a4paper,11pt]{article}
\usepackage{geometry,tikz}
\usepackage{a4wide}
\usepackage{graphicx}
\usepackage{epsf}
\usepackage{amsmath}
\usepackage{amssymb}
\usepackage[normalem]{ulem}

\usepackage{cite}\newcommand\orcidroldao{{\href{https://orcid.org/0000-0003-3978-532X}{\orcidicon}}}
\usepackage{cite}\newcommand\orciddudal{{\href{http://orcid.org/0000-0003-1326-6011}{\orcidicon}}}
\usepackage{cite}\newcommand\orcidtoniato{{\href{https://orcid.org/0000-0003-0969-9359}{\orcidicon}}}
\usepackage{cite}
\usepackage{cite}\newcommand\orcidsubhash{{\href{https://orcid.org/0000-0002-0113-0822}{\orcidicon}}}

\newcommand{\orcidicon}{%
	\begin{tikzpicture}
	\draw[lime, fill=lime] (0,0)
		circle [radius=0.16]
		node[white] {{\fontfamily{qag}\selectfont \tiny ID}};
	\draw[white, fill=white] (-0.0625,0.095)
		circle [radius=0.007];
	\end{tikzpicture}	\hspace{-2mm}
}

\usepackage{multirow}
\usepackage{subfigure}
\usepackage{appendix}
\usepackage{tikz}
\usepackage{amsfonts,amsthm,euscript,braket,xcolor}
\usepackage{breqn}
\newcommand{\be}{\begin{equation}}
	\newcommand{\ee}{\end{equation}}

\newcommand{\Rmnum}[1]{\expandafter\@slowromancap\romannumeral #1@}
\newcommand{\bea}{\begin{eqnarray}}
	\newcommand{\eea}{\end{eqnarray}}

\newcommand\scalemath[2]{\scalebox{#1}{\mbox{\ensuremath{\displaystyle #2}}}}
\numberwithin{equation}{section}

\usepackage{chngcntr}
\counterwithin{table}{section}
\counterwithin{figure}{section}

\usepackage{ulem}
\usepackage[colorlinks=true, linkcolor=blue, citecolor=blue, urlcolor=blue]{hyperref}
\begin{document}
	
	
	\title{\bf Holographic QCD model for heavy and exotic mesons at finite density: A self-consistent dynamical approach}
	\author{ \textbf{\textbf{Bruno Toniato\orcidtoniato\!\!$^{a,b}$}\thanks{bruno.toniato@ufabc.edu.br (corresponding author)}, \textbf{David Dudal\orciddudal\!\!$^{b}$}\thanks{david.dudal@kuleuven.be}, \textbf{Subhash Mahapatra\orcidsubhash\!\!$^{c}$}\thanks{mahapatrasub@nitrkl.ac.in},}, \\ \textbf{\textbf{Roldao da Rocha\orcidroldao\!\!$^d$}
\thanks{roldao.rocha@ufabc.edu.br}, Siddhi Swarupa Jena\orciddudal\!\!$^{c}$}\thanks{519ph2015@nitrkl.ac.in}
		\\\\\textit{{\small $^a$ Federal University of ABC, Center of Natural and Human Sciences, }}\\
        \textit{{\small Santo Andr\'e, 09210-170, Brazil}}\\
        \textit{{\small $^b$ KU Leuven Campus Kortrijk--Kulak, Department of Physics, Etienne Sabbelaan 53 bus 7657,}}\\
		\textit{{\small 8500 Kortrijk, Belgium}}\\
		        \textit{{\small $^c$ Department of Physics and Astronomy, National Institute of Technology Rourkela,}}\\
        \textit{{\small Rourkela - 769008, India}}\\
        \textit{{\small $^d$ Federal University of ABC, Center of Mathematics,}}\\
        \textit{{\small Santo Andr\'e, 09210-170, Brazil}}
}

\date{}
\maketitle
\begin{abstract}
	We present a self-consistent dynamical holographic QCD model to investigate the mass spectra and melting behavior of heavy and exotic mesons at finite temperature and finite density. Our approach is based on the Einstein-Maxwell-Dilaton (EMD) framework and incorporates an elsewhere already introduced, albeit by hand, phenomenological non-quadratic dilaton profile.
    This allows one to capture the non-linear Regge trajectories of heavy-flavor mesons and model certain exotic states. We show how to construct such models by actually solving the coupled Einstein, Maxwell, and dilaton field equations, ensuring mathematical self-consistency to replace any ad-hoc input. At finite temperature, we analyze the confinement-deconfinement transition via a Hawking-Page phase transition. We compute the spectral functions, revealing the sequential melting of quarkonia as the temperature is increased. Extending to finite density, we explore the impact of baryon chemical potential on meson stability, showing significant modifications in spectral peaks and effective potentials that indicate a more rapid melting of mesonic states as the chemical potential increases in the deconfined phase. The dual of the small/large black hole transition now indicates towards a first order phase transition line ending at a second order critical point. Interestingly, the spectral functions smoothly cross this phase transition line.
\end{abstract}

\section{Introduction}
\label{sec1}

AdS/QCD can implement non-perturbative aspects of the confinement mechanism in QCD, which states that quarks are confined inside hadrons. Weakly coupled gravity in an AdS${}_5$ bulk is dual to non-perturbative QCD, quantitatively describing strong interactions and several hadronic features \cite{Karch:2006pv,Branz:2010ub,Colangelo:2008us,Brodsky:2014yha,Erlich:2005qh,Sakai2005Apr,Sakai2005Nov,Rougemont:2023gfz}. AdS/QCD stands on AdS/CFT, where the large-$N_c$ limit in the SU($N_c$)   
Yang-Mills theory with four supersymmetry charges in 4-dimensional spacetime corresponds to type-IIB superstring theory, with the near-horizon limit describing the AdS${}_5 \times S^5$ effective metric of a stack of $N_c$ nonextremal Dirichlet 3-branes \cite{Maldacena:1997re,Witten:1998qj,Gubser:1998bc,Aharony:1999ti,Ammon:2015wua}. A consistent strongly coupled theory describing QCD can be accommodated on the AdS${}_5$ boundary at low energies, and the extra dimension across the AdS bulk is the QCD energy scale and high [low] energy regards the UV [IR] limit. In the IR limit, the so-called bottom-up AdS/QCD duality can properly describe important QCD features through appropriate warped factors in the metrics that equip AdS and dilaton backgrounds. 
 In the UV regime of QCD, computational perturbative methods in the weak-coupling regime quantitatively fit experimental data. Nevertheless, the IR regime is strongly coupled, making chiral symmetry breaking and confinement emerge.   As a result, any outcome achieved in a weakly coupled theory is more easily manageable and can be put in correspondence to solutions in the dual theory in non-perturbative regimes, where in general calculations are too intricate to solve even in Euclidean lattice QCD, thinking here about either the finite density case with its sign problem or dynamical questions like decays or transport properties that require a way to analytically continue the numerical lattice data to Minkowski space.
 
As QCD does not enjoy an exact conformal symmetry, it is still possible to borrow CFT to settle it as a bottom-up approach, constructing the dual theory when constraints 
are imposed to fit phenomenological results \cite{Afonin:2022hbb,dePaula:2009za,Ballon-Bayona:2017sxa,Ballon-Bayona:2021ibm,Karapetyan:2023sfo,Bartz:2018nzn,Ballon-Bayona:2023zal,Aref'eva2018May,Arefeva:2024vom,Erdmenger:2020flu,Gursoy:2007cb,Gursoy:2007er,Gursoy:2010fj,He2013Apr,Alho:2013hsa,Alanen:2009xs}. The soft-wall AdS/QCD model is modeled by adding a dilaton coupled to Einstein--Hilbert gravity, therefore, introducing a smooth cutoff in AdS, making both confinement and Regge trajectories \cite{Csaki,Gherghetta:2009ac,Bartz:2024dgd} possible. The AdS metric is deformed 
due to the action of both the chiral and the gluon condensations. Within this framework, linear confinement and chiral symmetry breaking can be explained.  
AdS/QCD corroborates universal features of hadrons, from the spectrum of light-flavor mesons to heavy-flavor quarkonia \cite{Braga:2017oqw,Ferreira:2019inu,daRocha:2021imz,Ferreira:2019nkz,Braga:2018fyc,Braga:2023qee,Colangelo:2018mrt,Ferreira:2020iry,Bernardini:2018uuy,Barbosa-Cendejas:2018mng,daRocha:2021ntm,daRocha:2024sjn}.  One of the main issues in hadron spectroscopy is to find some universal solvable model describing mesons and baryons as a function of quark masses. 
Taking into account the masses of the quarks makes non-linear Regge trajectories be addressed by suitable deformations of the dilaton. This procedure is appropriate to describe exotic mesons in AdS/QCD  \cite{MartinContreras:2020cyg,Karapetyan:2021ufz,MartinContreras:2023oqs}. 	

One can 
employ bottom-up AdS/QCD in the context of Einstein-Maxwell-dilaton gravity to better scrutinize exotic mesons.  As proposed in \cite{DeWolfe:2010he,DeWolfe:2011ts}, this model was studied to construct AdS black holes presenting first-order phase transitions ending at a critical point at finite temperature and finite chemical potential. These solutions comply with baryon susceptibilities and match QCD lattice data at vanishing chemical potential. The paper \cite{Rougemont:2015ona} showed that this model could match lattice QCD data for the equation of state of the quark-gluon plasma (QGP) near
the crossover transition with baryon chemical potential 
up to 400 MeV. Other approaches in \cite{Noronha:2009ud,Noronha:2010hb,He:2013qq,Ballon-Bayona:2020xls,Gubser:2008yx,Bohra:2019ebj,Bohra:2020qom,Dudal:2021jav,
Cai:2012xh,Yang:2015aia,Ferreira-Martins:2019wym, Jena:2022nzw,Shukla:2023pbp,Shukla:2024qlf} use the Einstein-Maxwell-dilaton gravity as a dual to describe phenomena in QCD. Ref.~\cite{Dudal:2017max} used such a setup to study a first-order phase transition from a small to a large black hole phase, respectively, associated with ``almost confinement'' and deconfinement phases, thereby allowing a mechanism to discuss nearly confined physics at finite temperature using the small black hole phase. Here,  we analytically and self-consistently solve the gravity equations of motion in terms of the warp factor and apply the solutions to describe exotic mesons, discussing and addressing our results as well as comparing them to the recent literature, thereby making a major improvement on the soft-wall holographic QCD models and their phenomenological relevance.  

Quadratic dilatons are usually reported in AdS/QCD to model the mesons mass spectra of light-flavor mesonic states as linear functions of the radial quantum number  \cite{MartinContreras:2020cyg,Song,Rinaldi:2020ybv}. When heavy quarks are considered as constituents for mesons, the Regge trajectories linearity breaks apart, even though the heavy mesons spectra are precisely described  \cite{Afonin:2014nya,jk}.
Heavy quarks can compose the so-called heavy-quark exotica, comprising tetraquarks, pentaquarks,  hadrocharmonia, hybrids mesons, and hadronic molecules  \cite{Guo,Lebed1,Oncala:2017hop}. 
Among the diversity of detected particles, exotic mesons remain a partially detached conundrum in QCD, and efforts to interpret and model  their fundamental properties are still needed. Exotic mesonic states do not obey the standard  $q\bar q$ structure,  where $q$ denotes a quark \cite{Lebed1,Jaffe1}. 
Generally, mesons with quantum numbers not allowed by the standard $q\bar{q}$ paradigm are denominated exotic mesons \cite{Jaffe1}.

The $q\bar{q}$ composition relies on the fact that no mesons with either isospin or strangeness greater than one exist.  The set of mesons already detected and reported in the Particle Data Group (PDG) \cite{ParticleDataGroup:2024cfk} are implemented by bound $q\bar q$ states clustered into $SU(N_f)$ multiplets,  with corresponding parity $P = (-1)^{L+1}$ and charge conjugation $C = (-1)^{L+S}$, for $S$ denoting, as usual, the meson spin, $L$ standing for the (orbital) angular momentum, and $J$ denoting the total angular momentum. Exotic mesons do not inhabit the $q\bar q$ quark model, since several resonances, as the ones corresponding to $J^{PC} = 0^{--}$, and also the ones with $J^{PC} = 0^{+-}, 1^{-+}, 2^{+-}$ such that $P = (-1)^J$ and $C = (-1)^{J+1}$, cannot be depicted in the standard constituent quark model \cite{Lebed1}. Nevertheless, some mesons can still comprise  $q\bar q$ meson states but bounded to one or more valence gluons and are named hybrid exotic mesons states \cite{Meyer}.  
They support the non-perturbative description of confined meson states in QCD, since valence gluons 
amalgamating into mesons are allowed.
Hybrid mesons $\bar{q}q$-pairs bond by color gluonic flux tubes, allowing other quantum numbers that are for prohibited in the $q\bar{q}$ model, as the case of exotic meson states
 $\pi_1$, with $I^G(J^{PC})=1^-(1^{-+})$, and bottomonium-type states $Z_b$, with $I^G(J^{PC})=1^+(1^{+-})$. Ref.~\cite{Bellantuono:2014lra} approached the  $\pi_1$ hybrid exotic meson using the gauge/gravity duality, demonstrating that they deconfine at lower temperatures when compared to other 
 $\bar{q}q$ light-flavor ordinary mesons.
Exotic mesons  include also charged resonances in the  charmonium-type $Z_c$ family with $I^G(J^{PC})=1^+(1^{+-})$, which has additional valence quarks, as the $Z_c(4430)$ state, in opposition to the neutral charmonium $c\bar c$ state. The  quarkonium-type meson state $Z_c(3900)^+$ was one of the earlier-discovered exotic mesons and consists of a  $c\bar c u\bar d$ state \cite{BESIII:2015cld}. A tetraquark state can be constituted by a hadronic molecule, with four valence quarks merging into either two heavy quarkonia or one heavy quarkonium bounded to a light-flavor mesonic state. The $X(3872)$ exotic state can be considered a hadronic molecule \cite{ParticleDataGroup:2024cfk}. In addition, color singlets can encompass  $qq\bar{q}\bar{q}$ and $Qq\bar{Q}\bar{q}$ tetraquarks, where $Q$ denotes hereon either a charm or a bottom heavy-flavor quark  \cite{Lebed1}. 
Besides, multiquark color 
singlet states of type $qq \bar q\bar q$  encompass tetraquarks and also hadronic molecules \cite{Liu}. Tetraquark states composed by a $c\bar c$ part are distinct from standard charmonia due to multiplets that can include charged resonances, as the $c\bar c u\bar d$, and also non-vanishing strangeness, as the $c\bar c d\bar s$, or even both, as the $c\bar c u\bar s$ state \cite{Godfrey:2008nc}.
Similar analogs can be proposed for bottomonium-type states. Tetraquarks have been thoroughly studied \cite{Lu:2020qmp}, being the four-charm-tetraquark and the  $\bar{c}du\bar{s}$ state recently detected in LHCb collaboration \cite{Fang:2022mks,LHCb:2021uow}. The quark-gluon plasma (QGP) originates from heavy-ion collisions as a prolific source of hot (deconfined) quarks, which can then reconfine into hadrons, as the QGP freezes out. Measurements of exotic states in heavy-ion collisions can provide relevant new tests of production and transport models involving exotic states. Mainly regarding heavy quark exotic states,  enhancement and suppression effects set in and are brought into play in the QGP. 

Exotic mesons also encode hadroquarkonium states, which are composed of a heavy-quark core $Q\bar Q$  inside a light-flavor  $q\bar q$ vector meson cluster. 
Each quark composite forms a color singlet, glued together through weak color van der Waals forces. From the phenomenological point of view, the standard charmonium decay, mainly consisting of the $J/\psi$, $\psi(2S)$, and $\chi_c$ resonances, makes the charmonium core dissociate from the light-flavor meson cluster, causing strong measurable effects \cite{Campanella:2018xev}.  
A relevant example consists of the $Z_c(4430)$ meson decay to $\psi(2S)$, in comparison to the $\psi(1S)=J/\psi$ state, which can be realized as a yield of a 
 $c\bar c$ core in $Z_c(4430)$. 
The family $Y(4008)$, $Y(4230)$, $Y(4260)$, $Y(4360)$, and $Y(4660)$ represent heavy-quark exotica constituted by standard charmonia bond to light-flavor mesons.

Having sketched the experimental, phenomenological and theoretical relevance of exotic mesons, let us now give an overview of how this paper is built: In Sec. \ref{sec2}, we construct the self-consistent dynamical holographic QCD model based on the Einstein-Maxwell-Dilaton (EMD) framework. The formulation begins with the introduction of the EMD action in a five-dimensional AdS spacetime, incorporating a form factor function that leads to a breaking of the conformal symmetry and capturing of confinement effects. The model originates from solving the coupled Einstein, Maxwell, and dilaton field equations, ensuring consistency in the holographic dual description of QCD. Sec.~\ref{sec3} shows the calculation of the heavy and exotic meson mass spectra at zero temperature and chemical potential. We numerically solve the Schrödinger-like equation for the vector mesons, obtaining the mass spectra for some heavy and exotic mesons. Comparisons with experimental data and theoretical predictions are provided, highlighting the role of non-linear Regge trajectories in capturing the heavy meson behavior. Sec.~\ref{sec4} extends the analysis to finite temperature by introducing, again self-consistently, a black hole background, describing the Hawking-Page phase transition as a holographic analog of confinement-deconfinement. We derive the blackening function, determine the temperature dependence of the background geometry, and compute the finite temperature equation of motion for the perturbation. In Sec.~\ref{sec5} we outline the numerical scheme, using the retarded Green's function, for computing the spectral functions; a fundamental quantity that provides insight into meson dissociation in a thermal medium. In Sec.~\ref{sec6} we extract the finite temperature spectral functions. As the temperature increases, the spectral peaks broaden and eventually disappear, signaling the dissolution of mesonic states, consistent with deconfinement taking place also at the level of particle states. Additionally, we investigate the temperature dependence of the effective potential of the Schrödinger-like equation. As the temperature increases, this potential becomes deformed and shallower, further supporting the conclusion that mesonic states become less stable and ultimately melt in a hot QCD environment. Sec.~\ref{sec7} extends the analysis of mesonic melting to a finite baryon density environment by introducing a nonzero chemical potential in the holographic QCD model. The spectral functions reveal that as the chemical potential increases, the spectral peaks shift, broaden, and disappear more rapidly, indicating a faster melting. We also discuss the phase diagram and investigate how the meson melting feels the first order phase transition line between confinement-deconfinement at finite density. The concluding remarks are presented in Sec.~\ref{sec8}. 

 \section{Einstein-Maxwell-Dilaton action} 
 \label{sec2}
 
The prescription followed in this work is to take the EMD action and construct a self-consistent holographic model to describe heavy and exotic vector mesons. In a five-dimensional spacetime, the EMD action is written as 
\begin{equation}\label{emd_action}
    S_{\text{EMD}} = -\frac{1}{16 \pi G_5} \int d^5x \, \sqrt{-g} 
    \left[ 
        R - \frac{f(\phi)}{4} F_{MN}F^{MN} 
        - \frac{1}{2} \partial_M \phi \partial^M \phi 
        - V(\phi) 
    \right]\,,
\end{equation}
where $F_{MN} = \partial_M A_N - \partial_N A_M$ is the field strength, $f(\phi)$ is a gauge kinetic function that couples the dilaton and the gauge field $A_M$, $V(\phi)$ is the dilaton field potential, and $G_5$ is Newton's gravitational constant in five dimensions. Varying the action (\ref{emd_action}), one  finds the following equations of motion:
\begin{eqnarray}\label{einstein_eq}
    R_{MN} - \frac{1}{2} g_{MN} R &=& T_{MN}\,,\\
\label{maxwell_eq}
    \nabla_M \left[ f(\phi) F^{MN} \right] &=& 0\,,\\\label{KG_eq}
    \partial_M \left[ \sqrt{-g} \partial^M \phi \right]
    - \sqrt{-g} 
    \left[ 
        \frac{\partial V}{\partial \phi} 
        + \frac{F^2}{4} \frac{\partial f}{\partial \phi} 
    \right] &=& 0\,,
\end{eqnarray}
where $F^2=F_{MN}F^{MN}$. Eqs.~(\ref{einstein_eq}) are the Einstein's field equations, whereas Eqs.~(\ref{maxwell_eq}) represent Maxwell equations, and Eq.~(\ref{KG_eq}) is the Klein-Gordon equation for the dilaton field. The energy-momentum tensor  takes the form
\begin{equation}\label{EM_tensor}
    T_{MN} = \frac{1}{2} \left[ \partial_M \phi \partial_N \phi - \frac{1}{2} g_{MN} \left(\partial \phi \right)^2 - g_{MN} V(\phi) \right]+\frac{f(\phi)}{2} \left[ F_{MP} F^{\;P}_{N} - \frac{1}{4} g_{MN} F^2 \right]\,.
\end{equation}

To have a self-consistent holographic model, Eqs.~\eqref{einstein_eq} -- \eqref{KG_eq}, which are coupled, must be solved simultaneously. This can be done using the potential reconstruction method \cite{Mahapatra:2020wym,Mahapatra:2018gig,Priyadarshinee:2023cmi,Priyadarshinee:2021rch,Daripa:2024ksg}. To do so, we take the following Ans\"atze for the metric, gauge, and dilaton fields,
\begin{align}
    \begin{split}
            ds^2 &= \frac{L^2 e^{2A(z)}}{z^2} 
            \left(-g(z) \, dt^2 + \frac{dz^2}{g(z)} + dx_i dx^i \right)\,, \\
            A_M &= (A_t = A_t(z), \, A_i = 0, \, A_z = 0), \\
    \phi &= \phi(z)\,,
    \end{split}
    \label{ansatz_metric}
\end{align}
where $L$ is the AdS radius, $A(z)$ is the warp factor and $g(z)$ is the blackening function. It is assumed that the nonzero component of the gauge field $A_t$ and the dilaton field $\phi$ depend only on the holographic coordinate $z \in [0,z_h]$. Here $z_h$ is the horizon radius defined by $g(z_h)=0$. The strongly coupled gauge theory lives at the asymptotic boundary of spacetime, which is defined by the $z\to0$ limit. If the Ans\"atze (\ref{ansatz_metric}) is plugged into Eqs.~(\ref{einstein_eq}) --  (\ref{KG_eq}), one gets the following expressions
\begin{align}
    \label{EMD_eqs}
    0 &= \phi'' + \phi' \left(\frac{g'}{g}  + 3A' -\frac{3}{z}  \right) 
    - \frac{L^2 e^{2A}}{g z^2} \frac{\partial V}{\partial \phi} 
    + A_t^{\prime 2} \frac{z^2 e^{-2A}}{2gL^2} \frac{\partial f}{\partial \phi}\,, \\
    0 &= A_t'' + A_t' \left(\frac{f'}{f} + A' -\frac{1}{z} \right)\,, \\
    0 &= g'' + g' \left(  3A' -\frac{3}{z} \right) 
    - A_t^{\prime 2} \frac{z^2 e^{-2A} f}{L^2}\,,\label{2.9} \\
    0 &= A'' + \frac{g''}{6g} 
    + A' \left(\frac{3g'}{2g}  -\frac{6}{z}  \right) 
    - \frac{1}{z} \left(\frac{3g'}{2g} -\frac{4}{z}  \right) 
    + 3A' + V \frac{L^2 e^{2A}}{3gz^2}\,,\label{2.10} \\ 
    0 &= A'' - A' \left( A' -\frac{2}{z} \right) 
    + \frac{\phi^{\prime 2}}{6}\,.
\end{align}

Given the five equations and four unknown functions, we choose Eq. (\ref{EMD_eqs}) to act as a constraint. As demonstrated in \cite{Dudal:2017max}, treating \( A(z) \) and \( f(z) \) as arbitrary functions allows us to determine all unknown functions in closed form. In particular, by applying the following boundary conditions
\begin{equation}
    \lim_{z \to 0} g(z) = 1 \,, \qquad\qquad  \lim_{z \to z_h} g(z) = 0,  \qquad  \qquad \lim_{z \to z_h} A_t(z) = 0\,,
\end{equation}
we obtain, in terms of $A(z)$ and $f(z)$, the following solutions
\begin{align}
    \label{EMD_sol1}
    g(z) &= 1 - 
    \frac{\int_0^{z} dx \, x^3 e^{-3A(x)} 
    \int_{C}^{x} dy \, \frac{y e^{-A(y)}}{f(y)}}
    {\int_0^{z_h} dx \, x^3 e^{-3A(x)}
    \int_{C}^{x} dy \, \frac{y e^{-A(y)}}{f(y)}}\,, \\[10pt]
    \phi'(z) &= \sqrt{6 \left( A^{'2} - A'' - \frac{2A'}{z} \right)}\,, \\[10pt]
    A_t(z) &= 
            -\frac{1}
        {\sqrt{\int_0^{z} dx \, x^3 e^{-3A(x)}
        \int_{C}^{x} dy \, \frac{y e^{-A(y)}}{f(y)}}} 
    \int_{z_h}^{z} dx \, \frac{x e^{-A(x)}}{f(x)}\,, \\[10pt]
    V(z) &= - \frac{3g z^2 e^{-2A}}{L^2} 
    \bigg(
        A'' + A' \left[ 3A' + \frac{3g'}{g} - \frac{6}{z}  \right] 
        - \frac{1}{z} \left[\frac{3g'}{2g}  -\frac{4}{z} \right] 
        + \frac{g''}{6g} 
    \bigg)\,.
    \label{EMD_sol2}
\end{align}

The expressions (\ref{EMD_sol1}) -- (\ref{EMD_sol2}) can be put back into Eqs.~(\ref{einstein_eq}) -- (\ref{KG_eq}) to verify that the equations are satisfied and the system is mathematically self-consistent. In \cite{Dudal:2017max}, the integration constant $C$ can be fixed in terms of the chemical potential $\mu$. If $\mu$ is set to zero, we have the following analytical solutions to Eqs.~\eqref{EMD_sol1} -- \eqref{EMD_sol2}:
\begin{align}
    \label{EMD_muzero}
    g(z) &= 1 - 
    \frac{\int_0^{z} dx \, x^3 e^{-3A(x)}}
         {\int_0^{z_h} dx \, x^3 e^{-3A(x)}}\,, \\[10pt]
    \phi'(z) &= \sqrt{6 \left( A^{'2} - A'' - \frac{2A'}{z} \right)}\,, \\[10pt]
    A_t(z) &= 0\,, \\[10pt]
    V(z) &= - \frac{3g z^2 e^{-2A}}{L^2} 
    \bigg(
        A'' + A' \left[ 3A' + \frac{3g'}{g} - \frac{6}{z}  \right] 
        - \frac{1}{z} \left[\frac{3g'}{2g}  -\frac{4}{z} \right] 
        + \frac{g''}{6g} 
    \bigg)\,.
\end{align}

Without loss of generality, we take the AdS radius to be $L=1$ for the rest of this work.

Note that the above solution corresponds to a black hole having a horizon at $z=z_h$. Another solution to the EMD field equations also exists, which corresponds to thermal-AdS (without a horizon). This solution, which does not involve a black hole, can be obtained by taking the limit $z_h \rightarrow \infty $, resulting in $g(z)=1$. As we will discuss later on, the black hole and thermal-AdS phases exchange dominance in the free energy phase diagram and undergo a Hawking/Page type first-order phase transition as the temperature is varied. In particular, the thermal-AdS/black hole phases are favored at low/high temperatures respectively. In the dual field theory language, the thermal-AdS phase corresponds to a confined phase, whereas the black hole phase corresponds to the deconfined phase. In the following, we will fix our holographic model parameters at zero (or finite) temperature by working in the thermal-AdS (or the black hole) background.

To fully solve the EMD equations, the functions \(f(z)\) and \(A(z)\), which are initially arbitrary, must be specified. In the next section, we base our choice for \(f(z)\) on a pre-existing holographic QCD model from the literature \cite{MartinContreras:2020cyg, MartinContreras:2021bis} that features a deformed dilaton field designed to model heavy and exotic mesons. On the other hand, according to \cite{Dudal:2017max}, \(A(z)\) is chosen to be a simple function that, in this EMD prescription, can show a clear confinement-deconfinement phase in the boundary theory. In doing so, we hope to extract certain quantities (mass spectra and spectral functions) for several QCD heavy and exotic states. This will include the spectral functions at non-zero temperatures and density, from which behavior we can gather further insight into the melting of the states and, subsequently, of the QCD deconfinement phase transition in the quark-gluon plasma.

\section{Heavy and exotic mass spectra}
\label{sec3}
In this section, the mass spectroscopy of exotic mesons with heavy flavor will be scrutinized at zero temperature and zero chemical potential.

\subsection{Choice for $A(z)$ and $f(z)$}\label{modelsec}
The function \(f(z)\) will first be determined by analyzing the mass spectra of heavy and exotic vector mesons at zero temperature and chemical potential. As was already shown in other works \cite{Mamani:2022qnf, Dudal:2017max}, a particularly good choice for this function is 
\begin{equation}
    f(z) = e^{-\left[A(z)+B(z) \right]}\,,
\end{equation}
where $B(z)$ is chosen to match the mesonic mass spectra. In the context of the original soft-wall model \cite{Herzog:2006ra}, $B(z)$ is defined to be the dilaton field $\Phi(z)$: $B(z) \equiv \Phi(z) = (az)^2$, where $a$ is a constant that defines the slope of the Regge trajectory.\footnote{Note that the dilaton field $\Phi(z)$ used in the soft-wall model should not be confused with the dilaton field $\phi(z)$ used in our holographic model in Eq.~(\ref{emd_action}).}

Instead of using the original dilaton field profile, we follow here the approach presented in \cite{MartinContreras:2020cyg, MartinContreras:2021bis}, which introduces a non-quadratic dilaton field profile. Linear Regge trajectories effectively describe the hadronic mass spectrum in the light sector, where constituent quark masses are small. However, the linearity breaks down for hadrons containing a \(s\)-quark or heavier quarks. As argued in \cite{MartinContreras:2020cyg, MartinContreras:2021bis}, Bethe--Salpeter type analyzes indicate that heavy hadrons follow the scaling of trajectories as \(n^{3/2}\), where \(n\) is the excitation number.

This behavior suggests that the Regge trajectory depends on the mass of the constituent quarks, with linearity observed for light-flavored mesons and nonlinearity for heavier ones. Consequently, the dilaton field is also expected to vary with quark masses. 

Following \cite{MartinContreras:2020cyg}, we introduce the averaged constituent quark mass, $\bar{m}$. The input quark masses are taken to be effectively dressed by the QCD non-perturbative binding energy and are borrowed from \cite{Scadron:2006dy}: $m_u = 0.336 \, \text{GeV}$, $m_d = 0.340 \, \text{GeV}$, $m_s = 0.486 \, \text{GeV}$, $m_c = 1.550 \, \text{GeV}$, and $m_b = 4.730 \, \text{GeV}$.

Next, we introduce a modified dilaton profile of the form
\begin{equation}
    \Phi(z) = (\kappa z)^{2-\alpha}\,,
\end{equation}
where the exponent $\alpha$ encodes the effects of the constituent quark mass. For $\alpha = 0$, one recovers the original quadratic dilaton and the corresponding linear Regge trajectory. In \cite{MartinContreras:2020cyg} both $\alpha$ and $\kappa$ vary with the average quark masses
\begin{align}
    \alpha(\bar{m}) & \equiv \alpha = a_\alpha - b_\alpha e^{-c_\alpha \bar{m}^2}\,, \\
    \kappa(\bar{m}) & \equiv \kappa = a_\kappa - b_\kappa e^{-c_\kappa \bar{m}^2}\,,
\label{kappam}\end{align}
with parameters $a_\alpha = 0.054$, $b_\alpha = 0.8484$, $c_\alpha = 0.4233$, $a_\kappa = 15.2085$, $b_\kappa = 14.8082$ and $c_\kappa = 0.0524$ \cite{MartinContreras:2020cyg}. Larger values of $\alpha$ and $\kappa$ correspond to heavier states. Thus, if we know (or model) a meson with some combination of constituent quarks, we can find the $\alpha$ and $\kappa$ parameters that suit the problem at hand best. 

In addition to the non-quadratic behavior introduced in \cite{MartinContreras:2020cyg}, \cite{MartinContreras:2021bis} presents small-$z$ corrections to $\Phi(z)$. Consequently, we adopt the dilaton profile proposed in \cite{MartinContreras:2021bis} and identify their $\Phi(z)$ dilaton field with our $B(z)$ function, which is incorporated into the arbitrary function $f(z)$ as
\begin{equation}\label{f_definition}
f(z) = e^{-\left[A(z)+B(z) \right]} = e^{-\left[A(z) + (\kappa z)^{2-\alpha} + Mz + \tanh{\left(\frac{1}{Mz} - \frac{\kappa}{\sqrt{\Gamma}}\right)}\right]} \,.
\end{equation}
in which $Mz$ and $\tanh{\left(\frac{1}{Mz} - \frac{\kappa}{\sqrt{\Gamma}}\right)}$ are the small-$z$ corrections to $\Phi(z)$, see also \cite{Braga:2017bml}. 

The other term, $A(z)$, can be chosen as a simple quadratic function, whose consequences were studied in \cite{Dudal:2017max}. In this model, such $A(z)$ can show the confined-deconfined phases in the boundary theory, mimicking QCD. Therefore here we adopt the following $A(z)$:
\begin{equation}
    A(z) = -\frac{\mathcal{C}}{8} z^2 \,,
\end{equation}
where $\mathcal{C} = 1.16 \, \text{GeV}^2$ is fixed to yields the Hawking--Page geometric transition at around $T \sim 270 \, \text{MeV}$ at zero chemical potential \cite{Dudal:2017max}, based on the lattice estimate of \cite{Lucini:2003zr}.

\subsection{Equation of motion for the vector mesons}
Given that all parameters and functions are fixed, we can first compute the mesonic mass spectra at zero temperature. A five-dimensional gauge field describes the vector mesons in the dual field theory. From the EMD action (\ref{emd_action}), we can write
\begin{equation} \label{Meson_Action}
S = -\frac{1}{16 \pi G_5} \int d^5x \sqrt{-g} \frac{f(\phi)}{4} F_{MN}F^{MN}\,.
\end{equation}

Varying the action (\ref{Meson_Action}) we find the following equations of motion
\begin{equation} \label{EOM_1}
\partial_M \left( \sqrt{-g} f(\phi) F^{MN} \right) = 0\,.
\end{equation}

In the zero temperature limit, $g(z) \rightarrow 1$ and the spacetime interval becomes
\begin{align}
    \label{metric_T0}
    ds^2 &= \frac{L^2 e^{2A(z)}}{z^2} 
    \left( 
        -\, dt^2 + dz^2 + dx_i dx^i 
    \right)\,.
\end{align}

By plugging (\ref{metric_T0}) into (\ref{EOM_1}) and choosing the radial gauge $A_z = 0$, supplemented with the transversal Landau (Lorenz) gauge $\partial_\nu A^\nu = 0$, Eq.~(\ref{EOM_1}) turns into
\begin{equation} \label{EOM_2}
\frac{z e^{-A(z)}}{L f(z)} \partial_z \left(\frac{L e^{A(z)} f(z)}{z} \partial_z A^\mu \right) + \Box A^\mu = 0\,,
\end{equation}
which can be led into a Schr\"odinger-like form by first implementing a Fourier transform 
\begin{equation} 
\label{fourier}
A^\mu(z, x^\nu) = \int \frac{d^4k}{(2\pi)^4} e^{i k_\alpha x^\alpha} A^\mu(z, k)\,,
\end{equation}
and performing the following change of variables   
\begin{align} \label{B_def}
    A^\mu &= e^{-H} \psi^\mu\,, \\
    H &= \frac{1}{2} \log\left[\frac{L e^{A(z)} f(z)}{z}\right]\,.
\end{align}

This turns Eq.~(\ref{EOM_2}) into
\begin{equation} \label{Mass_Spectra_T0}
-\partial_z^2 \psi(z) + U(z) \psi(z) = m_n^2 \psi(z)\,,
\end{equation}
for 
\begin{equation} \label{Potential_T0}
U(z) = \left(\partial_z \left[\frac{1}{2} \log\left(\frac{L e^{A(z)} f(z)}{z}\right)\right]\right)^2 
+ \partial_z^2 \left[\frac{1}{2} \log\left(\frac{L e^{A(z)} f(z)}{z}\right)\right]\,,
\end{equation}
where we identify the eigenmasses with the relation $k^2=m_{n}^{2}=-\omega^2+p^2$. For a massive perturbation in the bulk theory, the AdS/CFT dictionary dictates that a mass term of the form $M^2 L^2 = \left(\Delta - S \right) \left(\Delta + S - 4 \right)$, where $S$ is the spin and the angular momentum taken to be zero, must be added to the potential $U(z)$. Importantly, $\Delta$ is the scale dimension that determines the scaling behavior of the bulk field at the boundary \cite{Witten:1998qj}. From the quantum boundary theory perspective, $\Delta$ is the scaling dimension of the operator that creates hadrons. We can see right away that for isovector mesons, such as the charmonium (\( c\bar{c} \))  and bottomonium (\( b\bar{b} \)), $M^2 L^2 = 0$ as $\Delta=3$ and $S=1$.

A notable application of this nonquadratic dilaton model is the exploration of hadronic states that deviate from the conventional $q\bar{q}$ framework. As noted above, in holographic QCD, the distinction between different types of hadrons is captured by the scaling dimension $\Delta$ of the operator responsible for their creation. Consequently, the model provides a pathway to study exotic hadrons---those that do not conform to the standard $q\bar{q}$ structure---by identifying the appropriate scaling dimension $\Delta$ and estimating the parameter $\bar{m}$. Depending on the quark and/or meson content of the bound state, it is clear there can be a degeneracy in terms of $\Delta$. Even so, the estimation of the corresponding $\bar{m}$ can be achieved through a weighted averaging procedure, as detailed in \cite{MartinContreras:2020cyg}
\begin{equation}\label{exp13}
\bar{m}_{\mathrm{multi}} = \sum_{i=1}^{N} \left( P_i^{\mathrm{quark}} \bar{m}_i^{\mathrm{quark}} + P_i^{\mathrm{meson}} m_i^{\mathrm{meson}} \right)\,,\qquad \sum_{i=1}^{N} \left( P_i^{\mathrm{quark}}  + P_i^{\mathrm{meson}}  \right)=1\,.
\end{equation}

The exotic state is then modeled as a collection of \( N \) constituent quarks and/or mesons. Each constituent is assigned a weight \( P_i \) associated with its mass \( m_i \). These masses are then utilized to determine the constants \( \alpha \) and \( \kappa \). We have per case also included the corresponding weight factors and scale dimension $\Delta$ that we borrowed from \cite{MartinContreras:2020cyg}. In this work, we consider the exotic states $Z_c$ and $\pi_1$. $Z_c$ is modeled as a tetraquark which in turn is constructed as a ``hadronic molecule", two mesons bounded by a residual colorless QCD interaction, formed by a $J/\psi$ and a $\rho$: $\bar{m}_{tetra} = 0.283 m_{J/\psi} + 0.717 m_{\rho} $. Whereas $\pi_1$ is modeled as a ``hybrid meson", a state consisting of valence quarks and a constituent gluon: $\bar{m}_{hybrid} = 0.497 m_{u} + 0.497 m_{d} + 0.006 m_{G} $, where $m_G$ is the constituent gluon mass (taken to be $m_G=0.7\, \text{GeV}$ \cite{Hou:2001ig}). Both of these exotic states are massive perturbations with $\Delta=5$ for the hybrid meson and $\Delta=6$ for the tetraquark.

Below, we provide some Tables~\ref{tcharm} -- \ref{thybrid} with the heavy mesons' masses, including charmonium and bottomonium, as well the considered exotic states. The masses are numerically obtained by solving the eigenvalue problem given by Eq.~(\ref{Mass_Spectra_T0}) for the shown parameters.

\begin{table}[h!]
\centering
\begin{tabular}{||c||c|c|c|c||}
    \hline\hline
    \multicolumn{5}{|c|}{Charmonium $J/\psi$, $I^G (J^{PC}) = 0^+(1^{--})$} \\
    \hline\hline
    $n$ & State & $m_{\text{exp}}$ (\text{MeV}) & $m_{\text{th}}$ (\text{MeV}) & Relative error $\delta M$ \\
    \hline
    1 & $J/\psi$ & $3096.900 \pm 0.006$ & $3140.1$ & $1.39 \%$ \\\hline
    2 & $\psi(2S)$ & $3686.097 \pm 0.011$ & $3656.9$ & $0.79 \%$ \\\hline
    3 & $\psi(4040)$ & $4039.6 \pm 4.3$ & $4055.8$ & $0.40 \%$ \\\hline
    \hline
    \multicolumn{5}{||c||}{Experimental linear fit: $M^2 = 3.364(1.914 + n)$, $R^2 = 0.999508$} \\\hline
    \multicolumn{5}{||c||}{Experimental nonlinear fit: $M^2 = 10.947(-0.282 + n)^{0.399}$, $R^2 = 1.$} \\\hline
    \multicolumn{5}{||c||}{Theoretical linear fit: $M^2 = 3.295(2.015 + n)$, $R^2 = 0.999942$} \\\hline
    \multicolumn{5}{||c||}{Theoretical nonlinear fit: $M^2 = 7.001(0.685 + n)^{0.654}$, $R^2 = 1.$} \\\hline
\end{tabular}
\caption{\footnotesize Mass spectrum and Regge trajectories of the $J/\psi$ charmonium for $\kappa = 1.8 \, \text{GeV}$, $M = 1.7 \, \text{GeV}$, $\sqrt{\Gamma} = 0.53 \, \text{GeV}$, and $\alpha = 0.54$.\label{tcharm}}
\end{table}

\begin{table}[h!]
\centering
\begin{tabular}{||c||c|c|c|c||}
    \hline\hline
    \multicolumn{5}{|c|}{Bottomonium $\Upsilon$, $I^G (J^{PC}) = 0^+(1^{--})$} \\
    \hline\hline
    $n$ & State & $m_{\text{exp}}$ (\text{MeV}) & $m_{\text{th}}$ (\text{MeV}) & Relative error $\delta M$ \\
    \hline
    1 & $\Upsilon (1S)$ & $9460.4 \pm 0.09\pm0.04$ & $9506.5$ & $0.49 \%$ \\\hline
    2 & $\Upsilon (2S)$ & $10023.4 \pm 0.5$ & $9892.9$ & $1.30 \%$ \\\hline
    3 & $\Upsilon (3S)$ & $10355.1 \pm 0.5$ & $10227.2$ & $1.24 \%$ \\
    \hline\hline
    \multicolumn{5}{||c||}{Experimental linear fit: $M^2 = 8.864(9.175 + n)$, $R^2 = 0.9999$} \\\hline
    \multicolumn{5}{||c||}{Experimental nonlinear fit: $M^2 = 91.165(-0.113 + n)^{0.153}$, $R^2 = 1.$} \\\hline
    \multicolumn{5}{||c||}{Theoretical linear fit: $M^2 = 7.111(11.727 + n)$, $R^2 = 0.999997$} \\\hline
    \multicolumn{5}{||c||}{Theoretical nonlinear fit: $M^2 = 49.120(3.559 + n)^{0.402}$, $R^2 = 1.$} \\\hline
\end{tabular}
\caption{\footnotesize Mass spectrum and Regge trajectories of the $\Upsilon$ bottomonium for $\kappa = 9.9 \, \text{GeV}$, $M = 2.74 \, \text{GeV}$, $\sqrt{\Gamma} = 1.92 \, \text{GeV}$, and $\alpha = 0.863$.\label{tbottom}}
\end{table}

\begin{table}[h!]
\centering
\begin{tabular}{||c||c|c|c|c||}
    \hline\hline
    \multicolumn{5}{|c|}{Tetraquark $Z_c$, $I^G (J^{PC}) = 1^+(1^{+-})$} \\
    \hline\hline
    $n$ & State & $m_{\text{exp}}$ (\text{MeV}) & $m_{\text{th}}$ (\text{MeV}) & Relative error $\delta M$ \\
    \hline
    1 & $Z_c (3900)$ & $3887.2 \pm 2.3$ & $3890.1$ & $0.07 \%$ \\\hline
    2 & $Z_c (4200)$ & $4196 \substack{+35 \\ -32}$ & $4189.5$ & $0.15 \%$ \\\hline
    3 & $Z_c (4430)$ & $4478 \substack{+15 \\ -18}$ & $4444.6$ & $0.74 \%$ \\
    \hline\hline
    \multicolumn{5}{||c||}{Experimental linear fit: $M^2 = 2.471(5.112 + n)$, $R^2 = 1.$} \\\hline
    \multicolumn{5}{||c||}{Experimental nonlinear fit: $M^2 = 3.554(4.234 + n)^{0.874}$, $R^2 = 1.$} \\\hline
    \multicolumn{5}{||c||}{Theoretical linear fit: $M^2 = 2.311(5.564 + n)$, $R^2 = 0.999992$} \\\hline
    \multicolumn{5}{||c||}{Theoretical nonlinear fit: $M^2 = 6.836(2.665 + n)^{0.612}$, $R^2 = 1.$} \\\hline
\end{tabular}
\caption{\footnotesize Mass spectrum and Regge trajectories of the $Z_c$ tetraquark, for $\kappa = 1.75 \, \text{GeV}$, $M = 1.44 \, \text{GeV}$, $\sqrt{\Gamma} = 0.30 \, \text{GeV}$, and $\alpha = 0.539 $. \label{ttetra}}
\end{table}

\begin{table}[h!]
\centering
\begin{tabular}{||c||c|c|c|c||}
    \hline\hline
    \multicolumn{5}{|c|}{Hybrid meson $\pi_1$, $I^G (J^{PC}) = 0^-(1^{+-})$} \\
    \hline
    $n$ & State & $m_{\text{exp}}$ (\text{MeV}) & $m_{\text{th}}$ (\text{MeV}) & Relative error $\delta M$ \\
    \hline
    1 & $\pi_1 (1400)$ & $1354 \pm 25$ & $1354.7$ & $0.05 \%$ \\\hline
    2 & $\pi_1 (1600)$ & $1660 \substack{+15 \\ -11}$ & $1618.1$ & $2.52 \%$ \\\hline
    3 & $\pi_1 (2015)$ & $2014 \pm 20 \pm 16$ & $1849.9$ & $8.15 \%$ \\
    \hline\hline
    \multicolumn{5}{||c||}{Experimental linear fit: $M^2 = 1.111(0.593 + n)$, $R^2 = 0.99913$} \\\hline
    \multicolumn{5}{||c||}{Experimental nonlinear fit: $M^2 = 1.214(0.484 + n)^{0.953}$, $R^2 = 1.$} \\\hline
    \multicolumn{5}{||c||}{Theoretical linear fit: $M^2 = 0.793(1.308 + n)$, $R^2 = 0.999997$} \\\hline
    \multicolumn{5}{||c||}{Theoretical nonlinear fit: $M^2 = 0.816(1.268 + n)^{0.987}$, $R^2 = 1.$} \\\hline
\end{tabular}
\caption{\footnotesize Mass spectrum and Regge trajectories of the $\pi_1$ hybrid meson, for $\kappa = 0.468 \, \text{GeV}$, $M = 0.20 \, \text{GeV}$, $\sqrt{\Gamma} = 0.12 \, \text{GeV}$, and $\alpha = 0.034$. \label{thybrid}}
\end{table}

From the shown Tables \ref{tcharm} -- \ref{thybrid}, we can conclude that a nonlinear Regge trajectory as parameterized in \cite{MartinContreras:2020cyg}, 
\begin{align}\label{nu_index}
    M_n^2 &= a(n+b)^\nu\,,\\
    \nu (\Bar{m}) &\equiv \nu = a_\nu + b_\nu e^{-c_\nu \bar{m}}\,,
\end{align}
tends to better fit the experimental masses. Our computed masses, in general, agree well with the experimental data as well. We stress here again that our predictions for these masses, and associated nonlinear Regge behavior, are a consequence of the modeling in Subsection~\ref{modelsec}.

\begin{figure}[htb!]
    \centering
    \begin{minipage}[t]{0.49\textwidth}
        \centering
        \includegraphics[width=\textwidth]{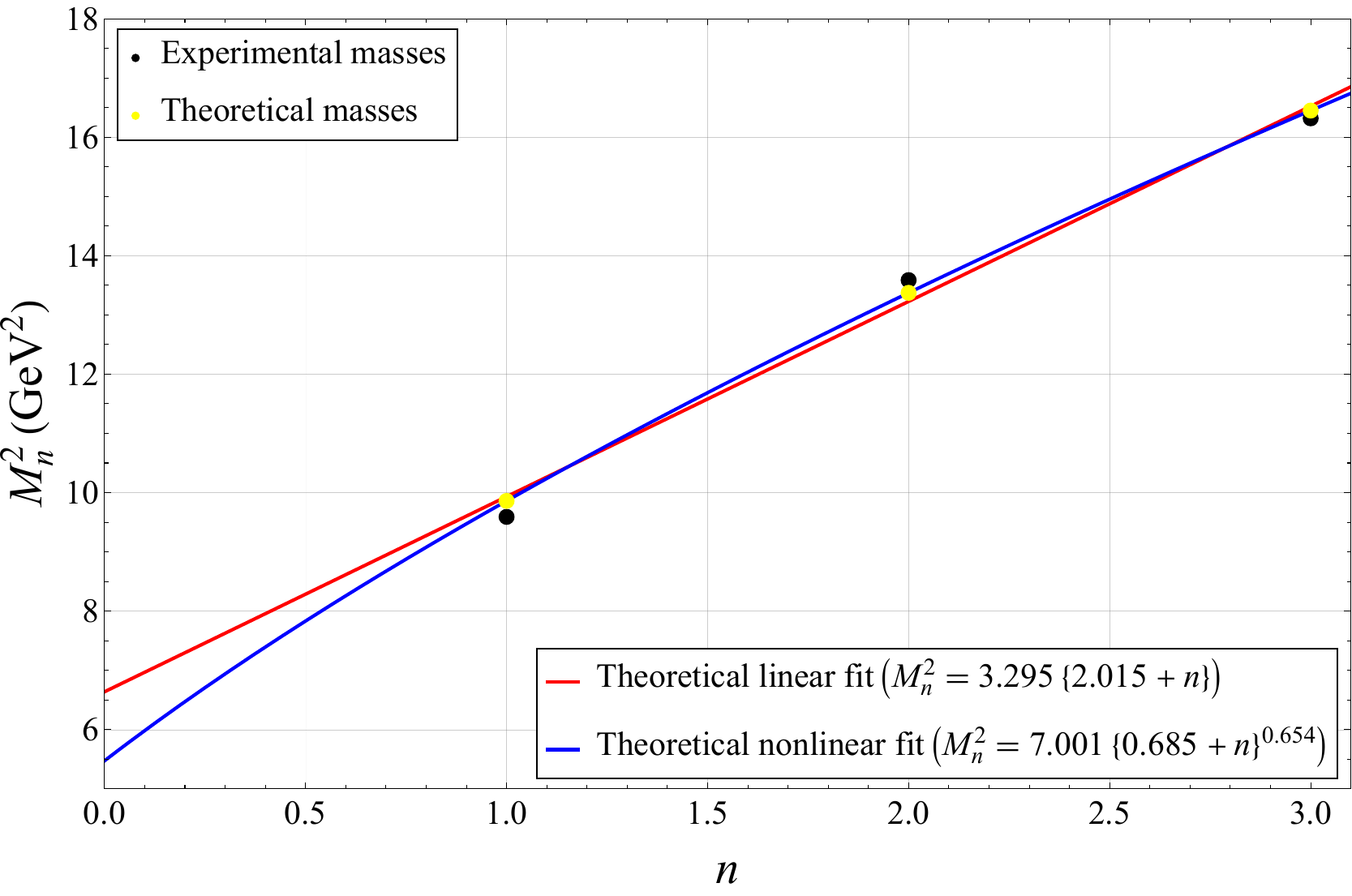}
        \caption{\small Linear and nonlinear fits for the charmonium.}
        \label{fig:RTcharm}  
    \end{minipage}%
    \hfill
    \begin{minipage}[t]{0.49\textwidth}
        \centering
        \includegraphics[width=\textwidth]{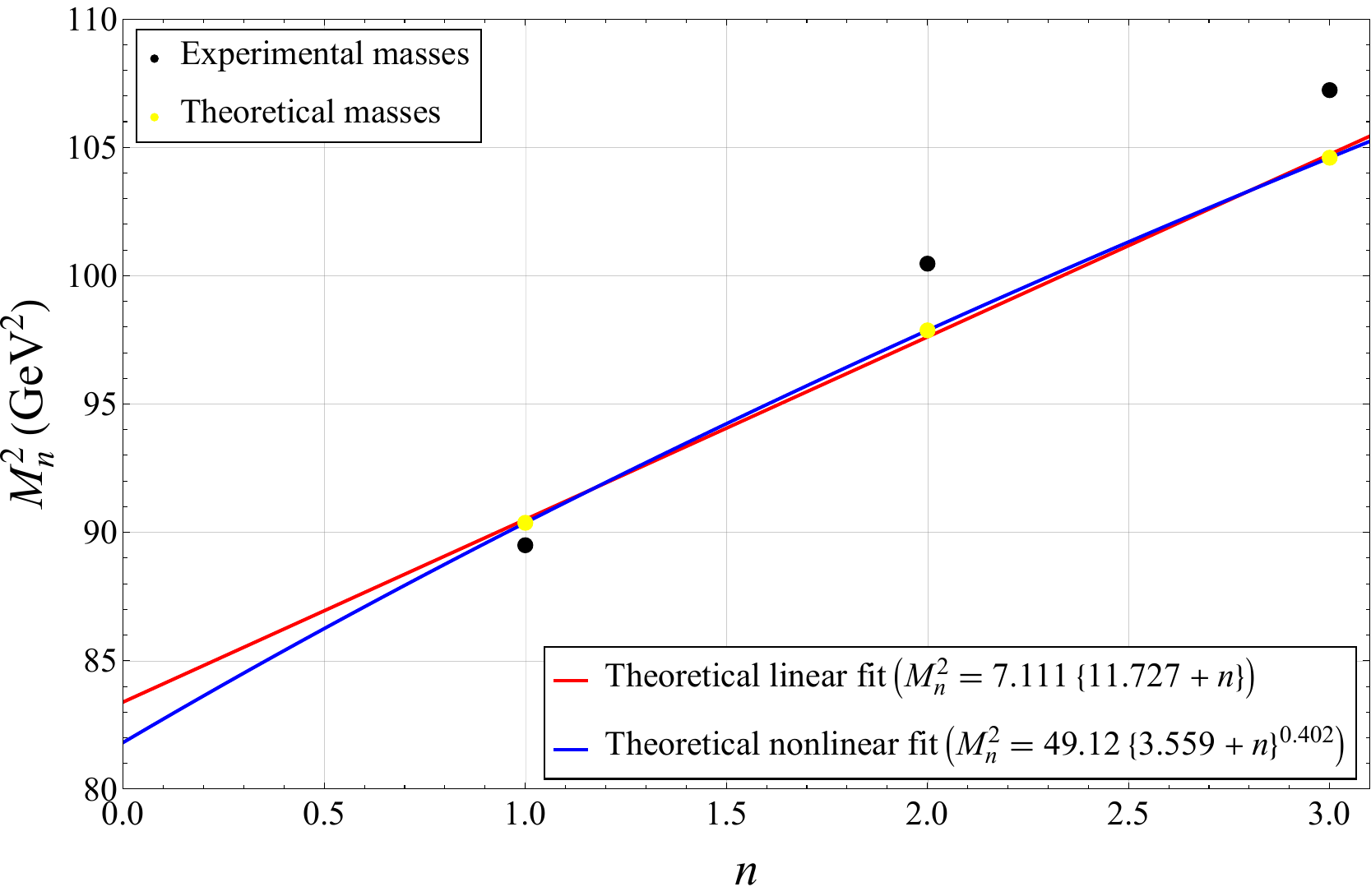}
        \caption{\small Linear and nonlinear fits for the bottomonium.}
        \label{fig:RTbottom} 
    \end{minipage}
\end{figure}

\begin{figure}[htb!]
    \centering
    \begin{minipage}[t]{0.49\textwidth}
        \centering
        \includegraphics[width=\textwidth]{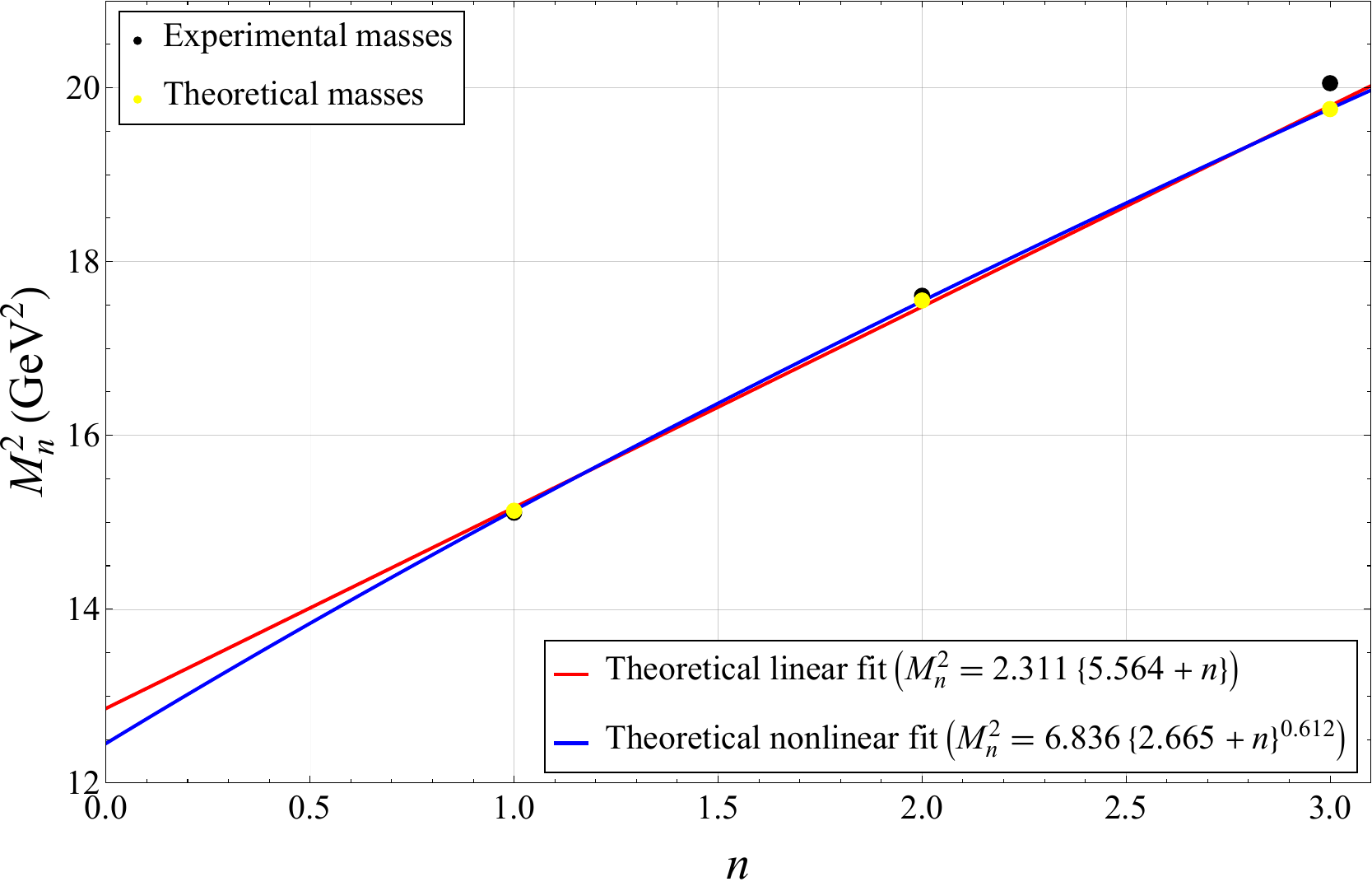}
        \caption{\small Linear and nonlinear fits for the tetraquark.}
        \label{fig:RTtetra}  
    \end{minipage}%
    \hfill
    \begin{minipage}[t]{0.49\textwidth}
        \centering
        \includegraphics[width=\textwidth]{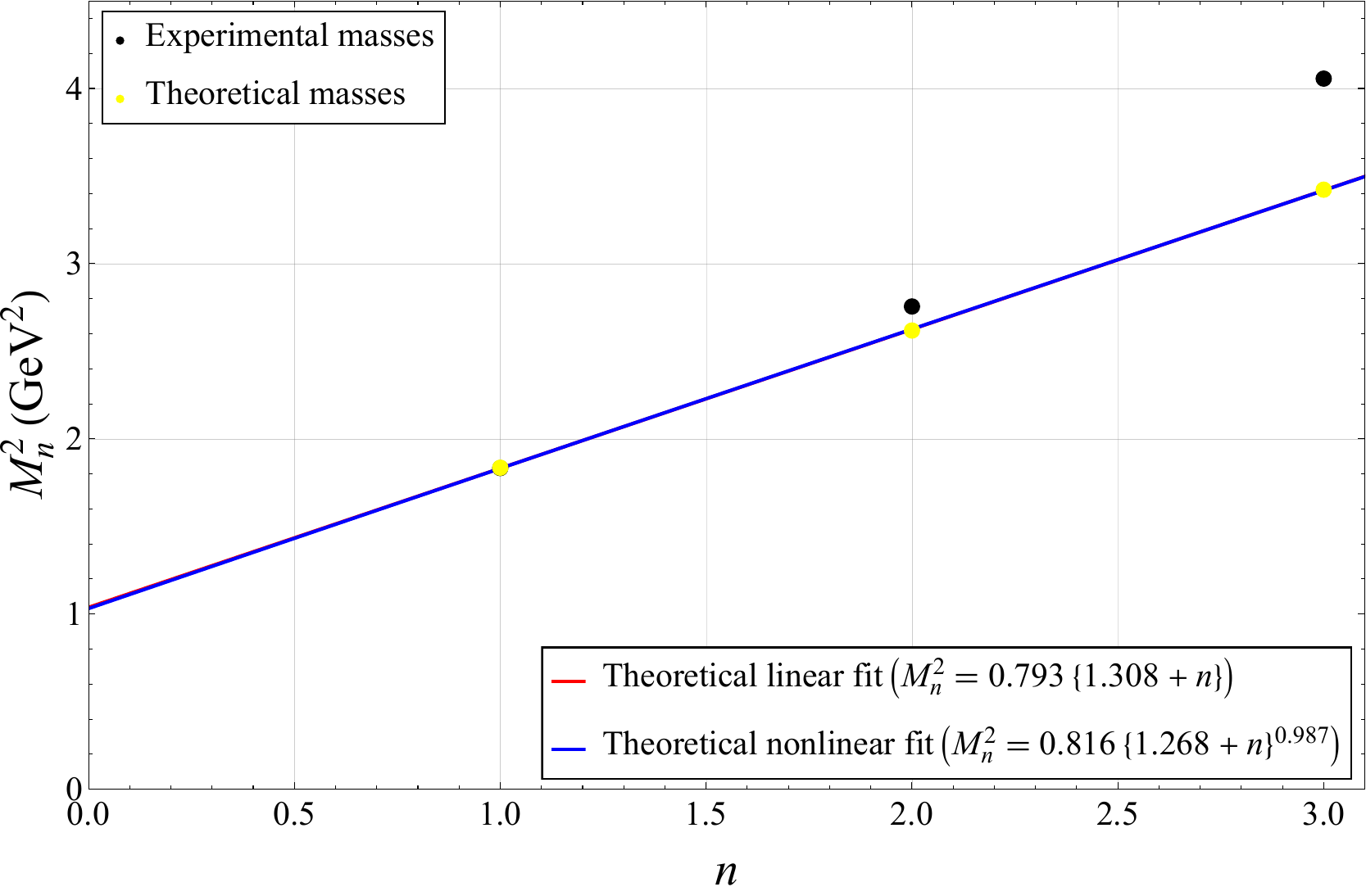}
        \caption{\small Linear and nonlinear fits for the hybrid meson.}
        \label{fig:RT hybrid}  
    \end{minipage}
\end{figure}

In the Figs.~\ref{fig:RTcharm} -- \ref{fig:RT hybrid} we can better visualize the difference between a linear and a nonlinear Regge trajectory. We note that as the particle mass decreases, like the considered hybrid meson, the linear and nonlinear trajectories become nearly identical. 

We conclude then that the resulting equation of motion (\ref{Mass_Spectra_T0}) therefore provides a mathematically self-consistent way of formulating the approximation of the mass spectra for heavy mesons and certain exotic states. This can be seen as the numerical results obtained for the overall masses are in good agreement with the experimental values and those reported in \cite{MartinContreras:2020cyg}. 

The effective potential \( U(z) \), shown in (\ref{Potential_T0}), for the considered states is presented below, Figs.~\ref{fig:T0potcharm} and \ref{fig:T0pottetra}, to showcase the dynamics of these hadronic structures within the holographic framework. Derived from the Schrödinger-like equation, this potential provides insights into the behavior, stability, and mass spectra of tetraquark states under the chosen model parameters. In the next sections we see that this potential is deformed by increasing the temperature $T$ and chemical potential $\mu$.

\begin{figure}[htb!]
    \centering
    \begin{minipage}[t]{0.49\textwidth}
        \centering
        \includegraphics[width=\textwidth]{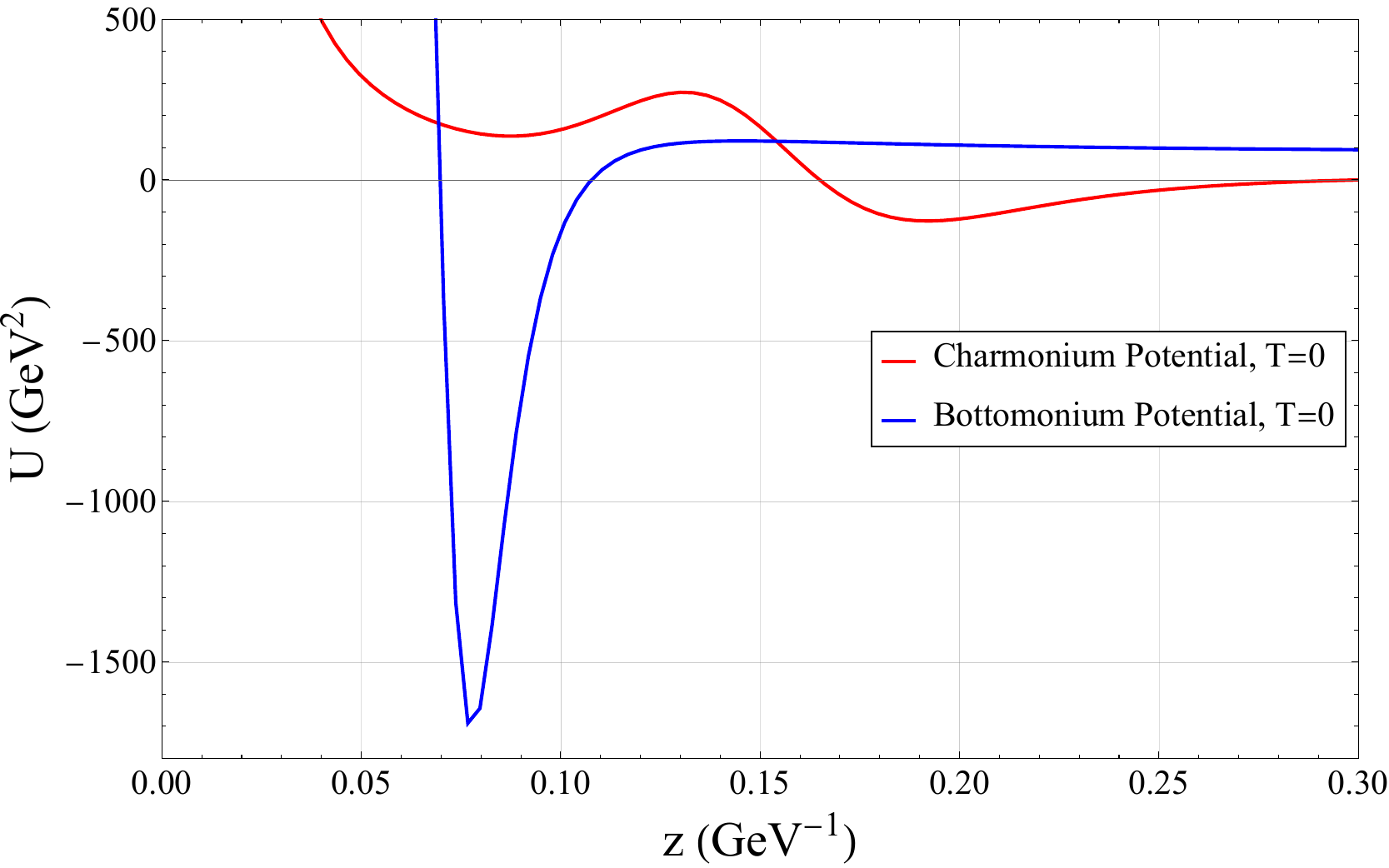}
        \caption{\small Charmonium and bottomonium potentials at $T=0$.}
        \label{fig:T0potcharm}
    \end{minipage}%
    \hfill
    \begin{minipage}[t]{0.49\textwidth}
        \centering
        \includegraphics[width=\textwidth]{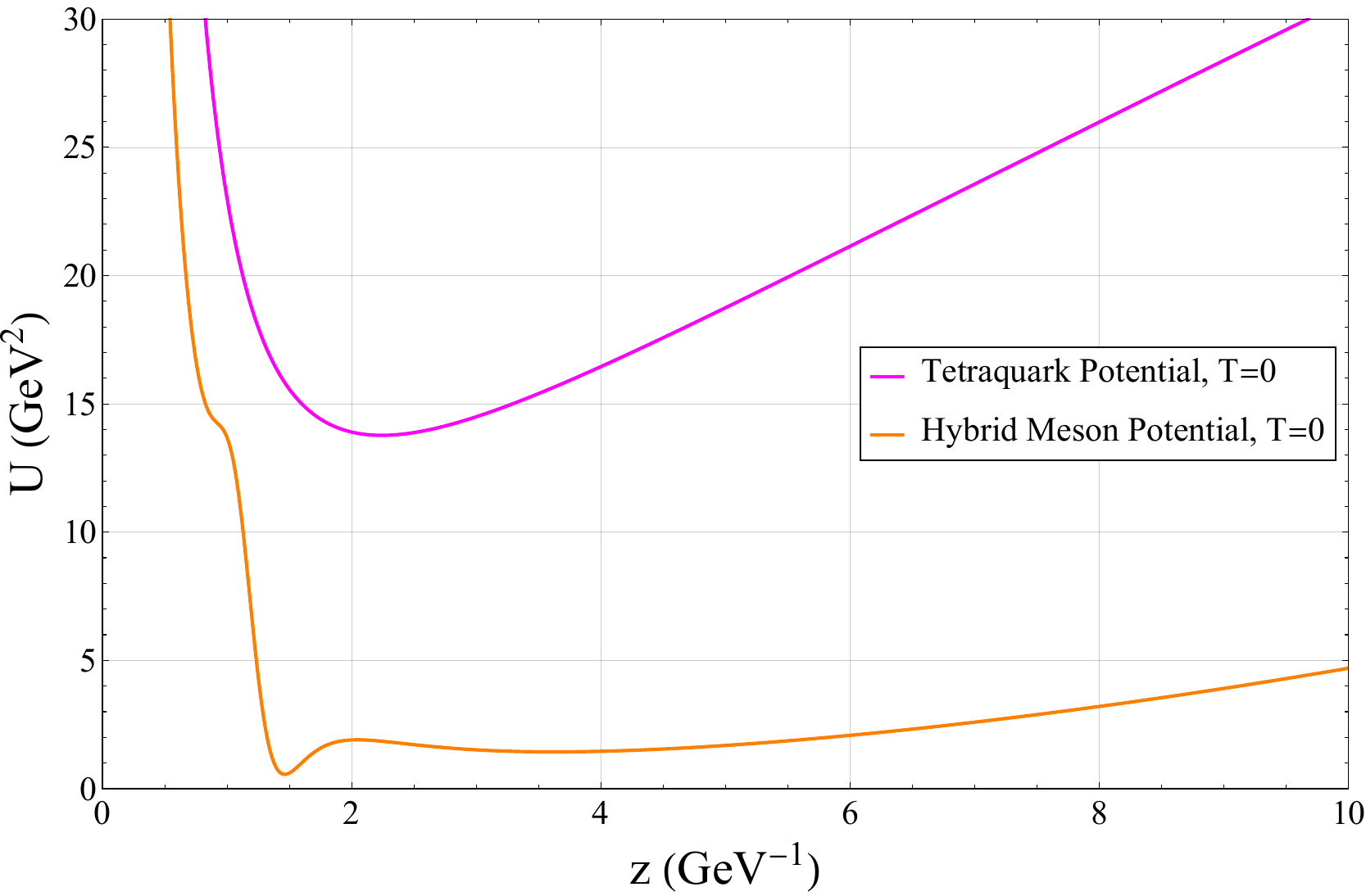}
        \caption{\small Tetraquark and hybrid meson potentials at $T=0$.}
        \label{fig:T0pottetra}
    \end{minipage}
\end{figure}

\begin{figure}[htb!]
    \centering
    \begin{minipage}[t]{0.49\textwidth}
        \centering
        \includegraphics[width=\textwidth]{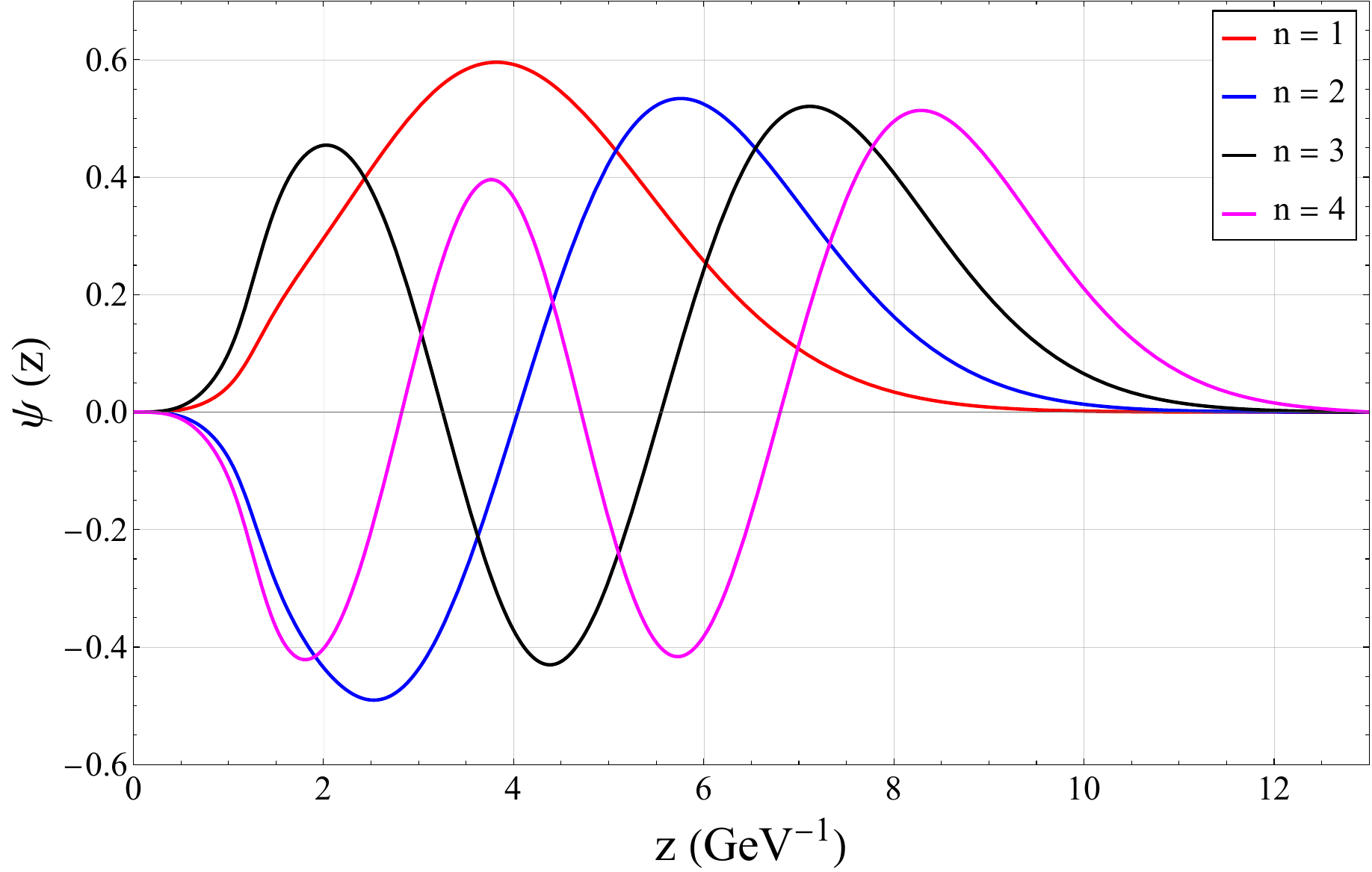}
        \caption{\small Hybrid meson wavefunctions.}
        \label{fig:T0wavehybrid}
    \end{minipage}%
    \hfill
    \begin{minipage}[t]{0.49\textwidth}
        \centering
        \includegraphics[width=\textwidth]{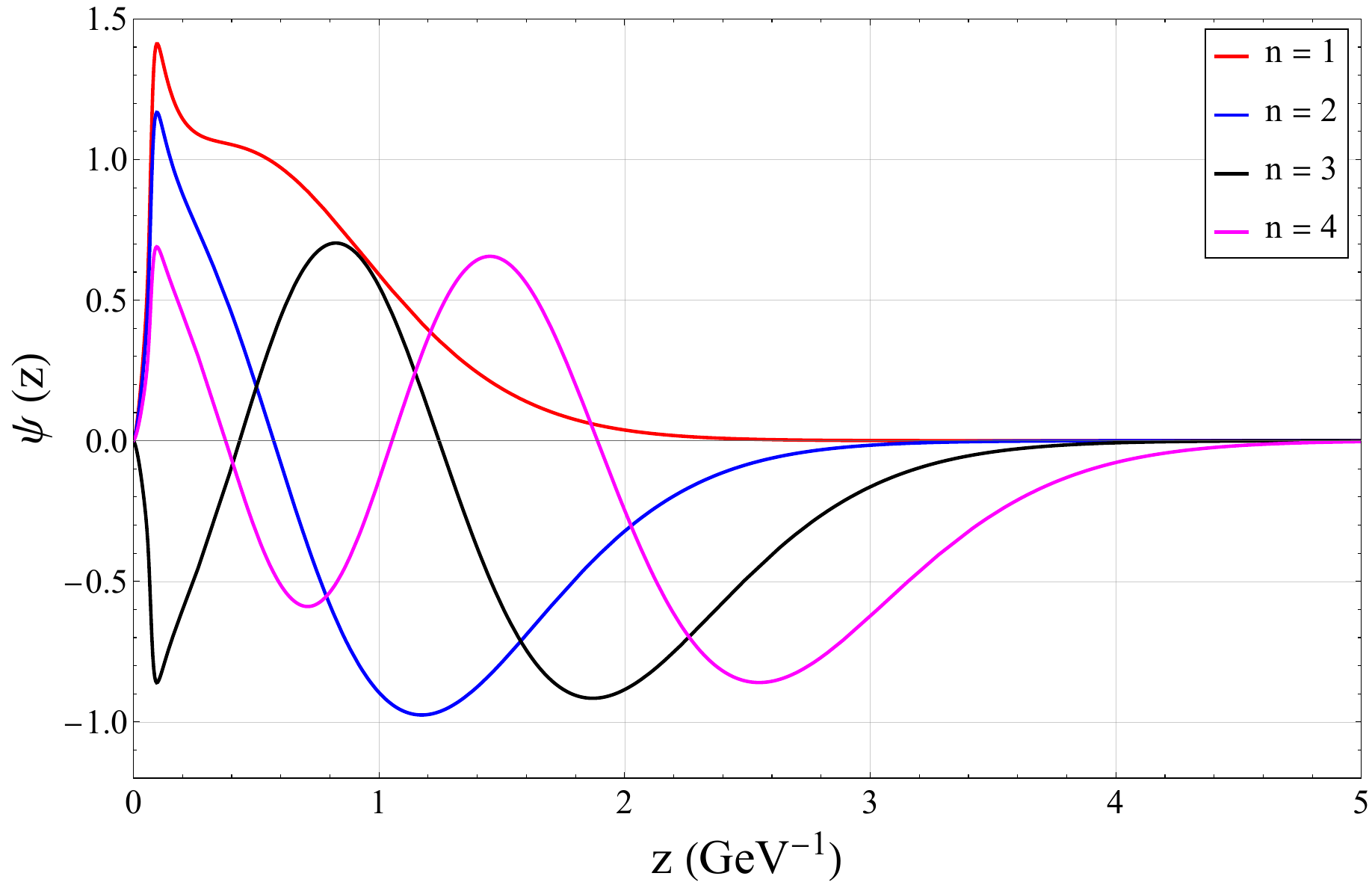}
        \caption{\small Bottomonium wavefunctions.}
        \label{fig:T0wavebottom}
    \end{minipage}
\end{figure}

Below we also plot the associated wavefunctions, Figs.~\ref{fig:T0wavehybrid} and \ref{fig:T0wavebottom}, for the hybrid meson and bottomonium states, found after solving Eq.~(\ref{Mass_Spectra_T0}) numerically.

We can see right away that as the values of the model parameters increase (low values for the hybrid meson and high values for bottomonium), the wavefunctions tend to go to zero faster and get more distorted near the origin. In the numerical solution to the eigenvalue problem (\ref{Mass_Spectra_T0}), we employed Dirichlet boundary conditions, where we set the wave function to vanish at $z=0$ and $z=\infty$.

\section{Metric and equation of motion at finite temperature}
\label{sec4}

To analyze a system at finite temperature (where we set \( \mu = 0 \) in this section), the blackening function \( g(z) \) must first be determined for the chosen profile \( A(z) = -\frac{\mathcal{C}}{8} z^2 \). From Eq.~(\ref{EMD_muzero}), it is evident that \( f(z) \) does not contribute to \( g(z) \). By evaluating the integral, we obtain
\begin{equation} \label{g_T_mu_0}
g(z) = 1 + \frac{e^{\frac{3 \mathcal{C} z^2}{8}} \left(8 - 3 \mathcal{C} z^2\right) - 8}{e^{\frac{3 \mathcal{C} z_h^2}{8}} \left(3 \mathcal{C} z_h^2-8 \right) + 8}\,.
\end{equation}

The black hole temperature is calculated as
\begin{equation} \label{BH_T_mu0}
T = -\frac{g'(z_h)}{4 \pi} = \frac{9  z_h^3 \mathcal{C}^2 e^{\frac{3 \mathcal{C} z_h^2}{8}}}{16 \pi \left[8 + e^{\frac{3 \mathcal{C} z_h^2}{8}} \left(3 \mathcal{C} z_h^2-8\right)\right]}\,.
\end{equation}

\begin{figure}[h!]
    \centering
    \includegraphics[width=0.75\textwidth]{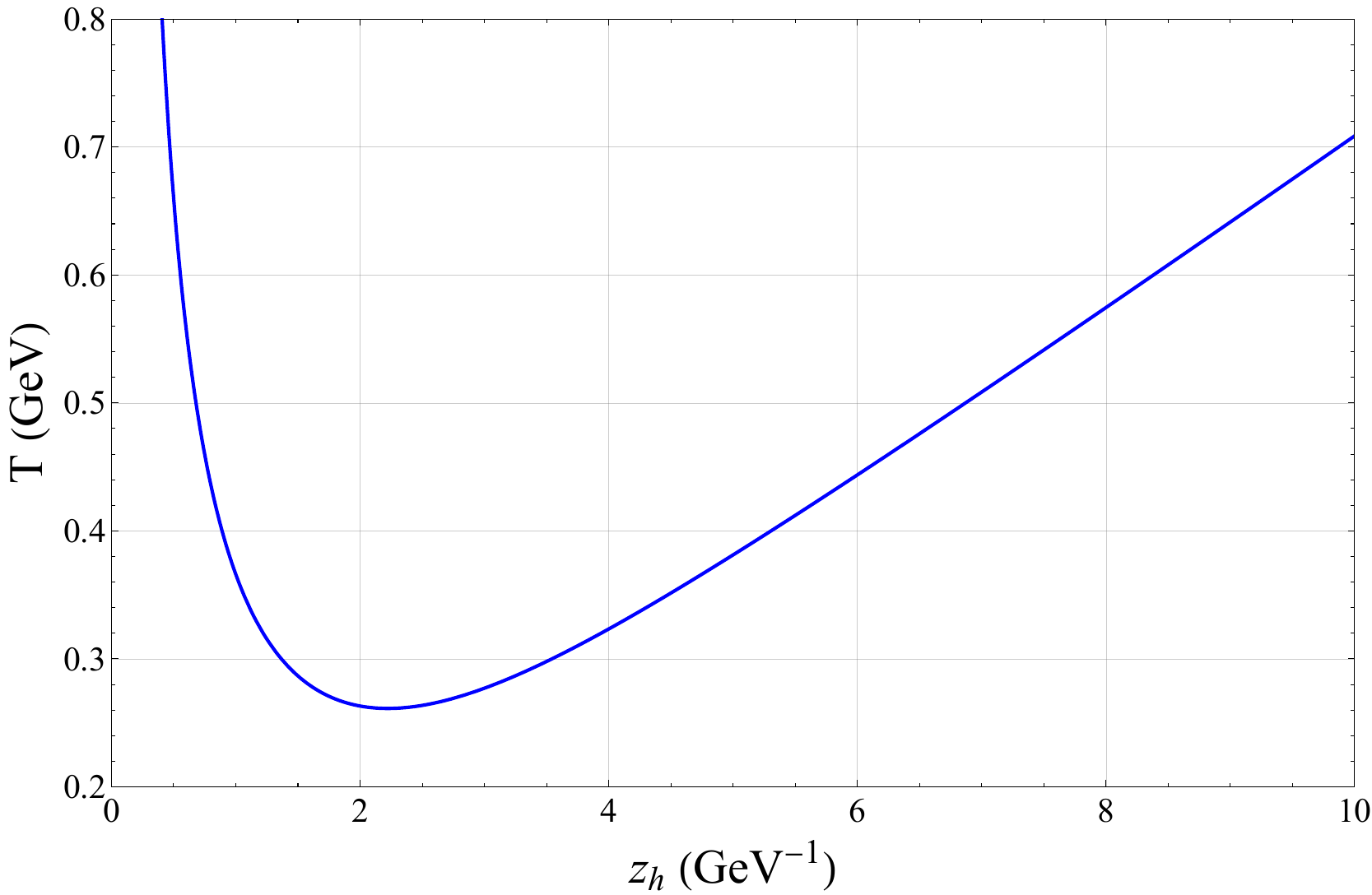}
    \caption{\small The Hawking temperature \( T \) as a function of horizon radius \( z_h \).}
    \phantomsection
    \label{fig:Tempmu0}
\end{figure}

We present below, Fig.~\ref{fig:Tempmu0}, the Hawking temperature $T$ as a function of the black hole horizon $z_h$ for the model parameter $\mathcal{C} = 1.16 \, \text{GeV}^2$. Here, the thermal AdS phase exists at all temperatures. However, there is a minimum temperature below which a black hole solution does not exist. Above this minimum temperature, two black hole solutions exist: a large black hole and a small one. The large black hole (small $z_h$) has a temperature that decreases as the inverse horizon radius $z_h$ increases. It is thermodynamically stable due to its positive specific heat. In contrast, the small black hole (large $z_h$) has a temperature that rises with the inverse horizon radius. It is thermodynamically unstable because of its negative specific heat. 

\begin{figure}[h!]
    \centering
    \includegraphics[width=0.75\textwidth]{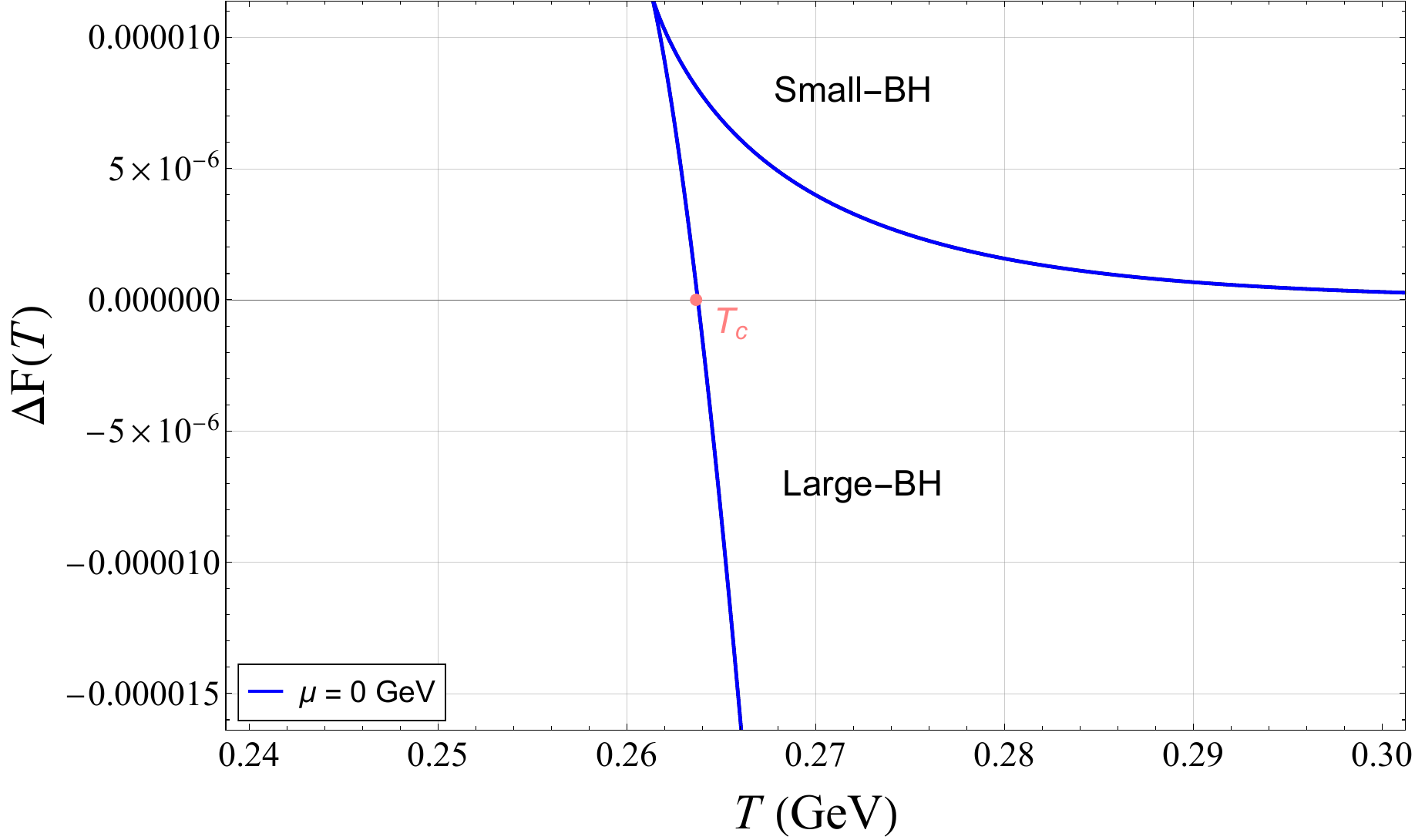}
    \caption{\small Free energy difference for $\mu=0$.}
    \phantomsection
    \label{fig:freemu0}
\end{figure}

The corresponding free-energy diagram is shown in Fig.~\ref{fig:freemu0}. Here, the free energy difference $\Delta F$ between the black hole and thermal-AdS phases is plotted. We see that depending upon the temperature the large black hole and thermal AdS phases exchange dominance as the temperature varies. In particular, the large black hole phase has the lowest free energy at large temperatures, whereas thermal-AdS has the lowest free energy at low temperatures. Accordingly, there is a first-order Hawking-Page phase transition between these two phases. For $\mu=0$, this phase transition appears at $T_{c}=0.264~\text{GeV}$. The thermal AdS phase corresponds to the confined phase on the boundary theory, while the black hole phase corresponds to the deconfined phase. In later sections, we will see how adding a nonzero chemical potential will affect this transition temperature.  Note that the small black hole phase always exhibits higher free energy than the large black hole and thermal-AdS phases. Accordingly, it corresponds to the global maxima of the solutions and is thermodynamically favored at all temperatures. Consequently, in the subsequent sections, the large black hole phase will be employed to discuss the finite temperature behavior of various observables.

To determine the perturbation equation of motion in the presence of a black hole background geometry, a generic \( g(z) \), we put back the factors of $g(z)$ in the equation of motion (\ref{EOM_2}) to find

\begin{equation} \label{FT1}
\frac{z e^{-A(z)}g(z)}{L f(z)} \partial_z \left(\frac{L e^{A(z)} f(z) g(z)}{z} \partial_z A^\mu \right) + \Box A^\mu = 0\,.
\end{equation}
We proceed in the particle's rest frame, where the four-momentum is given by \( q^\mu = (\omega, 0, 0, 0) \). Employing the radial gauge \( A_z = 0 \), we consider plane wave solutions of the form
\begin{equation}
A^\mu (z, x^\nu) = e^{i q_\alpha x^\alpha} A^\mu(z, \omega) = e^{-i \omega t} A^\mu(z, \omega)\,.
\end{equation}
By applying a Fourier transform, as showed in (\ref{fourier}), and doing the change of variable $A^\mu=e^{-H}\psi^\mu$ to (\ref{FT1}) we transform the finite temperature equation of motion into 
\begin{equation} \label{EOM_T}
-\partial_{r_*}^2 \psi^\mu(z) + V(z) \psi^\mu(z) = \omega^2 \psi^\mu(z)\,,
\end{equation}
where the tortoise coordinate $\partial_{r_*} = -g(z) \partial_z$ was applied to (\ref{FT1}) in order to redefine it into a Schr\"odinger-like form. The associated effective potential is given by 
\begin{equation} \label{Potential_T}
V(z) = \left(\partial_{r_*} \left[\frac{1}{2} \log\left(\frac{L e^{A(z)} f(z)}{z}\right)\right]\right)^2 
+ \partial_{r_*}^2 \left[\frac{1}{2} \log\left(\frac{L e^{A(z)} f(z)}{z}\right)\right]\,,
\end{equation}
where \( H \) is the same function as defined in Eq.~(\ref{B_def}). With the equation of motion formulated at finite temperature, we are equipped to calculate the spectral functions of heavy and exotic mesons, enabling the analysis of their melting behavior. Considering that (\ref{EOM_T}) is generic in $g(z)$, this is also the equation employed to study the states in a finite density plasma ($\mu>0$) as the chemical potential only alters the black hole metric.

One important comment to be made is with regards to the supplementary Landau gauge choice to derive the equation of motion for the perturbation (\ref{EOM_T}). As discussed in \cite{Dudal:2015kza}, if a nonzero spatial momentum is introduced, the system's isotropy is broken and the chosen gauge can become inadequate for a given metric. For that matters, if we did take $\vec q$, we would have found two distinct equations of motion, both for a transverse and longitudinal field propagation relative to the momentum, that would yield distinct predictions for the spectral functions. However, as $\vec q$ is taken to be $0$, the effective potentials for these equations coincide (we chose $\vec q=0$ beforehand in this work) and we are indeed left with Eq.~(\ref{EOM_T}).

\section{Procedure to compute the spectral functions} \label{sec5}

In this section we outline the numerical procedure to compute the spectral functions. This is done through some approximations, e.g.~via a tortoise coordinate and a Padé-like approximant of $B(z)$, to calculate the retarded Green's function.

\subsection{Tortoise coordinate approximation}
In this work, we adopt the procedure outlined in \cite{Jena:2024cqs} to compute the spectral functions. Typically, in a non-dynamical soft-wall model, where Einstein's equations are not necessarily satisfied, \( 1/g(z) \) can be explicitly integrated, allowing for a closed-form expression for \( r_* \). For instance, in the case of a simple blackening function \( g(z) = 1 - \frac{z^4}{z_h^4} \), the expression for \( r_* \) is given by
\begin{eqnarray}
	r_*(z) = \frac{z_h}{2} \left[\log\left({\frac{z_h-z}{z_h+z}}\right) -\arctan\left({\frac{z}{z_h}}\right) \right]\,.
\end{eqnarray}

However, the complexity of our expression for \( g(z) \) prevents us from obtaining a closed-form expression for \( r_* \). To proceed with the computations, we perform a series expansion of \( 1/g(z) \) up to a specified order and then integrate the resulting expression. Expanding up to the second order, as an example, this approach yields an approximate expression for \( r_* \) as
\begin{dmath}
  r_*(z) \approx  \frac{3 \mathcal{C} z_h^2+ 8 e^{-\frac{3 \mathcal{C} z_h^2}{8}}-8}{864 \mathcal{C}^2 z_h^6} z \left[z^2 \left(80-3 \mathcal{C}^2 z_h^4\right)-48 z z_h \left(\mathcal{C} z_h^2+10\right)+3 z_h^2 \left(\mathcal{C} z_h^2 \left[3 \mathcal{C} z_h^2+80\right]+432\right)\right]-384
   z_h^3 \log \left(1-\frac{z}{z_h}\right)\,.
\end{dmath}

\begin{figure}[h!]
    \centering
    \includegraphics[width=0.75\textwidth]{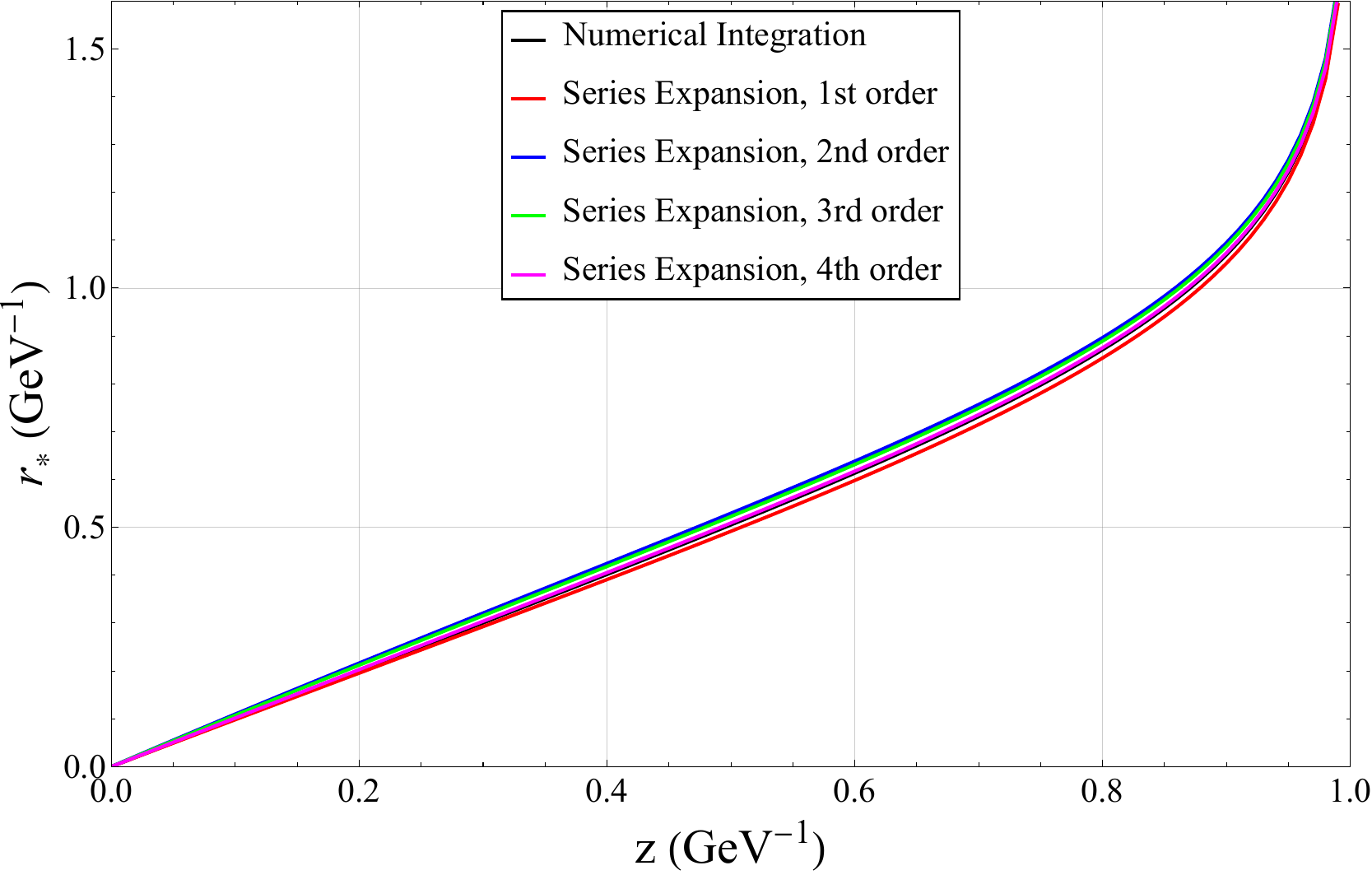}
    \caption{\small Tortoise coordinate \( r_* \) approximations.}
    \phantomsection
    \label{fig:Tortapprox}
\end{figure}

A comparison between the numerical approximation (integrating  \( 1/g(z) \) numerically) and the series expansion is plotted in Fig.~\ref{fig:Tortapprox} for the tortoise coordinate \( r_* \). We can see that even at a low order, the series expansion of the tortoise coordinate agrees quite well with the numerical result.

\subsection{Pad\'{e}-like approximant to $B(z)$}

Another important approximation adopted throughout this section to enhance the stability of the numerical scheme and to improve the Frobenius analysis is the Pad\'{e}-like approximation applied to the hyperbolic tangent term in the definition of \( f(z) \), contained in the $B(z)$ function, in Eq.~(\ref{f_definition}). The Pad\'{e} approximation is a method for approximating a function using a rational function, expressed as the ratio of two polynomials \cite{Baker_Graves-Morris_1996}. This technique often provides superior convergence and accuracy compared to standard polynomial approximations, particularly for functions with singularities or complex behavior. A Pad\'{e} approximant is characterized by the degrees of its numerator and denominator polynomials, denoted as \( [n, m] = P_n(x)/Q_m(x) \), where \( n \) and \( m \) specify the respective polynomial degrees. The Pad\'{e} approximant of order $[n,m]$ is unique and shares its Taylor expansion with the function it approximates up to order $m+n$. 

Given that \( \tanh{x} = \frac{e^{2x} - 1}{e^{2x} + 1} \), the Pad\'{e} approximant for \( e^x \) can be computed and substituted into the formula for \( \tanh{x} \) to find a good rational approximation to $\tanh{x}$ (what we call Pad\'{e}-like approximant). To achieve a more accurate approximation, we focus on deriving a Pad\'{e}-like approximant for \( \tanh{x} \) centered at a generic point \( x = a \). This process begins with the observation that the Taylor series expansion of \( e^x \) around \( x = a \) is given by
\begin{equation}
    e^x = e^a \sum_{k=0}^\infty \frac{1}{k!} (x - a)^k\,.
\end{equation}
Knowing this expansion allows us to simply extend the Pad\'{e} approximant for $e^x$, computed at $x=0$  \cite{05fc0d9b-f28e-3545-ab7b-94a0385f9451}. We find then that the Pad\'{e} approximant $[n, m]$ for $e^x$ centered at $x=a$ is given by
\begin{equation} \label{padeexp}
    e^x_{[n,m]} = \frac{P_n(x)}{Q_m(x)} = \frac{
e^a \sum_{j=0}^n \frac{(n+m-j)!}{n!} \binom{n}{j} (x-a)^j
}{
\sum_{j=0}^m \frac{(n+m-j)!}{n!} \binom{m}{j} (a-x)^j
}\,.
\end{equation}
As an example, we set $[n,m] = [2,2]$ to get
\begin{equation} \label{exp22}
    e^x_{[2,2]} = \frac{
e^a \left[
1 + \frac{1}{2}(x-a) + \frac{1}{12}(x-a)^2
\right]
}{
1 + \frac{1}{2}(a - x) + \frac{1}{12}(a - x)^2
}\,.
\end{equation}
We then apply Eq. (\ref{exp22}) to \( \tanh{x} = \frac{e^{2x} - 1}{e^{2x} + 1} \), yielding the $[2,2]$ Pad\'{e}-like\footnote{Note that the Pad\'{e}-approximant of a ratio is not the ratio of the Pad\'{e} approximants. However, for our eventual needs, this Pad\'{e}-like approximant serves perfectly well, see later in the text.} approximant for $\tanh{x}$, centered at $x=a$, as:
\begin{equation} \label{tanh22}
\tanh{x}_{[2,2]} = 1 + \frac{
8x(a+3) - 2a(a+6) - 8x^2 - 24
}{
e^a (a - 2x)\left[a - 2(3 + x)\right] + 12(1 + e^a - x) + 4x^2 - 4ax + a^2 + 6a
}\,.
\end{equation}

It is found that $a=3$ gives a good interpolation between the behavior of $B(z)$ near zero and for higher values of $z$. We also set $x \rightarrow \left(\frac{1}{Mz} - \frac{\kappa}{\sqrt{\Gamma}}\right)$ in (\ref{tanh22}) by looking at (\ref{f_definition}). Given so, we can write the $[2,2]$ Pad\'{e}-like approximant for $B(z)$ as
\begin{multline}\label{Pade22}
B(z) \approx B(z)_{[2,2]} = (\kappa z)^{2-\alpha} + Mz
\\
\!\!\!\!\!\!\!\scalemath{.95}{+ \frac{
 M^2 z^2 \left[3\Gamma ( e^3 -13) + 4 k^2 ( e^3 -1 ) - 24 \sqrt{\Gamma} k \right] + 8 \sqrt{\Gamma} Mz (3 \sqrt{\Gamma} + k - k e^3 )  + 4 \Gamma (e^3 -1)
}{
M^2 z^2 \left[3\Gamma (e^3 + 13)+ 4 k^2 (e^3 + 1) + 24 \sqrt{\Gamma} k \right] - 
8 \sqrt{\Gamma}  M z ( 3 \sqrt{\Gamma} + k + k e ^ 3) + 4 \Gamma (e^ 3 + 1)
}}.
\end{multline}

\begin{figure}[h!]
    \centering
    \includegraphics[width=0.75\textwidth]{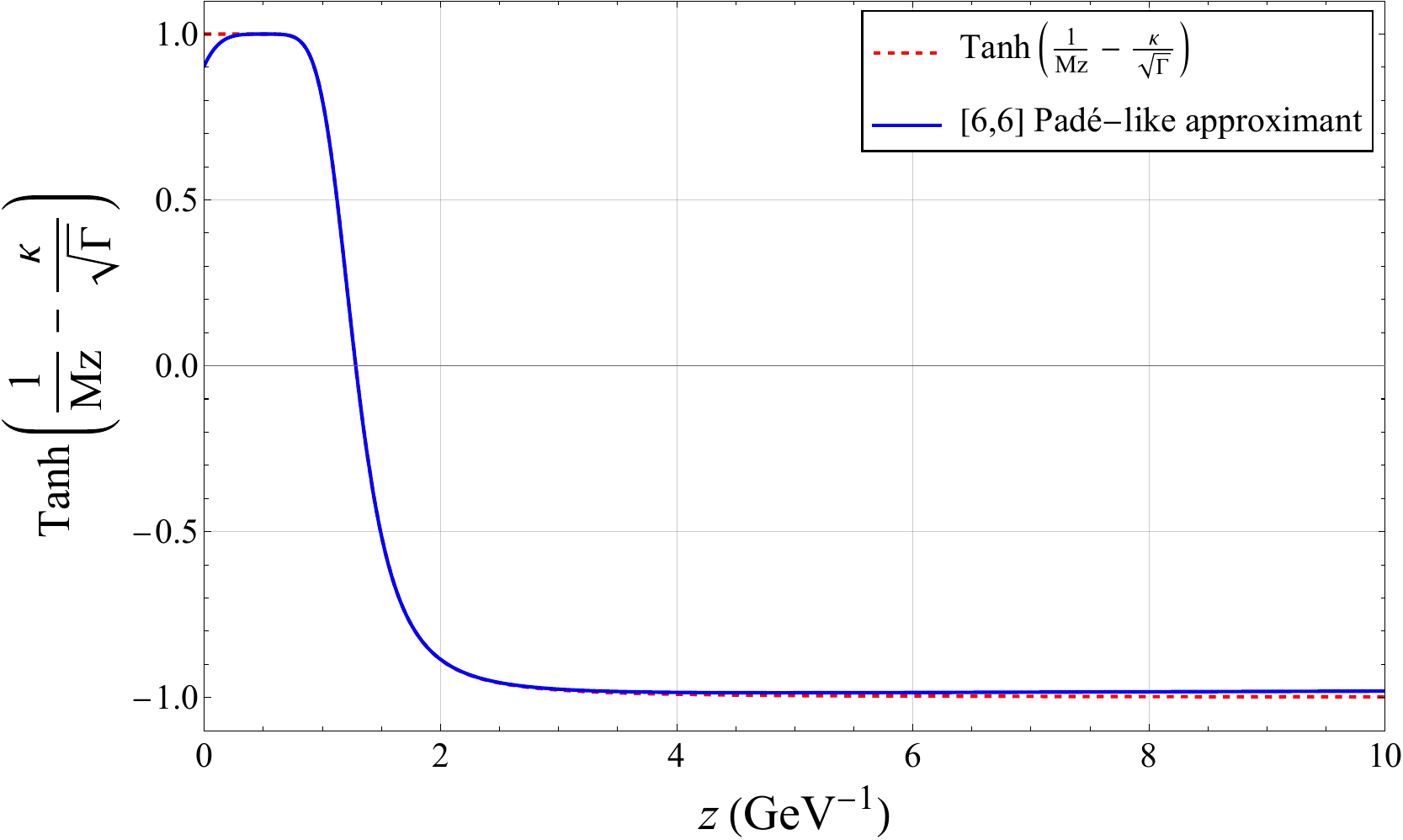}
    \caption{\small Pad\'{e}-like approximant to $\tanh{\left(\frac{1}{Mz}-\frac{\kappa}{\sqrt{\Gamma}}\right)}$ for hybrid meson parameters.}
    \phantomsection
    \label{fig:Padetan}
\end{figure}

In Fig.~\ref{fig:Padetan} we show this Pad\'{e}-like approximant, using the $[6,6]$ Pad\'{e} approximant to the exponential (centered at $a=3$), to our desired function $\tanh{\left(\frac{1}{Mz}-\frac{\kappa}{\sqrt{\Gamma}}\right)}$. We can clearly see that our constructed rational function does an excellent job at approximating $\tanh{\left(\frac{1}{Mz}-\frac{\kappa}{\sqrt{\Gamma}}\right)}$.

\begin{figure}[h!]
    \centering
    \includegraphics[width=0.75\textwidth]{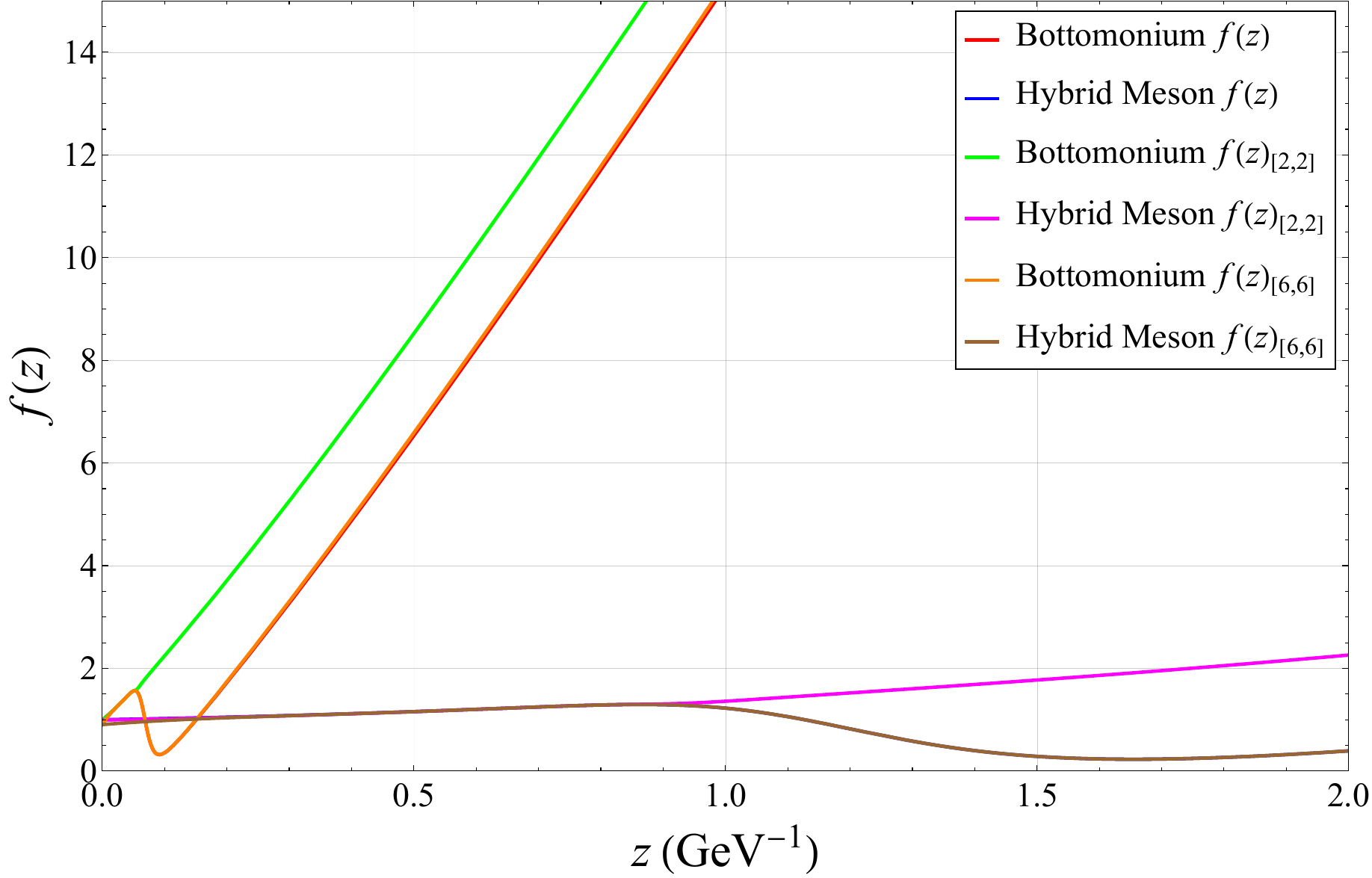}
    \caption{\small Padé-like approximants $B(z)_{[2,2]}$ and $B(z)_{[6,6]}$.}
    \phantomsection
    \label{fig:PadeB}
\end{figure}

Fig.~\ref{fig:PadeB} illustrates the Pad\'{e}-like approximants for $B(z)$ of the bottomonium and hybrid meson states parameters. 

In addition to \( B(z)_{[2,2]} \), the Pad\'{e}-like approximant \( B(z)_{[6,6]} \) was also computed following the exact same procedure as previously described in this work and subsequently plotted. It is evident that increasing the degrees of the polynomials \( P_n(x) \) and \( Q_m(x) \) significantly improves the quality of the approximations. The approximant \( B(z)_{[6,6]} \) accurately captures the behavior of \( B(z) \) across both lower and higher values of \( z \). Consequently, \( B(z)_{[6,6]} \) will be employed in the computations presented in the subsequent calculations. For reference, \( B(z)_{[6,6]} \) is given by
\begin{equation}\label{Pade66}
B(z) \approx B(z)_{[6,6]} = (\kappa z)^{2-\alpha} + Mz + \frac{\sinh{\left(\frac{3}{2}\right)} \, \mathcal{K}(z) - \cosh{\left(\frac{3}{2}\right)} \, \mathcal{J}(z)}{\sinh{\left(\frac{3}{2}\right)} \, \mathcal{J}(z) - \cosh{\left(\frac{3}{2}\right)} \, \mathcal{K}(z)}\,,
\end{equation}
where
\begin{eqnarray}
\!\!\!\!\!\!\!\!\!\!\!\!\!\!\!\!\!\!\!\!\mathcal{J}(z) &=& 42 \left(\frac{2 k}{\sqrt{\Gamma}}-\frac{2}{M z}+3\right) \left[\left(\frac{2 k}{\sqrt{\Gamma}}-\frac{2}{M z}+3\right)^4+240 \left(\frac{2 k}{\sqrt{\Gamma}}-\frac{2}{M z}+3\right)^2+7920\right]\,, \\
\!\!\!\!\!\!\!\!\!\!\!\!\!\!\!\!\!\!\!\!\!\!\mathcal{K}(z)\!&\!=\!&\left(\frac{2 k}{\sqrt{\Gamma}}\!-\!\frac{2}{M z}\!+\!3\right)^6\!+\!840 \left(\frac{2 k}{\sqrt{\Gamma}}\!-\!\frac{2}{M z}\!+\!3\right)^4\!+\!75600 \left(\frac{2
   k}{\sqrt{\Gamma}}\!-\!\frac{2}{M z}\!+\!3\right)^2\!+\! 665280\,.
\end{eqnarray}

\subsection{Power expansion near the horizon}

It is important to note that \( g(z_h) = 0 \), which implies that the equations of motion simplify to a free-particle form near the horizon. Consequently, the solutions close to the horizon can be expressed as ingoing and outgoing plane waves traveling into and out of the black hole interior as
\begin{eqnarray}
    \psi \sim \mathbb{C} e^{-i\omega r_*} + \mathbb{D} e^{i\omega r_*}\,,
\end{eqnarray}
where \( \mathbb{C} \) and \( \mathbb{D} \) are constants. Building on this, the solutions to the equations of motion near the horizon can be expanded in a power series as:
\begin{eqnarray}
    \psi_{\pm} = e^{\pm i\omega r_*} \left[ a_0^{\pm} + a_1^{\pm} (z_h - z) + a_2^{\pm} (z_h - z)^2 + \dots \right]\,,
\end{eqnarray}
where we set \( a_0^{\pm} = 1 \). By substituting this expansion into the equations of motion, we can determine the coefficients \(a_1^{\pm}\) and \(a_2^{\pm}\). However, due to the complexity introduced by the equation of motion and the tortoise coordinate, these coefficients are highly intricate, and their explicit forms are not provided here. It is sufficient to note that these coefficients depend on \(\omega\) and model parameters such as \(\mathcal{C}\), \(\sqrt{\Gamma}\), among others. For the purposes of subsequent computations, the solutions \(\psi_{\pm}\) are truncated at the second-order term, \((z_h - z)^2\).

It is evident that the ingoing and outgoing solutions form a basis for any wave function that may be considered. Consequently, renormalizable and non-renormalizable solutions can be expressed as linear combinations of the ingoing and outgoing solutions
\begin{align}    
    \psi_{2} &= \mathbb{C}_2 \psi_- + \mathbb{D}_2 \psi_+ \,, \\
    \psi_{1} &= \mathbb{C}_1 \psi_- + \mathbb{D}_1 \psi_+ \,.
\end{align}

\subsection{Frobenius analysis near the boundary}

In the limit \( z \to 0 \), the leading term in the power series solution is assumed to take the form \( z^\beta \), following the prescription for a Frobenius expansion near \( z = 0 \). The asymptotic analysis of Eq.~(\ref{EOM_T}) yields the following indicial equation
\begin{equation}
    \beta^2 - \beta - \frac{3}{4} = 0\,,
\end{equation}
which has solutions \( \beta = -\frac{1}{2} \) and \( \beta = \frac{3}{2} \). To solve the differential equation for the perturbation by evolving from the boundary to the horizon, it is also necessary to determine the derivative of \( \psi_i \) at the boundary, ensuring a consistent numerical setup. Obtaining the first derivative at the holographic boundary requires a detailed examination of the asymptotic structure of \( \psi_i \) via Frobenius analysis. 

Since the indicial equation produces two solutions differing by an integer, the larger root \( \beta = \frac{3}{2} \) corresponds to a regular series expansion, whereas the smaller root \( \beta = -\frac{1}{2} \) leads to a logarithmic series. These two solutions can be expressed as
\begin{align}
    \psi_1(z) &= z^{3/2} \sum_{k=0}^{+\infty} a_k z^k \,, \\
    \psi_2(z) &= K \ln(z) \psi_1(z) + z^{-1/2} \sum_{k=0}^{+\infty} b_k z^k \,.
 \label{power}
\end{align}

The explicit forms of \( a_k \) and \( b_k \) can be determined by substituting the series solutions into Eq.~(\ref{EOM_T}) and matching terms order by order. For normalization, we choose \( a_0 = b_0 = 1 \). The coefficients of these expansions depend on \( \omega \) and the parameters of the model, such as \( \Gamma \). As an example, the first coefficient of the expansion for \( \psi_1(z) \) and the $\ln(z)$ coefficient $K$ in  \( \psi_2(z) \), for the considered equation of motion, considering the full non-approximate $B(z)$, are given by
\begin{align}
    a_1 &= \frac{1}{32} \left(M^2 - 4 \omega^2 \right) \,, \\
    K &= \frac{1}{8} \left(M^2 - 4 \omega^2 \right) \,.
 \label{power1}
\end{align}

Generally, the coefficients depend on the momentum $q^2$, but we did set $q=0$ beforehand. These solutions, starting near the boundary, also form a basis for constructing wave functions, which can be written as
\begin{align}
    \psi_{+} &= \mathcal{A}_+ \psi_2 + \mathcal{B}_+ \psi_1 \,, \\
    \psi_{-} &= \mathcal{A}_- \psi_2 + \mathcal{B}_- \psi_1 \,.
\end{align}

\subsection{Computation of the retarded Green's function}

Both the near-boundary and near-horizon expansions provide a basis in which wave functions can be expressed as linear combinations. From the relations between \( \psi_{\pm} \), \( \psi_1 \), and \( \psi_2 \), it follows that the coefficients satisfy
\begin{gather}
 \begin{pmatrix} \mathcal{A}_{+} & \mathcal{B}_{+} \\ \mathcal{A}_{-} & \mathcal{B}_{-} \end{pmatrix}
 =
  \begin{pmatrix} \mathbb{C}_{2} & \mathbb{D}_{2} \\ \mathbb{C}_{1} & \mathbb{D}_{1} \end{pmatrix}^{-1}\,.
\end{gather}

Our primary interest lies in the computation of the spectral functions. Within this approach, this requires determining the retarded Green's function and subsequently extracting its imaginary part  \cite{Mamani:2022qnf,Jena:2024cqs}.

The retarded Green's function is computed by numerically integrating the equations of motion over a regular lattice extending from a point near the boundary, \( z = \epsilon \), to a point close to the horizon, \( z = z_h - \epsilon \). The initial conditions are derived from the solution near the boundary at \( z = 0 \), specifically \( \psi_1(\epsilon) \), obtained through Frobenius analysis, along with its derivative \( \partial_z \psi_1(\epsilon) \). Similarly, the second solution near the boundary, \( \psi_2(\epsilon) \), along with its derivative \( \partial_z \psi_2(\epsilon) \), is used to compute the other independent solution for the equations of motion using the same numerical scheme.

As a result, two numerical solutions for \( \psi \) are obtained. These functions are then evaluated near the horizon, \( z_h - \epsilon \), where the following relationship is constructed
\begin{gather}
\begin{pmatrix} \psi_a(z_h - \epsilon) \\ \partial_z \psi_a(z_h - \epsilon) \end{pmatrix}
=
\begin{pmatrix} \psi_-(z_h - \epsilon) & \psi_+(z_h - \epsilon) \\ \partial_z \psi_-(z_h - \epsilon) & \partial_z \psi_+(z_h - \epsilon) \end{pmatrix}
\begin{pmatrix} \mathbb{C}_a \\ \mathbb{D}_a \end{pmatrix}\,,
\label{coef}
\end{gather}
where \( a = 1, 2 \) correspond to the numerical results obtained from solving the equations of motion, and \( \psi_{\pm} \) represent the asymptotic solutions near the horizon as given by the power expansions in Eqs.~(\ref{power}) and (\ref{power1}). All quantities are evaluated at \( z_h - \epsilon \). The coefficients \( \mathbb{C}_a \) and \( \mathbb{D}_a \) are computed by inverting the above square matrix. 

Combining this with the coefficient relations in Eq.~(\ref{coef}), we arrive at
\begin{eqnarray} \label{green}
    \frac{\mathcal{B}_-}{\mathcal{A}_-} = - \frac{\mathbb{D}_2}{\mathbb{D}_1} = \frac{\partial_z \psi_-(z_h - \epsilon) \psi_2(z_h - \epsilon) - \psi_-(z_h - \epsilon) \partial_z \psi_2(z_h - \epsilon)}{\partial_z \psi_-(z_h - \epsilon) \psi_1(z_h - \epsilon) - \psi_-(z_h - \epsilon) \partial_z \psi_1(z_h - \epsilon)}\,.
\end{eqnarray}
This provides a stable numerical method for calculating the ratio \( \frac{\mathcal{B}_-}{\mathcal{A}_-} \), which is sufficient to compute the retarded Green's function and, consequently, the spectral function. As demonstrated in \cite{Miranda:2009uw, Mamani:2018uxf, Mamani:2022qnf}, the spectral function is directly related to \( \frac{\mathcal{B}_-}{\mathcal{A}_-} \) via the expression
\begin{eqnarray}
    \rho(\omega, q = 0) = - 2 \Im G^R(\omega, q = 0) \propto \Im \frac{\mathcal{B}_-}{\mathcal{A}_-}\,.
\end{eqnarray}
where \( \rho(\omega, q = 0) \) represents the spectral function calculated numerically using the described prescription (with \( q = 0 \) set initially), and \( \Im \) denotes the imaginary part.

The numerical integration of the equation of motion over the interval \( z \in [\epsilon, z_h - \epsilon] \), necessary to compute \( \psi_i(z_h - \epsilon) \) and \( \partial_z \psi_i(z_h - \epsilon) \) as they appear in the Green's function formula (\ref{green}), is carried out using Mathematica's \texttt{NDSolve} with the \texttt{StiffnessSwitching} numerical scheme. This method is specifically chosen to address the stiff nature of the problem, as explicit methods like the standard fourth-order Runge-Kutta or Euler methods proved inadequate for producing reliable results.

The remaining components in Eq.~(\ref{green}), specifically \( \psi_{\pm}(z_h - \epsilon) \) and \( \partial_z \psi_{\pm}(z_h - \epsilon) \), are obtained analytically based on the results of the Frobenius analysis and the associated power series expansions.

The small positive parameter \( \epsilon \) functions as both the lattice spacing and the boundary cutoff. For the computations performed, a value of \( \epsilon = 0.001 \) was used. The value of \( z_h \), corresponding to a specific temperature \( T \), is determined by solving the transcendental equation (\ref{BH_T_mu0}) for \( z_h \) using Newton's method.

\section{Melting of heavy and exotic states} \label{sec6}

In this section we apply the numerical scheme developed in Sec. \ref{sec5} to the model parameters. The resulting spectral functions are analyzed to determine the mesonic melting.

\subsection{Spectral functions at $\mu=0$}

The spectral function characterizes the distribution of states that can be excited by a quarkonium state as it dissolves in a hot and dense QCD medium. It contains vital information about the dissociation of quarkonium states and the formation of the quark-gluon plasma (QGP). Specifically, the spectral peaks represent the existence of bound states of heavy quarks and anti-quarks, such as charm and bottom quarks, within the QGP. The appearance, disappearance, or modification of these peaks offers valuable insights into the properties of the QGP. As demonstrated earlier, by taking the small \( \epsilon \) limit, holographically significant information about the spectral function can be obtained as it was done in \cite{Fujita:2009wc, Fujita:2009ca}.

\begin{figure}[htb!]
    \centering
    \begin{minipage}[t]{0.49\textwidth}
        \centering
        \includegraphics[width=\textwidth]{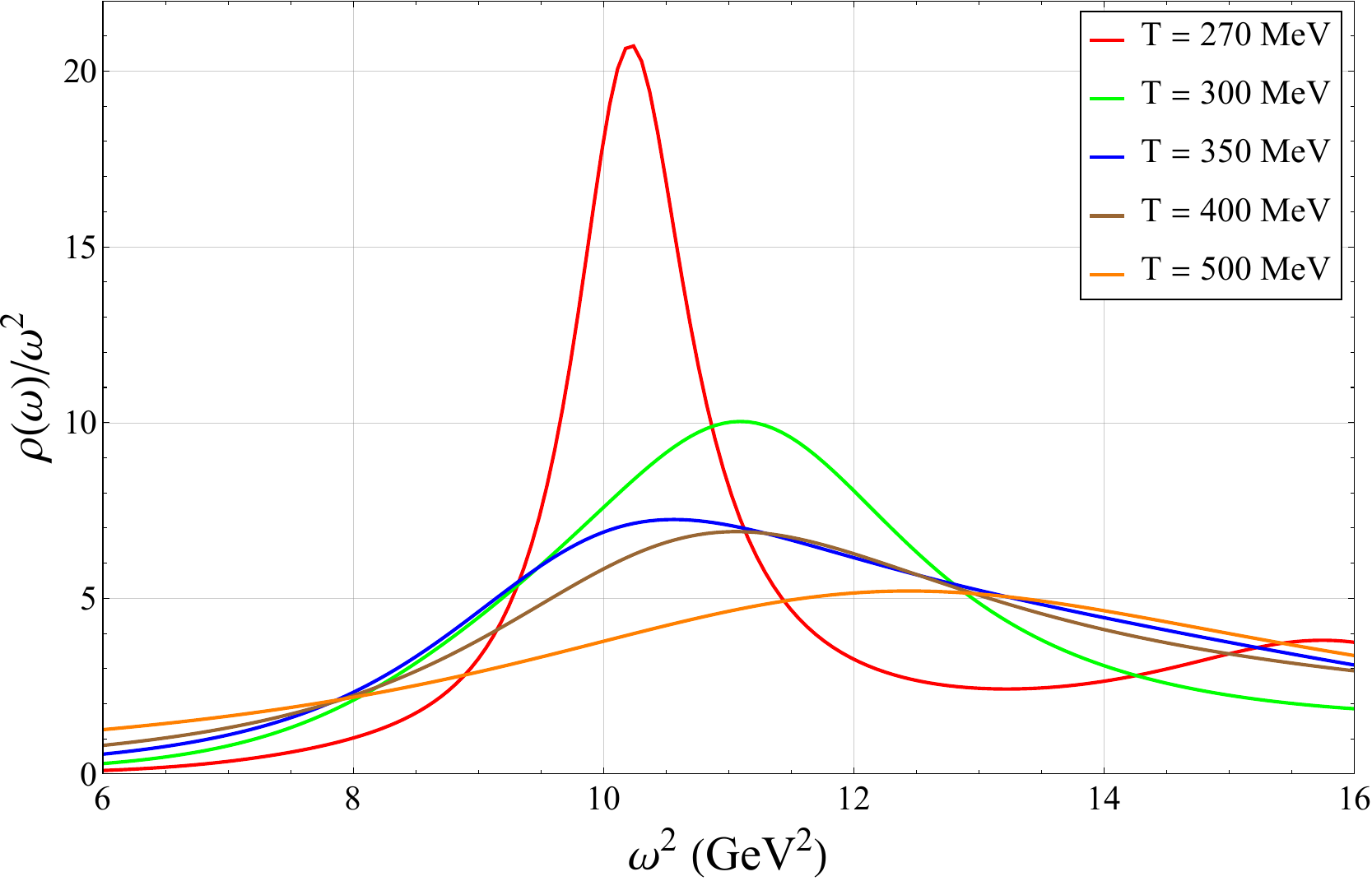}
        \caption{\small Charmonium spectral function.}
        \phantomsection
        \label{fig:SpecTcharm}
    \end{minipage}%
    \hfill
    \begin{minipage}[t]{0.49\textwidth}
        \centering
        \includegraphics[width=\textwidth]{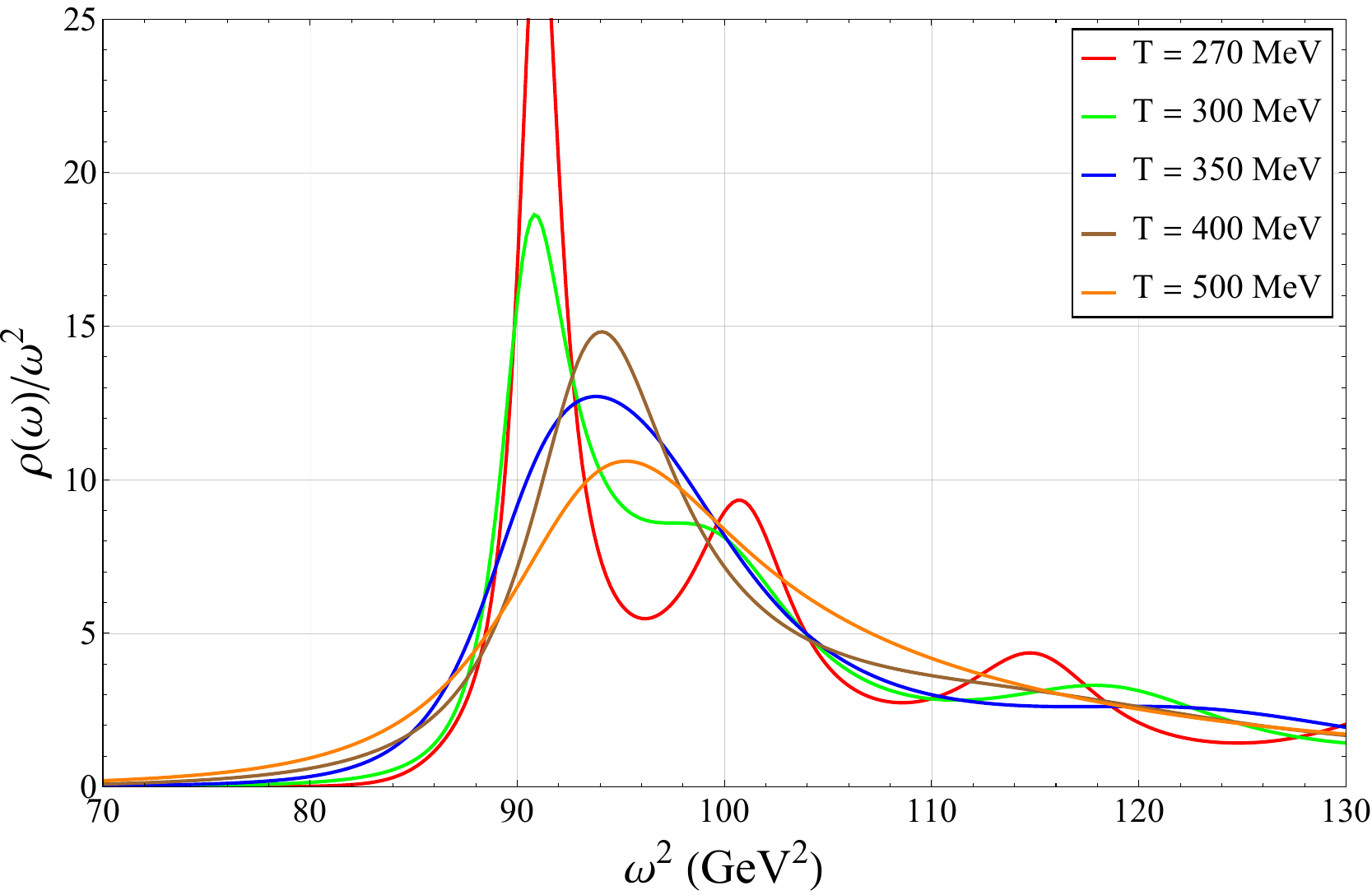}
        \caption{\small Bottomonium spectral function.}
        \phantomsection
        \label{fig:SpecTbottom}
    \end{minipage}
\end{figure}

\begin{figure}[htb!]
    \centering
    \begin{minipage}[t]{0.49\textwidth}
        \centering
        \includegraphics[width=\textwidth]{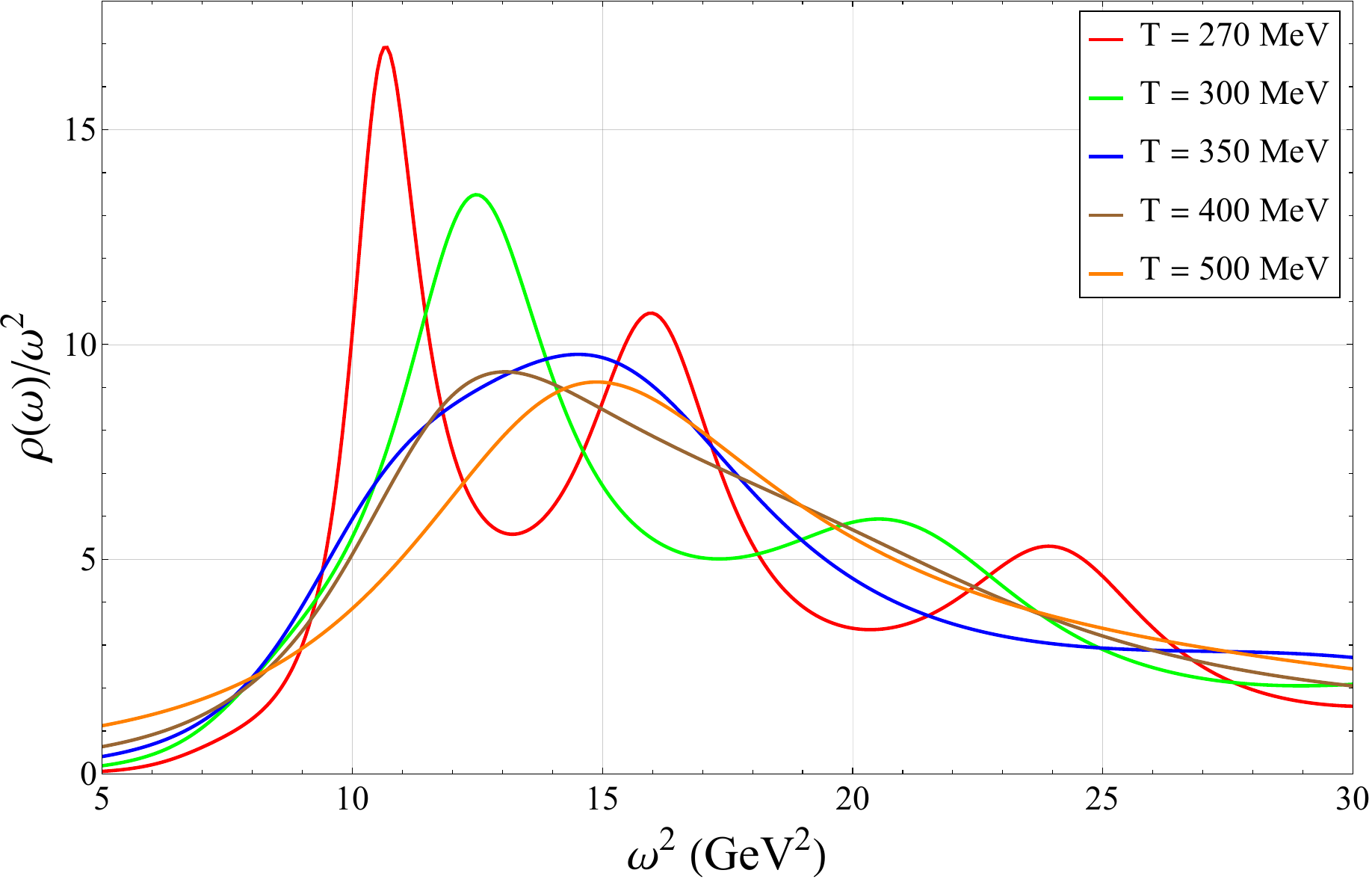}
        \caption{\small Tetraquark spectral function.}
        \phantomsection
        \label{fig:SpecTtetra}
    \end{minipage}%
    \hfill
    \begin{minipage}[t]{0.49\textwidth}
        \centering
        \includegraphics[width=\textwidth]{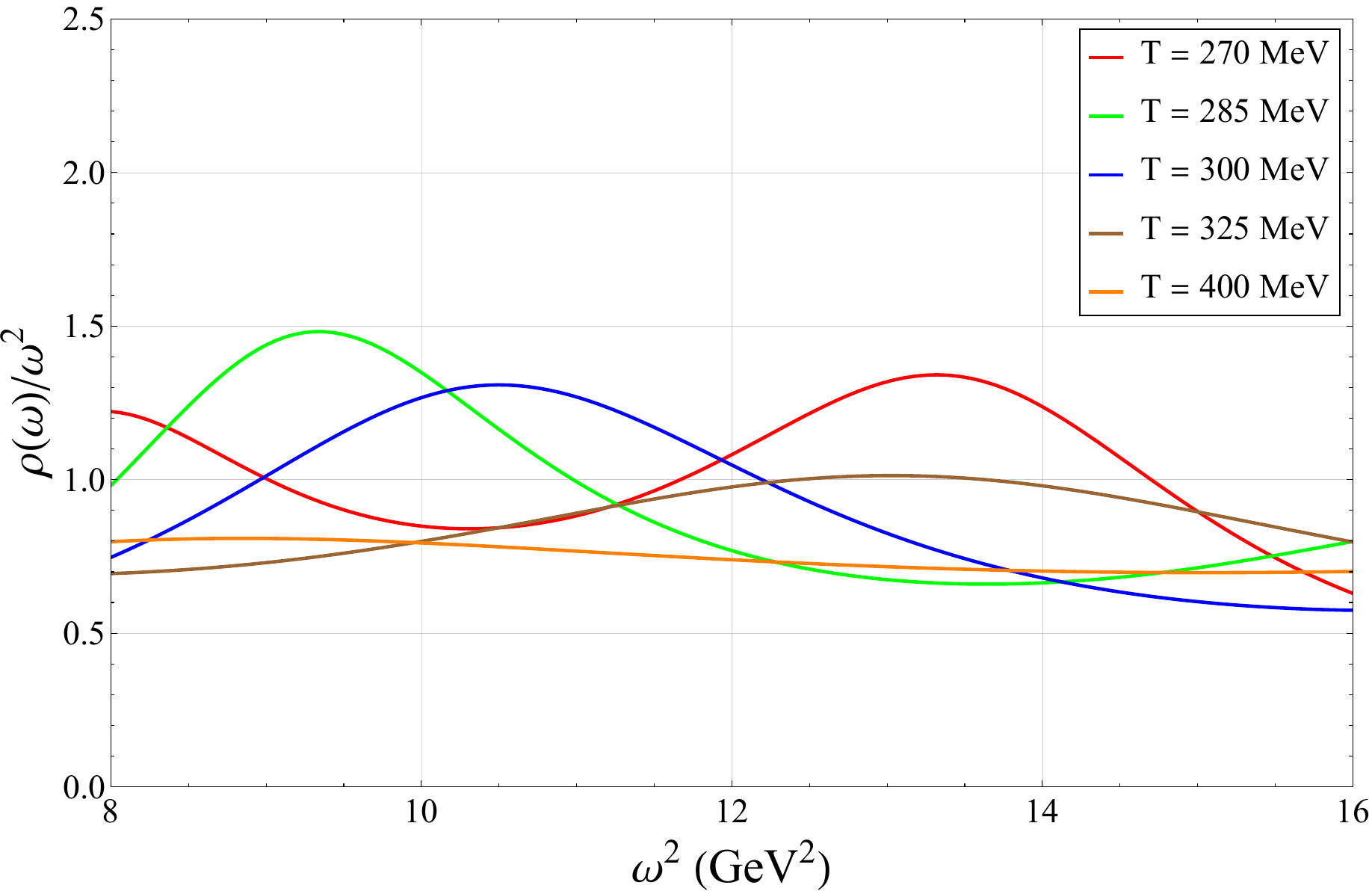}
        \caption{\small Hybrid meson spectral function.}
        \phantomsection
        \label{fig:SpecThybrid}
    \end{minipage}
\end{figure}

Experimental measurements and theoretical investigations of quarkonium production and suppression in heavy-ion collisions provide valuable tools for exploring the properties of the quark-gluon plasma (QGP) and understanding the behavior of strong interactions under extreme conditions. In this study, we focus on examining the melting of quarkonium states by analyzing the variations in the spectral peaks as a function of temperature. In particular, equipped with the numerical procedure described in the above section, we compute the spectral functions for charmonium, bottomonium, tetraquark, and hybrid states. The numerical results are presented in Figs.~\ref{fig:SpecTcharm} -- \ref{fig:SpecThybrid} for a range of temperatures.

We can clearly see the melting process through the spectral functions. As the temperature increases, the peaks tend to disappear and the function broadens out. This indicates the melting of the mesonic states into a hot medium. We can do a very rough estimate of the melting temperature for the charmonium, bottomonium, and tetraquark states to be around $T\sim 400-450~ \text{MeV}$ as their spectral function peaks disappear at these temperatures. However, we can clearly see that the hybrid meson melts at a lower temperature than the rest as the peaks disappear more quickly as we increase the temperature. For this state, we estimate this temperature to be around $T\sim 325-400~ \text{MeV}$.

\subsection{Temperature dependence of the effective potential $V(z)$ at $\mu=0$}

\begin{figure}[htb!]
    \centering
    \begin{minipage}[t]{0.49\textwidth}
        \centering
        \includegraphics[width=\textwidth]{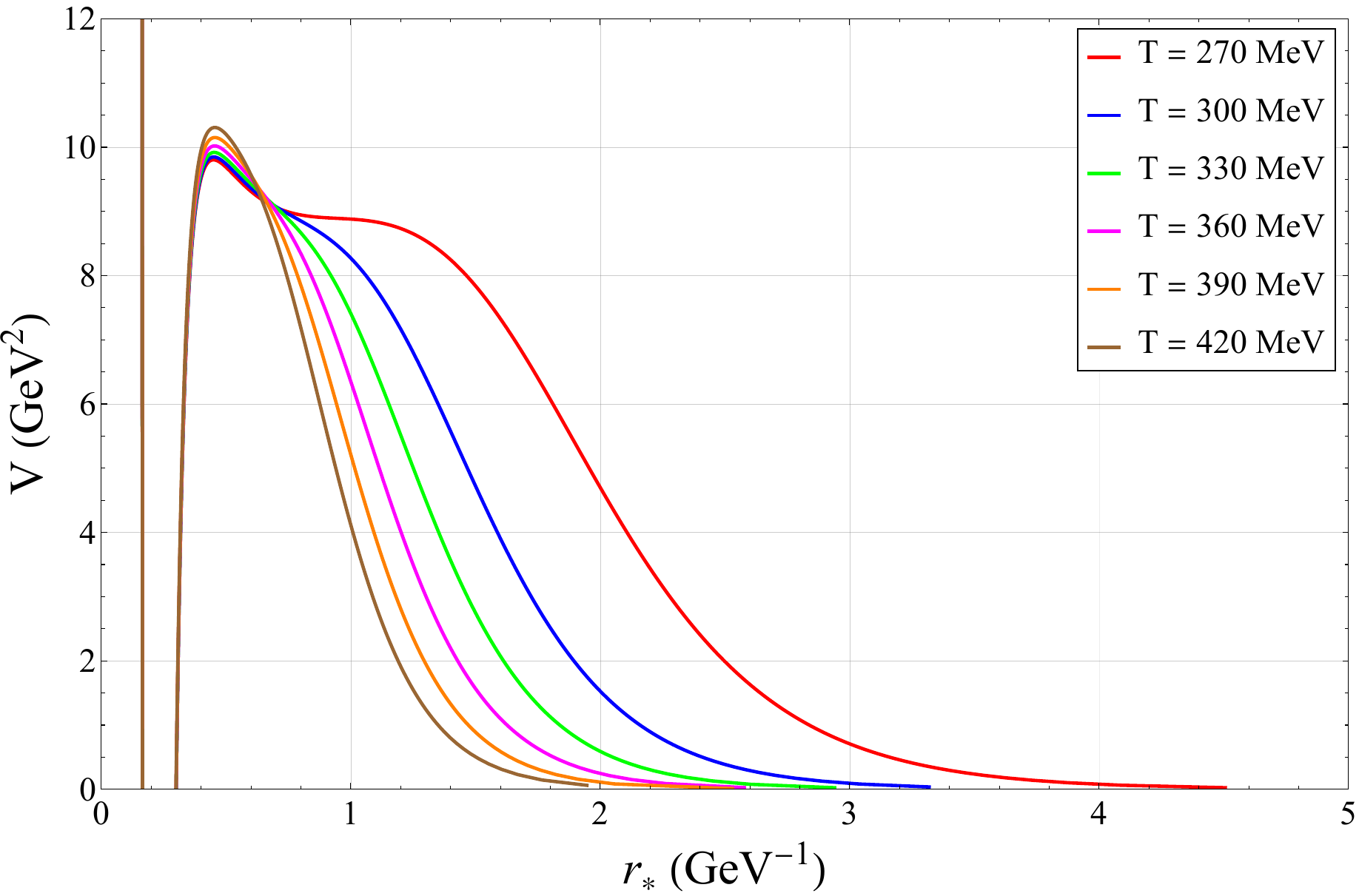}
        \caption{\small Charmonium effective potential $V(z)$.}
        \phantomsection
        \label{fig:Tpotcharm}
    \end{minipage}%
    \hfill
    \begin{minipage}[t]{0.49\textwidth}
        \centering
        \includegraphics[width=\textwidth]{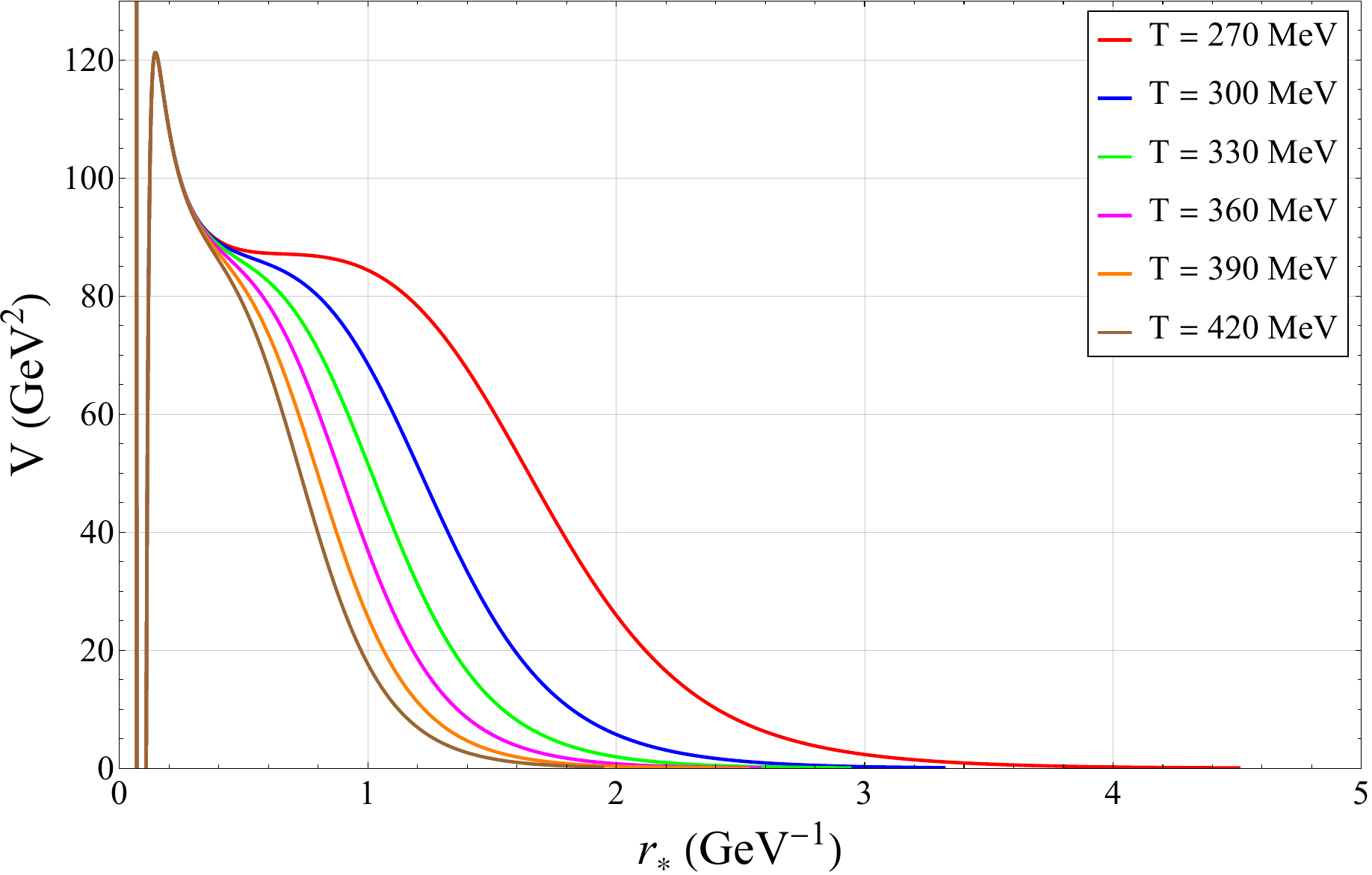}
        \caption{\small Bottomonium effective potential $V(z)$.}
        \phantomsection
        \label{fig:Tpotbottom}
    \end{minipage}
\end{figure}

\begin{figure}[htb!]
    \centering
    \begin{minipage}[t]{0.49\textwidth}
        \centering
        \includegraphics[width=\textwidth]{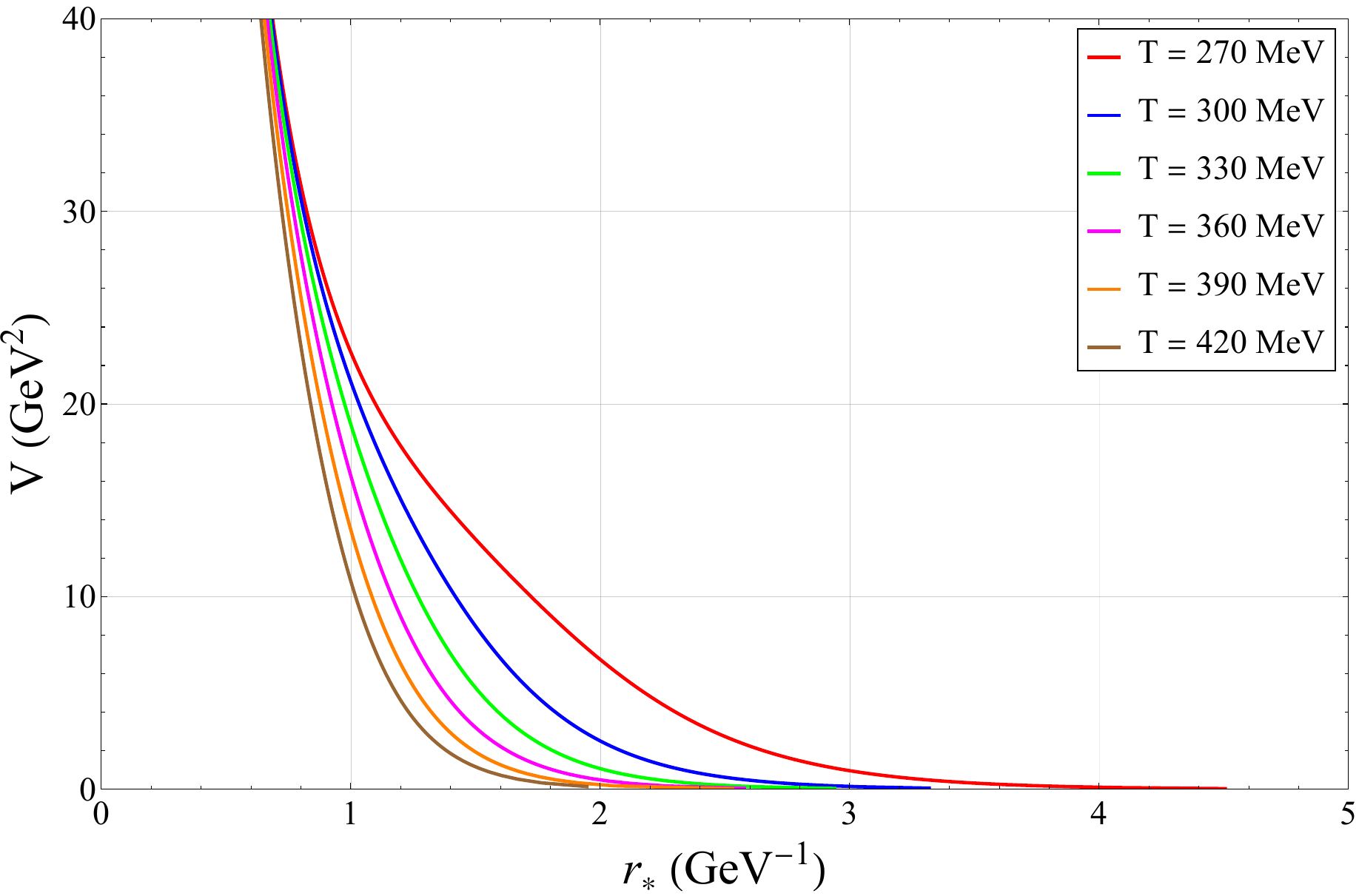}
        \caption{\small Tetraquark effective potential $V(z)$.}
        \phantomsection
        \label{fig:Tpottetra}
    \end{minipage}%
    \hfill
    \begin{minipage}[t]{0.49\textwidth}
        \centering
        \includegraphics[width=\textwidth]{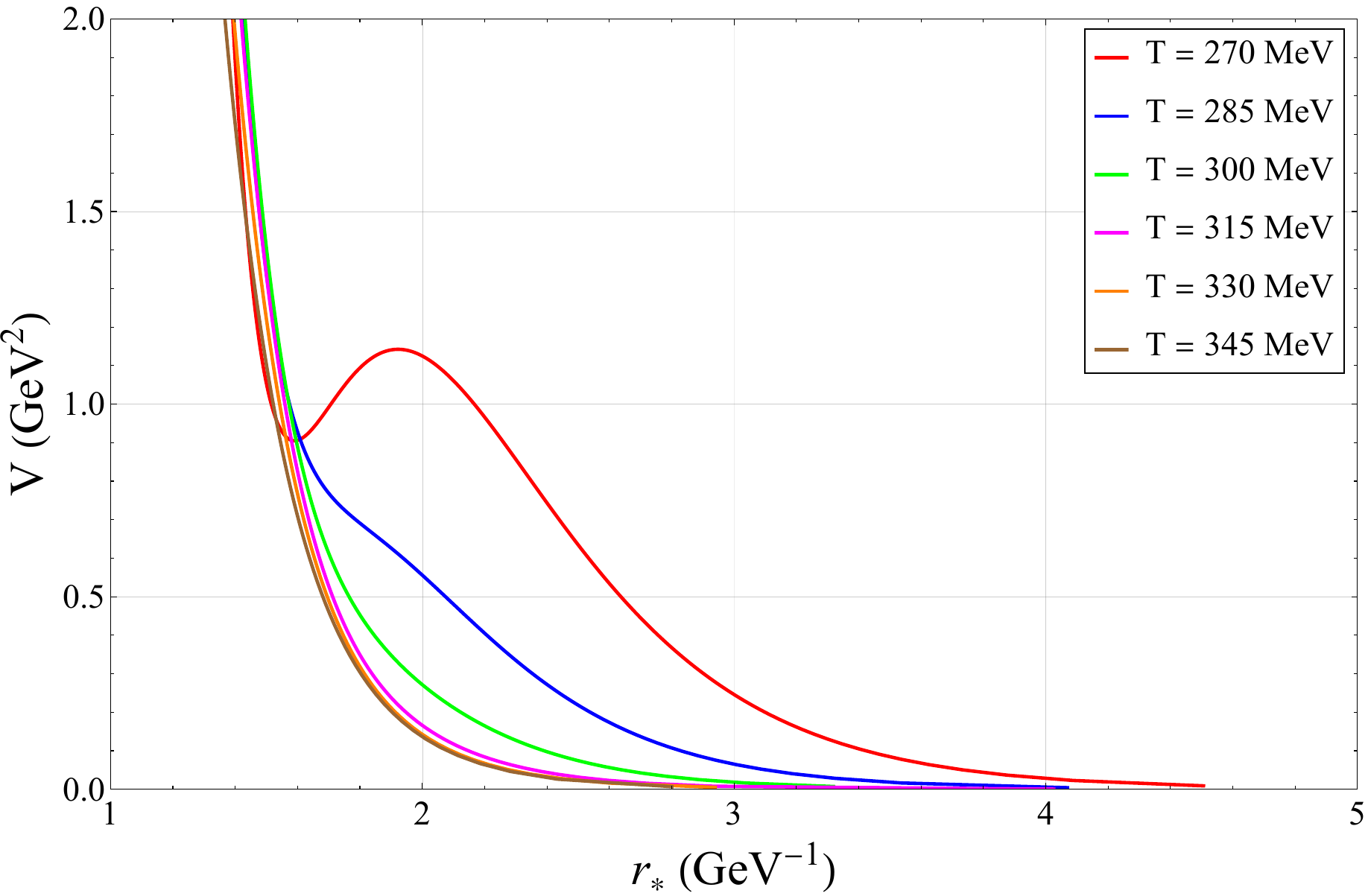}
        \caption{\small Hybrid meson effective potential $V(z)$.}
        \phantomsection
        \label{fig:Tpothybrid}
    \end{minipage}
\end{figure}

Another way to verify the melting process is to plot the effective potential for the Schrödinger-like equation for various temperatures and see how it changes as the temperature is increased. So in Figs.~\ref{fig:Tpotcharm} -- \ref{fig:Tpothybrid} we present the holographic potential, defined in Eq.~(\ref{Potential_T}), for some values of $T$, for the considered states. 

From the Figs.~\ref{fig:Tpotcharm} -- \ref{fig:Tpothybrid} we can conclude that, for temperatures close to the minimum reachable value in this model ($T \sim T_c$), the potential tends to display a single global minimum (except for the tetraquark), representing holographic confinement and the stability of mesonic states. As the temperature increases, the shape of the potential changes significantly, with the global minimum disappearing and the well becoming shallower, reflecting the weakening of confinement and thermal dissociation of mesons. This evolution of the potential highlights the holographic encoding of thermal melting, where the interplay between the dilaton structure and rising temperature governs the confinement and dissociation processes of mesonic states.

\section{Finite density plasma} \label{sec7}
Having explored the system at zero chemical potential and analyzed its thermal behavior, we are now set to investigate the dynamics of a finite density plasma. To incorporate finite density effects, we introduce the chemical potential $\mu$ by solving Eq.~(\ref{2.9}) under a specific boundary condition. The chemical potential serves as a critical parameter that modifies the holographic setup and allows the study of systems where baryonic or charge density plays a significant role.

\subsection{Computation of $A_t(z)$ at finite $\mu$}

Eq. (\ref{2.9}) governs the behavior of the temporal component of the gauge field, $A_t(z)$, which is directly related to the chemical potential in holography. By solving this equation with appropriate boundary conditions, we establish the profile of $A_t(z)$ across the bulk spacetime. The boundary conditions are as follows:
\begin{enumerate}
    \item $A_t(z)$ must vanish at the black hole horizon ($z = z_h$), ensuring a regular solution at this point.
    \item At the asymptotic boundary ($z \to 0$), the value of $A_t(z)$ is proportional to the chemical potential $\mu$, linking the bulk dynamics to the boundary field theory.
\end{enumerate}

The solution for $A_t(z)$ allows us to introduce finite density effects into the EMD holographic system, modifying the equations of motion and the resulting physical quantities. Specifically, the presence of a nonzero chemical potential impacts the blackening function $g(z)$. The change in the metric, in turn, influences the thermodynamic properties and phase structure of the system, giving us also different spectral functions found in the last section. After computing the gauge field $A_t(z)$, we can calculate the metric $g(z)$ by solving Eq.~(\ref{2.10}). 

Unfortunately, a general analytic solution of $g(z)$ could not be determined unless we restricted ourselves to low values of $\alpha$ and excluded the low-$z$ corrections to the function $f(z)$ presented in Eq.~(\ref{kappam}). This limitation arises due to the intricate coupling between the equations of motion, which becomes analytically intractable for higher values of $\alpha$ or when the full complexity of $f(z)$, including its low-$z$ corrections, is incorporated. Using this new definition, we find the equation of motion for $A_t(z)$, from Eq.~(\ref{2.9}):
\begin{align} 
    f(z) &= e^{-\left[-\frac{c}{8} z^2 + (\kappa z)^{2-\alpha} \right]} \,, \\
    0 &= A_t'' + A_t' \left( -\frac{1}{z} + \frac{f'}{f} + A' \right) \Rightarrow 0 = \left[{(\alpha -2) (\kappa z)^{2 - \alpha}  -1}\right] \, A_{t}' + z\,A_{t}'' \label{Atequation} \,.
\end{align}

The general solution to Eq.~(\ref{Atequation}) has been determined and is expressed as:
\begin{equation} \label{generalAt}
    A_t(z) = \mathcal{Y}_2 + \mathcal{Y}_1 \frac{1}{\alpha - 2}\left[(-1)^{\frac{2}{\alpha - 2}} z^2 (\kappa z)^{\frac{2 (2 - \alpha)}{\alpha - 2}} \, \Gamma\left(-\frac{2}{\alpha - 2}, -(\kappa z)^{2 - \alpha}\right)\right],
\end{equation}
where $\mathcal{Y}_{1,2}$ are integration constants and $\Gamma(u, v) = \int_v^{\infty} t^{u-1} e^{-t} \, dt$ is the incomplete gamma function. As a consistency check, we consider the general solution (\ref{generalAt}) in the limit $\alpha \rightarrow 0$ and apply the boundary conditions outlined earlier in this section, yielding:
\begin{equation}
    A_t(z) = \mathcal{Y}_2 + \mathcal{Y}_1 \frac{e^{\kappa^2 z^2}}{2 \kappa^2} \Rightarrow A_t(z) = \mu\frac{e^{\kappa^2 z_h^2} -e^{\kappa^2 z^2} }{e^{\kappa^2 z_h^2} -1}.
\end{equation}

We readily see that, in this limit, our general solution to $A_t(z)$ reproduces what was found in \cite{Mamani:2022qnf}, where a quadratic dilaton profile was implied. 

To apply the given boundary conditions to Eq.~(\ref{generalAt}), with $\alpha \neq 0$, we found it was better to introduce a small positive number $\delta$ (to impose the asymptotic boundary condition, where $z \rightarrow \delta$) and after the computations take the limit $\delta \rightarrow 0$. 
This procedure gives us the following constants of integration: 
\begin{equation}
\mathcal{Y}_2 = \frac{\mu}{1 - \frac{\Gamma\left(-\frac{2}{\alpha-2}, -(\kappa \delta)^{2 - \alpha}\right)}{\Gamma\left(-\frac{2}{\alpha-2}, -(\kappa z_h)^{2 - \alpha}\right)}} \,,   
\end{equation}
\begin{equation}
\scalemath{.99}{\mathcal{Y}_1 = \frac{- \mu (-1)^{-\frac{2}{\alpha-2}} (\alpha-2) (\kappa^2 z_h \delta)^{\frac{2\alpha}{ \alpha-2}}}{z_h^2 (\kappa z_h)^{\frac{4}{ \alpha -2}} (\kappa \delta)^{\frac{2\alpha}{\alpha-2}} \Gamma\left(-\frac{2}{\alpha-2}, -(\kappa z_h)^{2 - \alpha}\right) - \kappa^{\frac{4}{\alpha-2}} (\kappa z_h \delta)^{\frac{2\alpha}{\alpha-2 }} \Gamma\left(-\frac{2}{\alpha-2 }, -(\kappa \delta)^{2 - \alpha}\right)}} \,. 
\end{equation}

Substituting these constants into (\ref{generalAt}) and taking the limit $\delta \rightarrow 0$ gives the full solution to $A_t(z)$: 
\begin{align}
    A_t(z) &= \mu \lim_{\delta \rightarrow 0} \frac{\Gamma\left(-\frac{2}{ \alpha -2}, -(\kappa z)^{2 - \alpha}\right) - \Gamma\left(-\frac{2}{\alpha -2 }, -(\kappa z_h)^{2 - \alpha}\right)}{\Gamma\left(-\frac{2}{\alpha-2 }, -(\kappa \delta)^{2 - \alpha}\right) -\Gamma\left(-\frac{2}{\alpha-2}, -(\kappa z_h)^{2 - \alpha}\right)} \\ 
    &= \mu + \mu \frac{ \Gamma\left(-\frac{2}{\alpha -2 }\right) - \Gamma\left(-\frac{2}{\alpha -2 }, -(\kappa z)^{2 - \alpha}\right)}{\Gamma\left(-\frac{2}{ \alpha-2 }, -(\kappa z_h)^{2 - \alpha}\right) -\Gamma\left(-\frac{2}{ \alpha-2 }\right)} \,. \label{Atsolution}
\end{align} 

This full form of $A_t(z)$ is quite intricate and we need to approximate the incomplete gamma function in some way to introduce $A_t(z)$ in Eq.~\eqref{2.10} to solve for $g(z)$. To do so, we use the following series expansion (for small $u$ and $v$): 
\begin{equation}
    \Gamma(u, v) = \Gamma(u) - \sum_{n=0}^{\infty} \frac{(-1)^n v^{n + u}}{n! (u + n)}\,.
\end{equation}

\begin{figure}[h!]
    \centering
    \includegraphics[width=0.75\textwidth]{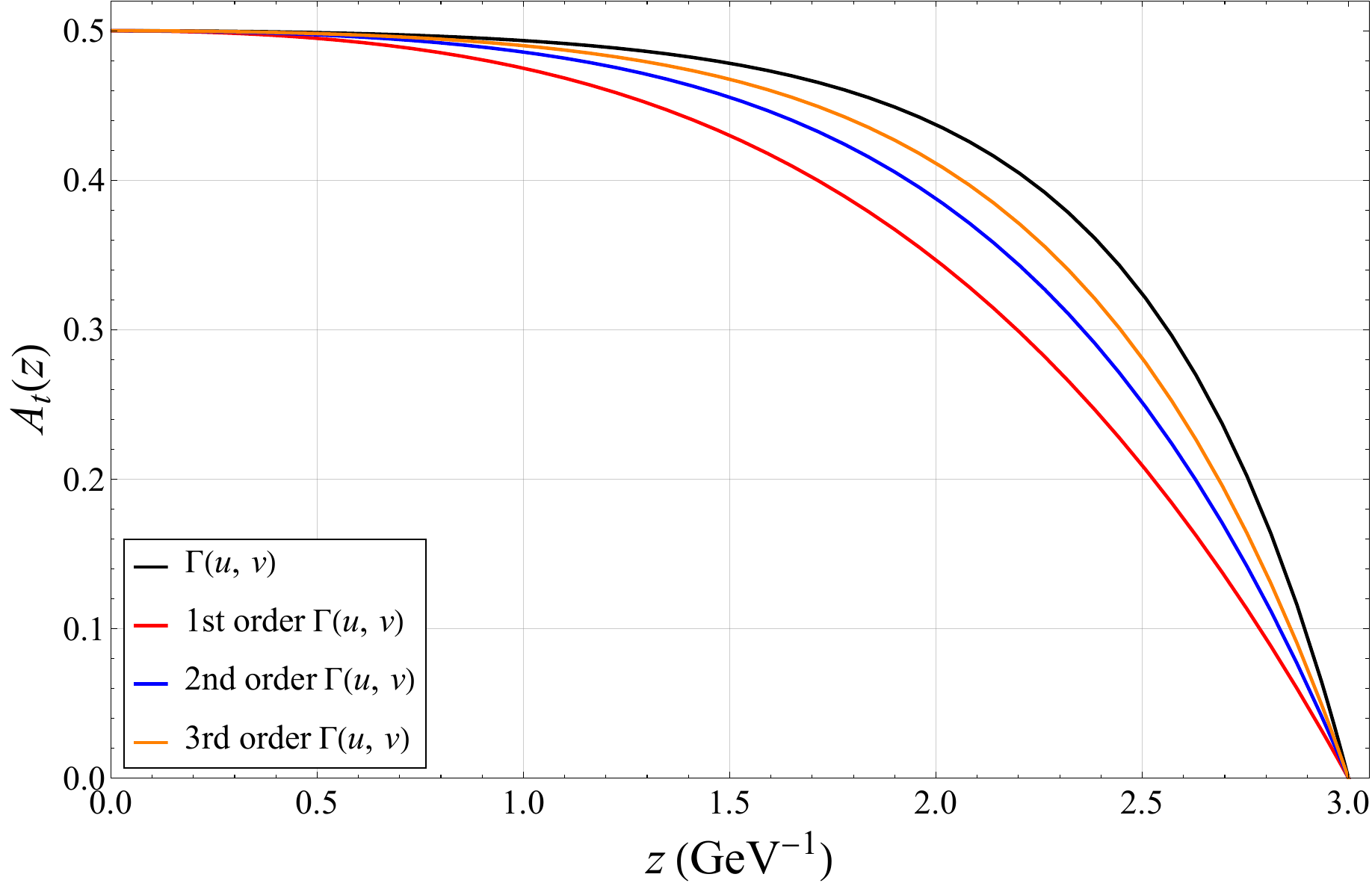}
    \caption{\small $A_t(z)$ approximation to 3rd order in $\Gamma(u, v)$ with parameters $z_h = 3 \, \text{GeV}^{-1}$, $\mu = 0.5 \, \text{GeV}$, $k = 0.8 \,  \text{GeV}$, and $\alpha = 0.5$.}
    \phantomsection
    \label{fig:GaugeA}
\end{figure}

In Fig.~\ref{fig:GaugeA}  we plot $A_t(z)$ for different orders of approximation to $\Gamma(u, v)$. We can see then that higher orders in $\Gamma(u, v)$ produce better approximations to $A_t(z)$. If we choose lower values for the model parameters, like the hybrid meson one, the first order is considered sufficient to continue the calculations. From now on we use the following $A_t(z)$:
\begin{equation} \label{At1st}
    A_t(z) = \mu -  \mu\frac{\left(\frac{z}{z_h}\right)^{2 - \alpha} \left[(\alpha -4)(\kappa z)^\alpha  - 2 \kappa^2 z^2 \right]}{ (\alpha-4 )(\kappa z_h)^\alpha -2 \kappa^2 z_h^2}.
\end{equation}

\subsection{Analytic formula for $g(z)$}

Using Eq.~(\ref{At1st}) to $A_t(z)$, the equation of motion for the metric, Eq.~\eqref{2.10}, becomes
\begin{equation} \label{gequation}
    0 = g'' - \frac{3 (4 + c z^2)}{4 z} g' - 4\mu^2 z^2 e^{\frac{3 c z^2}{8} - (\kappa z)^{2 - \alpha}} \frac{ z^{2 - 2\alpha} \left[\kappa^2 z^2 + (\kappa z)^\alpha\right]^2 z_h^{ 2\alpha-4} (\alpha-4 )^2 }{\left[ (\kappa z_h)^\alpha ( \alpha-4)-2 \kappa^2 z_h^2 \right]^2}.
\end{equation}

The boundary conditions of Eq.~(\ref{gequation}) are so that $g(z)$ must vanish at the horizon, $g(z_h) = 0$, and become unity at the holographic boundary, $g(0) = 1$. 
To compute Eq.~(\ref{gequation}) with such boundary conditions,  the Taylor series expansion of $\exp{\left[\frac{3 c z^2}{8} - (\kappa z)^{2 - \alpha}\right]}$ can be implemented to first order, to eliminate such exponential from the differential equation. After doing all these approximations, the problem remains barely analytically tractable. We could then find a solution to $g(z)$ that depend on the generic set of parameters $\mu$, $\alpha$, $\kappa$, $c$ and $z_h$. Even so, we should say that for larger values of these parameters, especially $\kappa$ (that correlates with bigger $\alpha$), the derived quantities may be inflicted with larger errors. We then focus, for the rest of this section, on computing the finite density corrections to the hybrid meson ($\kappa = 0.468 \, \text{GeV} $ and $\alpha=0.034$) spectral functions. The calculated $g(z)$ contains hundreds of terms. Just for reference, we show a few of those terms: 
\begin{align} \label{g_fullmu}
    g(z) &= \alpha \mu^2 e^{\frac{3 c z^2}{8}} \frac{   2^{21 - \frac{9\alpha}{2}} 3^{-6 + \frac{3\alpha}{2}} c^{-6 + \frac{\alpha}{2}}  \left(\frac{c}{k}\right)^\alpha k^6 z_h^{2\alpha - 4}  \Gamma\left(4 - \frac{3\alpha}{2}, \frac{3 c z^2}{8}\right)}{\left[2 k^2 z_h^2 - (k z_h)^\alpha (\alpha - 4)\right]^2} \nonumber \\
    &- \alpha^2 \mu^2  z^2  e^{\frac{3 c z^2}{8}} \frac{2^{10 - \frac{3\alpha}{2}} 3^{\frac{\alpha}{2} - 3} c^{\frac{\alpha}{2} - 3}  k^{2 - \alpha}\left(\frac{1}{k z_h}\right)^{-2\alpha}  \Gamma\left(3 - \frac{\alpha}{2}, \frac{3 c z^2}{8}\right)}{z_h^4 \left(2 k^2 z_h^2 - (k z_h)^\alpha (\alpha - 4)\right)^2} \\
    &+ \mu^2 z^{6 - 3\alpha} \frac{4096 k^{2 - \alpha}  (k z z_h)^{2\alpha} }{3 c z_h^4 \left[2 k^2 z_h^2 - (k z_h)^\alpha (\alpha - 4)\right]^2 (\alpha - 8) (\alpha - 6)} + \cdots \nonumber
\end{align}

\begin{figure}[h!]
    \centering
    \includegraphics[width=0.75\textwidth]{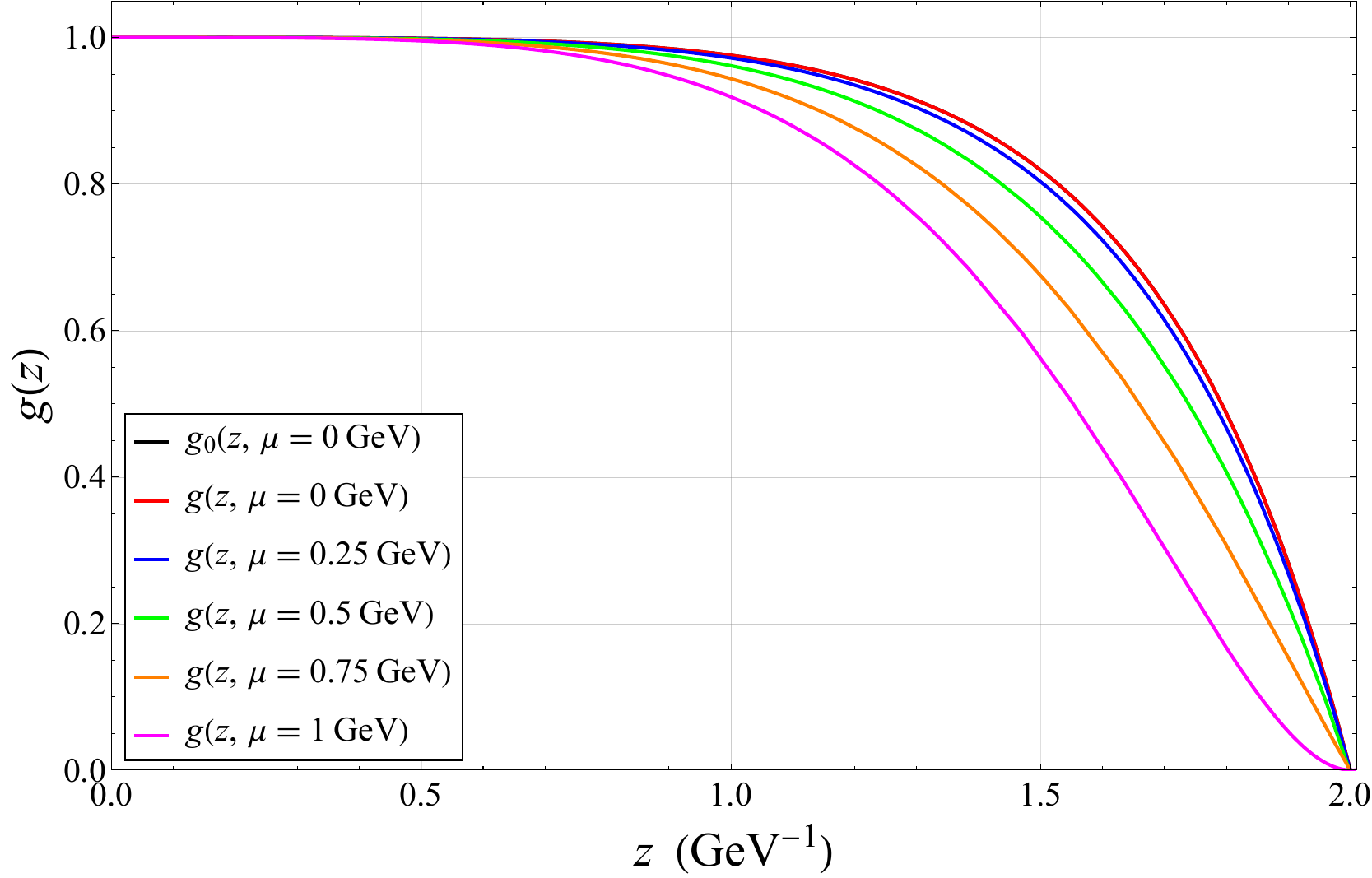}
    \caption{\small The radial profile of $g(z)$ for various $\mu$ using the hybrid meson parameters. Here $z_h=2$ is used.}
    \phantomsection
    \label{fig:gmu}
\end{figure}

Fig.~\ref{fig:gmu}  shows the radial profile of $g(z)$ for some values of chemical potential $\mu$. We define $g_0(z)$ as the metric calculated earlier at zero chemical potential, shown in Eq.~ (\ref{g_T_mu_0}). We refer to $g(z)$ as the now calculated metric shown in Eq.~(\ref{g_fullmu}). 

From Fig.~\ref{fig:gmu} we can clearly see that the calculated metric at nonzero chemical potential reproduces the metric calculated at $\mu = 0$ as their curves overlap. We notice the deformation of the curves as we increase $\mu$. We use this computed metric (\ref{g_fullmu}) in the next sections to analyze its thermodynamical proprieties and calculate how the $\mu$ affects the mesonic melting process.

\subsection{Thermodynamical analysis}
\begin{figure}[h!]
    \centering
    \includegraphics[width=0.75\textwidth]{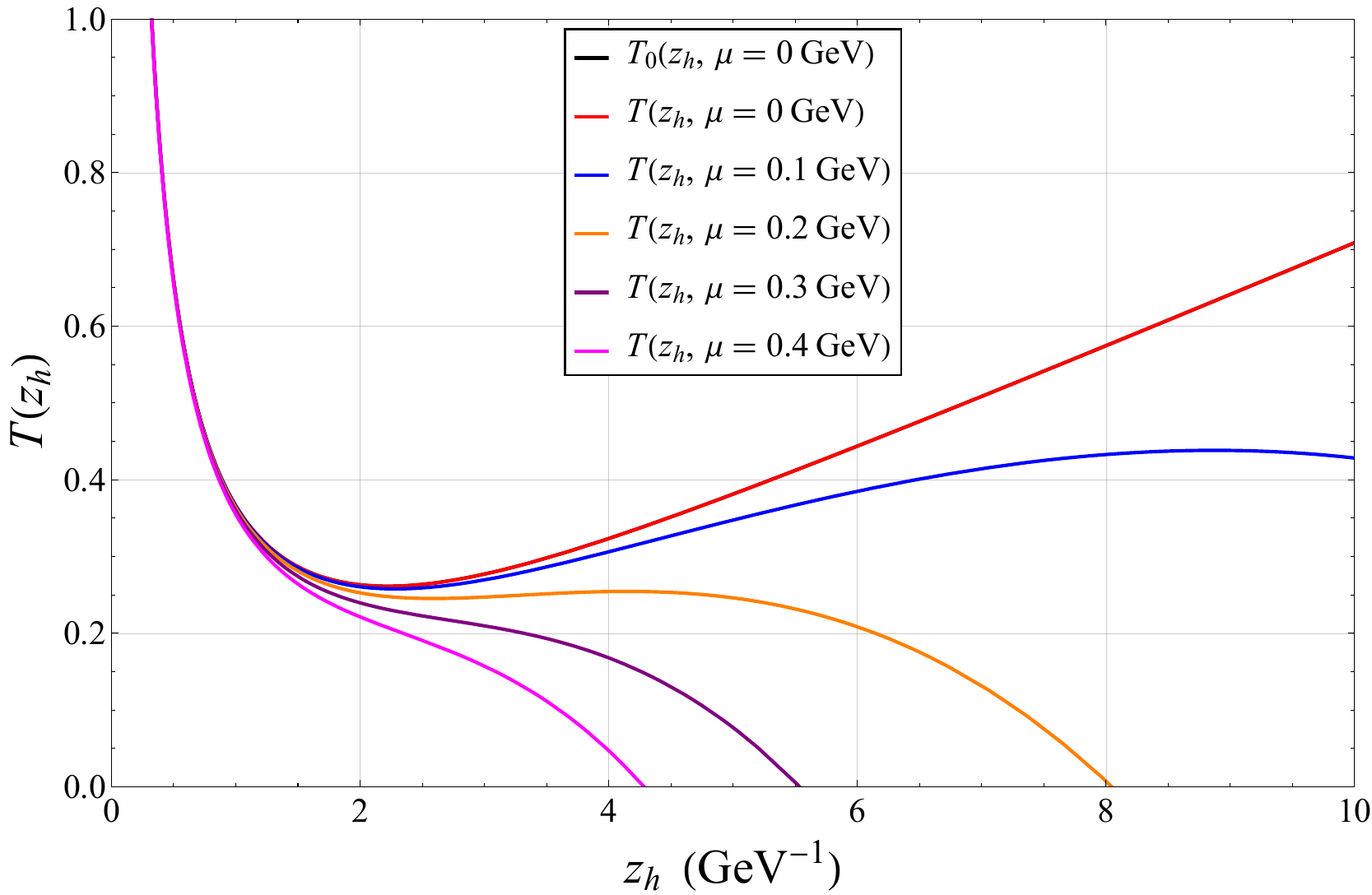}
    \caption{\small Temperature as a function of $z_h$ for various $\mu$ using the hybrid meson parameters.}
    \phantomsection
    \label{fig:tvsmu}
\end{figure}

\begin{figure}[h!]
    \centering
    \includegraphics[width=0.75\textwidth]{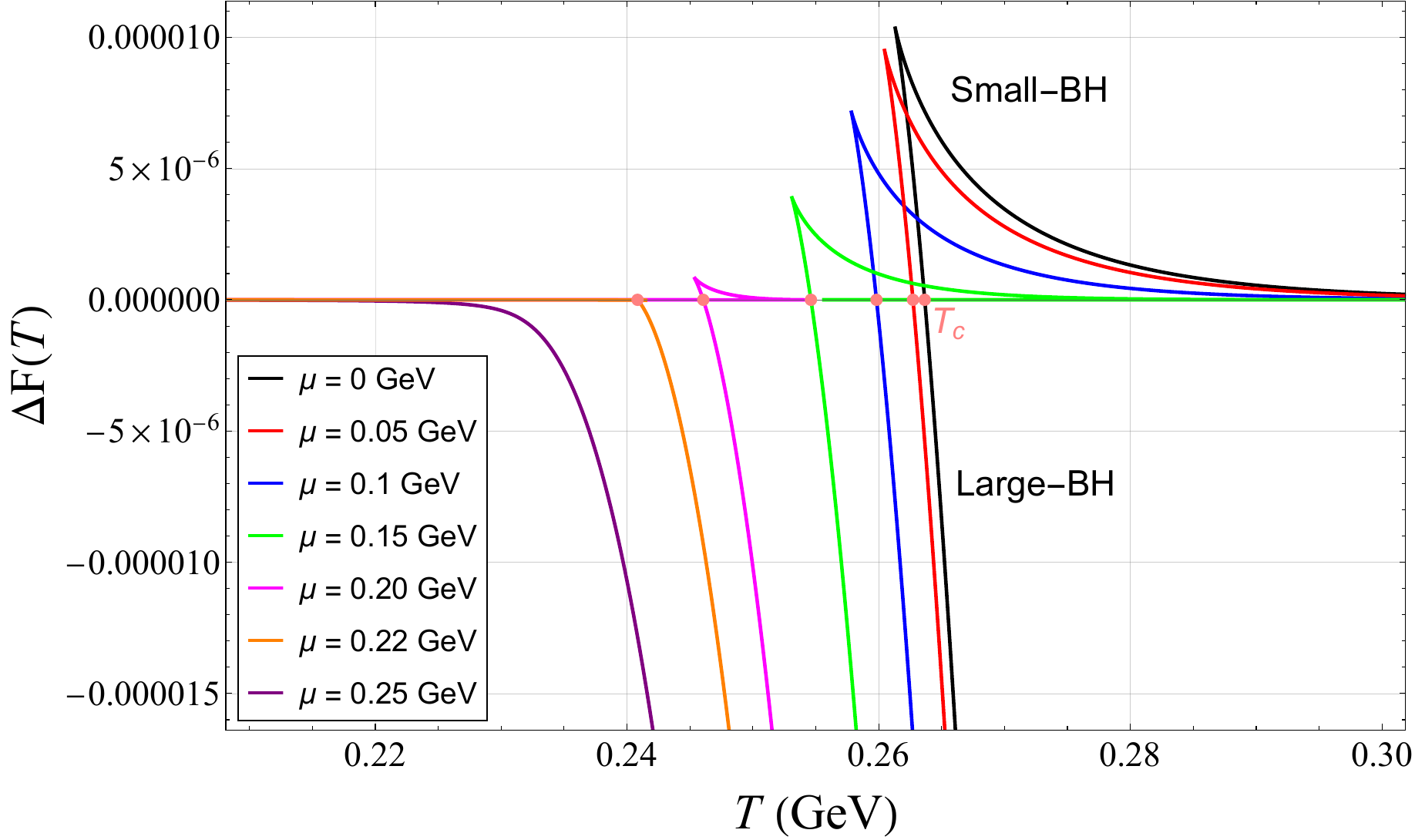}
    \caption{\small The thermal profile of the free energy difference $\Delta F (T)$ for various values of $\mu$. Here hybrid meson parameters are used.}
    \phantomsection
    \label{fig:freemu}
\end{figure}

With the expression of $g(z)$ at hand we can analyze the thermodynamical structure of the gravity solution at finite density. Our findings are quite similar to those of \cite{Dudal:2017max}. The results are shown in Figs.~\ref{fig:tvsmu} and \ref{fig:freemu}. Here, $T_0$ is defined as the temperature corresponding to the choice $g_0(z)$. Fig.~\ref{fig:tvsmu} shows that, again, that temperature agrees with the case of $g(z)$ if $\mu = 0$ is taken into account. 

The thermodynamic structure at finite chemical potential becomes drastically different from its uncharged counterpart. For small but finite $\mu$, there now exist three black hole branches---large, small, and intermediate---instead of two. For large and small black hole branches, the temperature decreases with $z_h$. These two branches exhibit positive specific heat and are thermodynamically stable. Whereas, for the intermediate branch, the temperature increases with $z_h$. This branch exhibits negative specific heat and, accordingly, it is thermodynamically unstable. As $\mu$ increases, the size of the intermediate branch shrinks and the large and small black hole branches come close to each other. At a certain critical value of the chemical potential $\mu=\mu_c$, the intermediate branch completely disappears and the small and large black hole branches merge to form a single black hole branch. For $\mu>\mu_c$, only one black hole branch remains which becomes extremal at some horizon radius. 

The corresponding free energy behavior is shown in Fig.~\ref{fig:freemu}. Here, the black hole free energy is normalized again with respect to the thermal-AdS. For $\mu<\mu_c$, the free energy exhibits a swallow-tail like structure, indicating a first-order phase transition. In particular, the free energy of large/small black hole phases is minimal for high/low temperatures. This suggests a first-order phase transition between the large and small black hole phases as the temperature is varied. The corresponding transition temperature turns out to be decreasing with $\mu$.  The intermediate phase, forming the base of the swallow-tail, always has higher free energy than the large and small black hole phases and is thermodynamically disfavored at all temperatures. At $\mu=\mu_c$, the large and small black hole phases merge together and the first-order phase transition between them ceases to exist. The $\mu_c$, therefore, defines a second-order critical point at which the first-order phase transition line between the small/large black hole phases stops. This, at least qualitatively, reflects what one expects for genuine (lattice) QCD at finite density/temperature \cite{deForcrand:2002hgr,Stephanov:2004wx,Borsanyi:2025dyp}.

\begin{figure}[h!]
    \centering
    \includegraphics[width=0.75\textwidth]{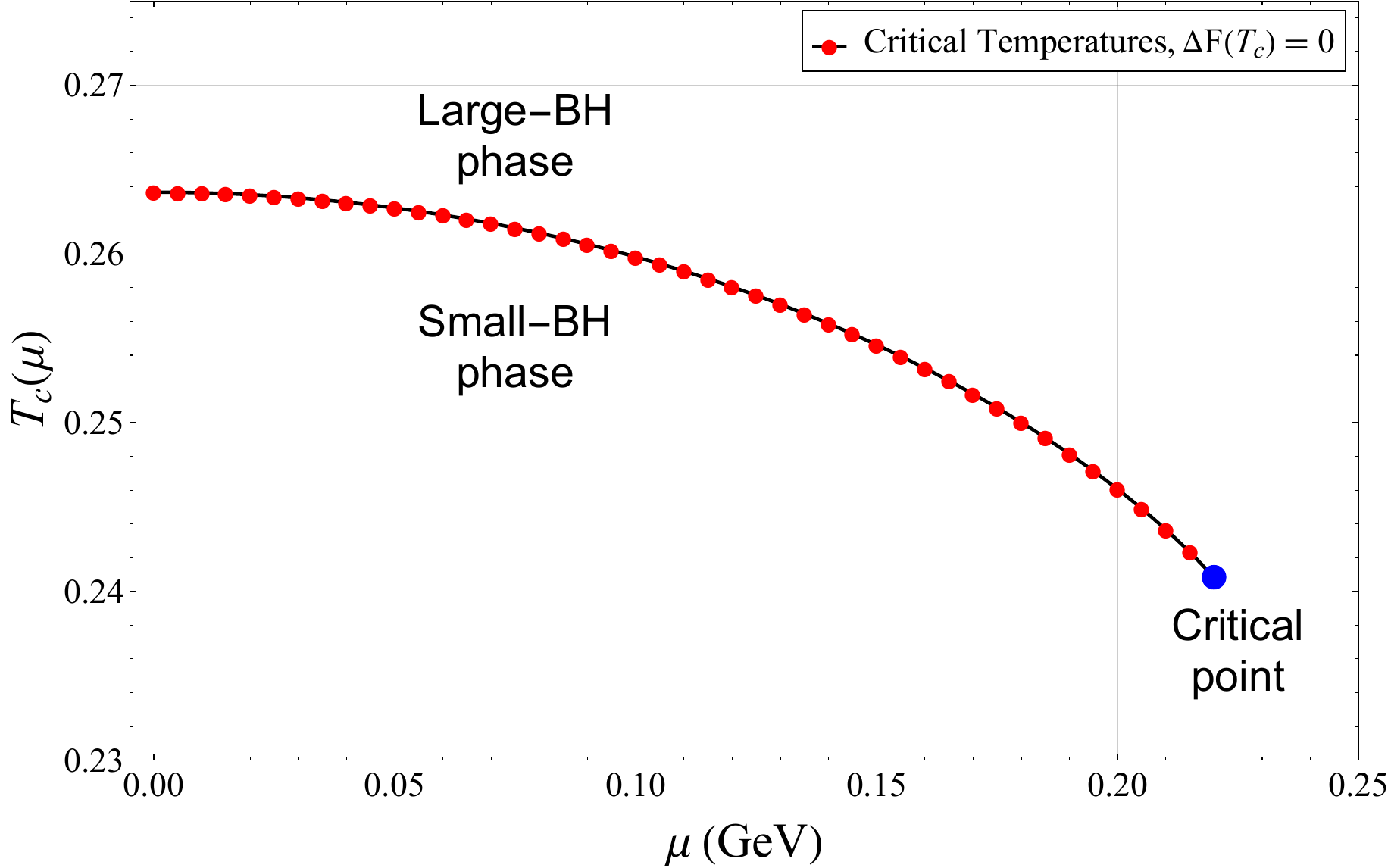}
    \caption{\small $T_{c}$ as a function of $\mu$: For small values of $\mu$, a first-order phase transition occurs between the large and small black hole phases. This first-order phase transition line terminates at a second-order critical point.}
    \phantomsection
    \label{fig:criticalphase}
\end{figure}

The above situation is analogous to the famous van der Waal's type phase transition observed in ordinary liquid-gas thermodynamic systems. For $\mu>\mu_c$, only one black hole phase exists which remains thermodynamically favored at all temperatures, i.e., its free energy is always negative. The overall thermodynamic phase structure at finite $\mu$ is shown in Fig.~\ref{fig:criticalphase}.   

In the dual boundary theory, the small and large black hole phases correspond to the specious-confined and deconfined phases \cite{Dudal:2017max}, respectively. In particular, the small black hole phase does not represent true confinement (which would typically be dual to a horizonless geometry at zero temperature) because its Polyakov loop expectation value, while extremely small, is still nonzero. It does, however, exhibit linear confinement at large distances and low temperatures. Due to this distinction, this phase is referred to as the specious-confined phase rather than simply the confined phase. On the gravity side, the presence of a small black hole in the specious-confined phase introduces a well-defined notion of temperature. Given this, it will be interesting to examine the thermal behavior of the melting process in both specious-confined and deconfined phases.

\subsection{Effective potential and spectral functions at finite $\mu$}

\begin{figure}[h!]
    \centering
    \includegraphics[width=0.75\textwidth]{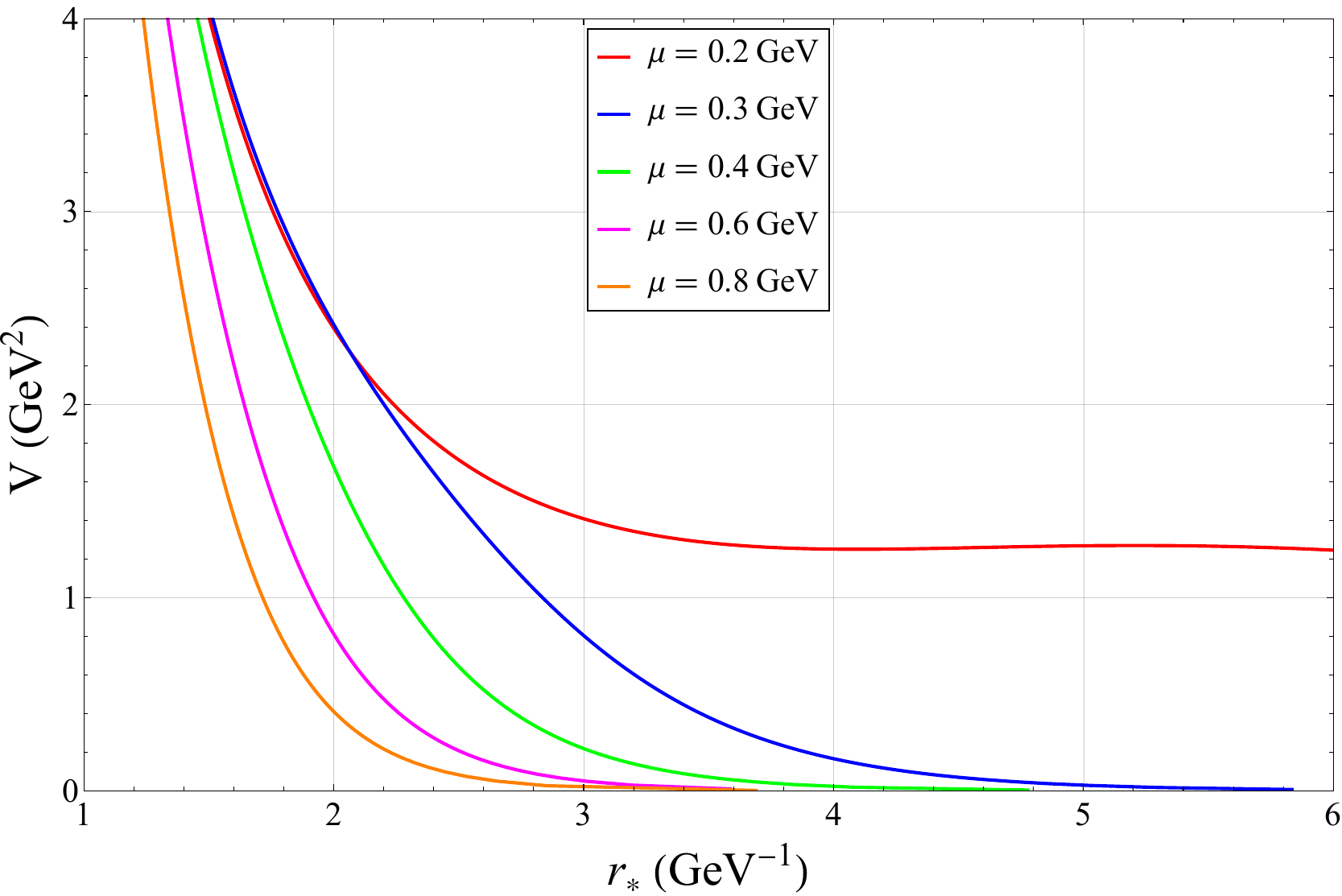}
    \caption{\small Effective potential $V$ at $T=270 \, \text{MeV}$ as a function of $r_*$ for various values of $\mu$. Here hybrid meson parameters are used.}
    \phantomsection
    \label{fig:potmu}
\end{figure}

As we now have a formula for $g(z)$ that explicitly depends on $\mu$, we can consider how a finite density configuration affects the melting process. For the hybrid meson parameters, we plug the newfound metric into the fluctuation's equation of motion (\ref{EOM_T}) to find how $\mu$ changes the effective potential (Fig.~\ref{fig:potmu}) for a given fixed temperature, $T=270 \, \text{MeV}$ in this case.

We observe that the increase in $\mu$ has a similar effect as the increase in temperature, evidenced by comparing Fig.~\ref{fig:potmu} with Figs.~\ref{fig:Tpotcharm} -- \ref{fig:Tpothybrid}. Given so, the effective potential indicates that the introduction of a chemical potential speeds up the melting process. 

\begin{figure}[h!]
    \centering
    \includegraphics[width=0.75\textwidth]{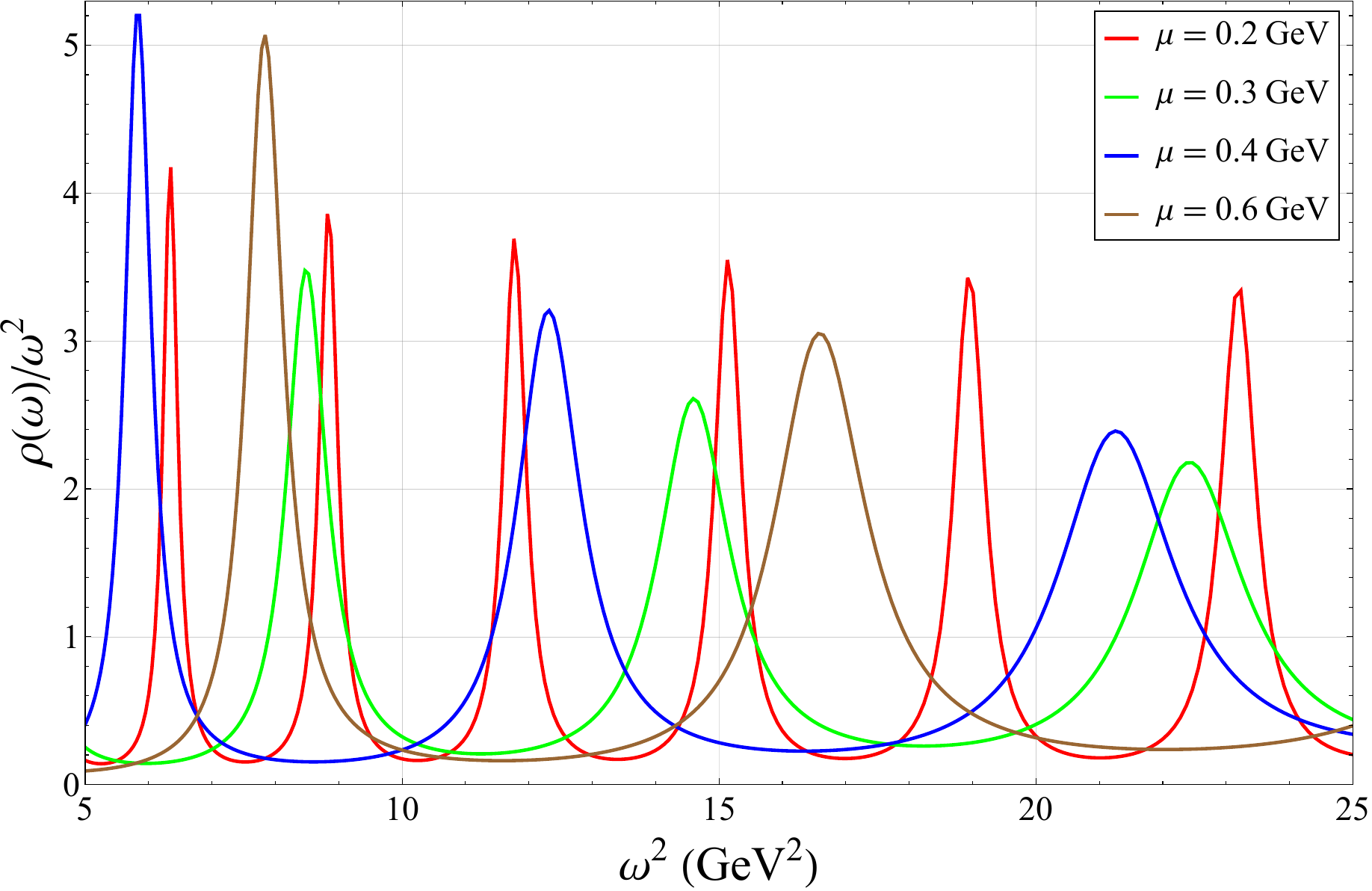}
    \caption{\small Spectral functions at $T=270~\text{MeV}$ for various values of $\mu$.}
    \phantomsection
    \label{fig:figmuspec}
\end{figure}

If we apply the numerical procedure outlined in Section \ref{sec5} to compute the spectral functions\footnote{In this case, the Pad\'{e}-like approximant becomes unnecessary as the function $f(z)$ does not contain the hyperbolic tangent term.} for this finite density system, we are able to compute and plot the spectral functions, for a fixed $T=270 \, \text{MeV}$ temperature, and see how the peaks change as $\mu$ becomes larger (Fig.~\ref{fig:figmuspec}). 

Fig.~\ref{fig:figmuspec} shows that, as $\mu$ increases, the peaks not only broaden out but also the distance between the consecutive peaks increases. This indicates, as the effective potential demonstrated, the melting process occurring more rapidly. This implies that the medium at higher baryon density also melts the resonances—much the same way a hot medium does at zero density. Even at a fixed temperature, raising $\mu$ can cause these once‐sharp excitations to lose their strength and ultimately dissolve into the continuum. The fact that one still sees fairly well‐defined peaks for small to moderate
$\mu$ means that the mesonic states are not immediately destroyed at low density, but eventually, at sufficiently large $\mu$, they do disappear, signaling the loss of a well‐defined bound state in a dense medium.

\begin{figure}[h!]
    \centering
    \includegraphics[width=0.75\textwidth]{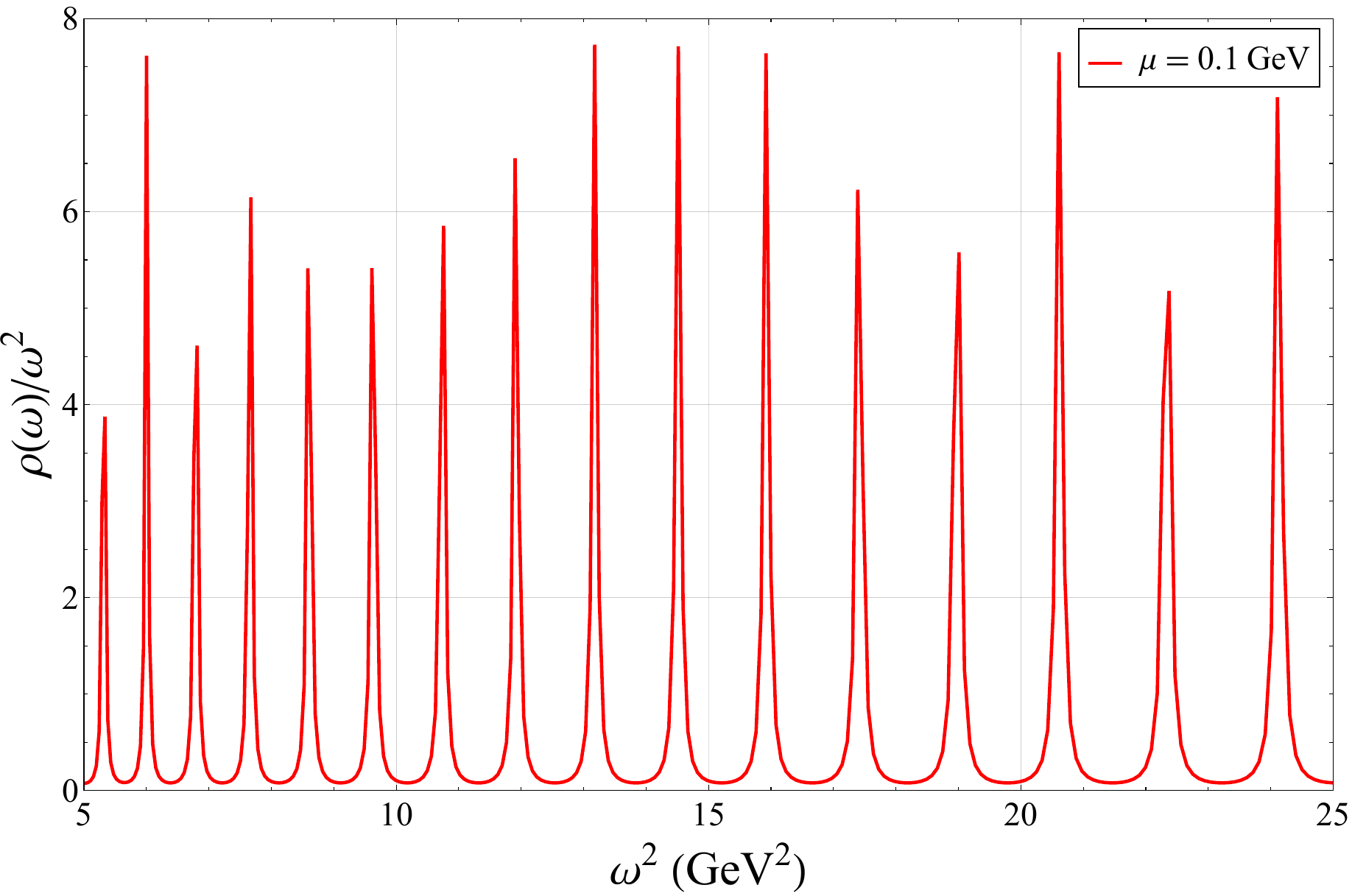}
    \caption{\small Spectral function in the small black hole geometry. Here $\mu=0.1~\text{GeV}$ and $T=0.1~\text{GeV}$ are used.}
    \phantomsection
    \label{fig:small}
\end{figure}

It is also interesting to investigate the spectral function in the spacious-confined phase, corresponding to a small black hole phase on the dual gravity side. For this purpose, we choose the parameters $\mu=0.1~\text{GeV}$ and $T=0.1~\text{GeV}$, well within the small black hole geometry (large $z_h$), and apply the already developed numerical procedure. The result is shown in Fig.~\ref{fig:small}.

We can clearly see that the spectral function exhibits a large number of very sharp and localized peaks in the specious-confined phase. This is what one should expect since this mimicking of confined QCD should generate sharp and well-defined states, compatible with one expects from genuine QCD at low scales. The same behavior is observed at small temperatures in the specious-confined phase for other values of $\mu$ as well. In the limit of zero temperature and chemical potential, the spectral function is a collection of delta functions centered at the particle's masses. 

The above specious-confined phase result is in contrast with what is seen in the deconfined phase. In Fig.~\ref{fig:figmuspec}, where we have a considerably smaller $z_h$, the peaks are more spurious and broader, compared to what is seen on Fig.~\ref{fig:small}. This signals the melting of the states in the finite density plasma. 

We conclude then, from the spectral functions computations, that the small black hole phase represents well defined, non-melted, states and the large black hole phase represents, at least for high enough temperatures and chemical potentials, non well defined, melted, states. 

One important thing to be noticed is that despite the large/small black hole geometry transition being regarded as a first order phase transition, we see that this does not correspond to substantial variations to the spectral functions near the transition point. We observe a gradual and smooth transition from the confined states to the melted states. In other words, to be able to see the clear differences in geometries---at least from the perspective of the spectral functions---that reproduce the different QCD matter phases, we have to go ``deep'' into the large/small geometries so that we examine a ``very small'' (low temperature) and ``very large'' (high temperature) black hole. 

\begin{figure}[h!]
    \centering
    \includegraphics[width=0.75\textwidth]{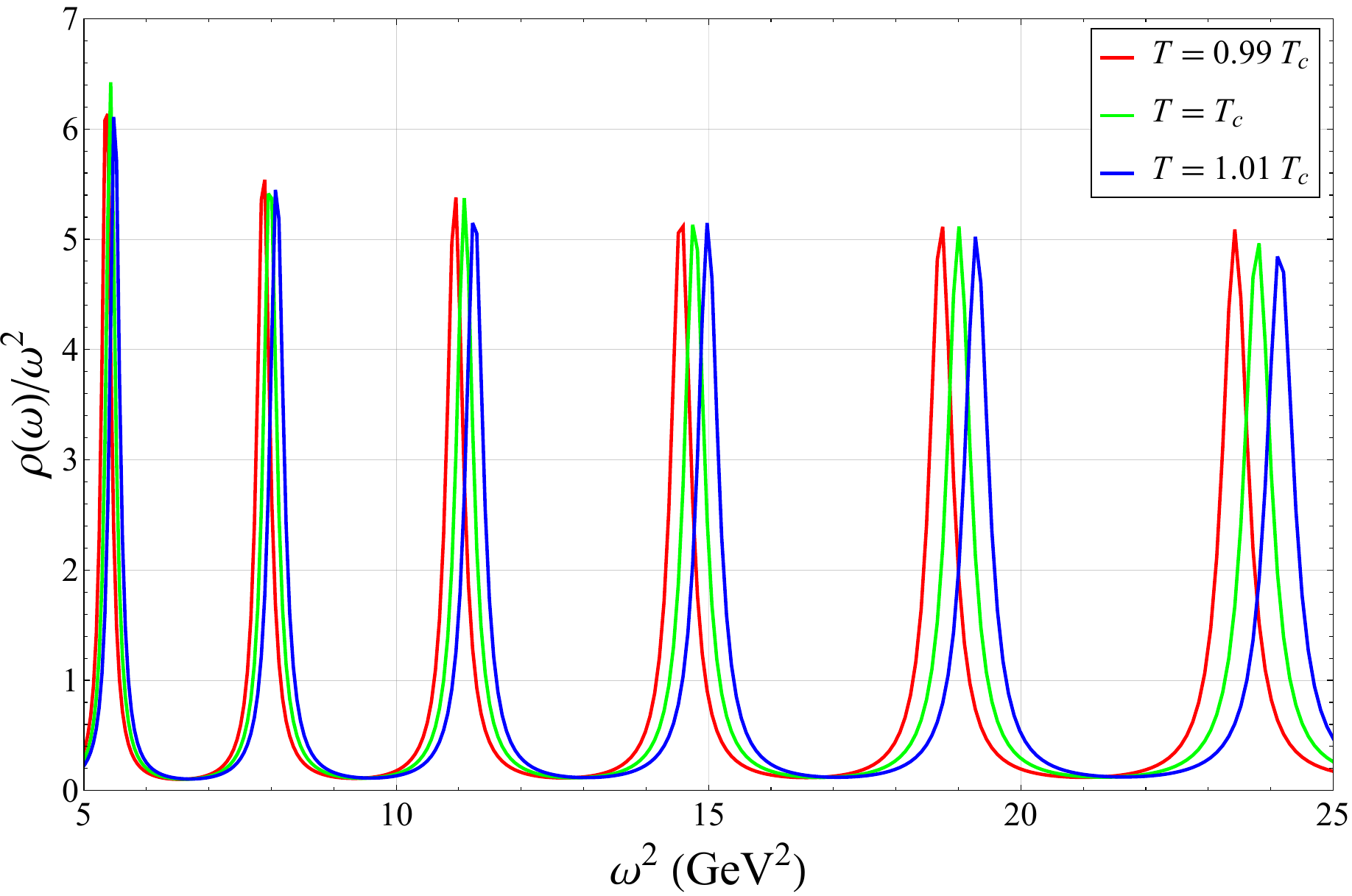}
    \caption{\small Spectral functions around the critical temperature for $\mu=0.1 ~\text{GeV}$, large and small black hole phase.}
    \phantomsection
    \label{fig:trans}
\end{figure}

To illustrate this, in Fig.~\ref{fig:trans}, we plot the spectral functions near the transition temperature for $\mu=0.1~\text{GeV}$, that corresponds to $T_{c} = 0.26~\text{GeV}$, in both large ($T=1.01 T_{c}$) and small ($T=0.99 T_{c}$) black hole configurations. 

One other point of interest is the critical end point (the blue point at the end of the phase structure curve, shown in Fig.~\ref{fig:criticalphase}) that defines where, in parameter space, the two black hole phases merge to form a single stable black hole that exist for all temperatures. 

\begin{figure}[htb!]
    \centering
    \begin{minipage}[t]{0.49\textwidth}
        \centering
        \includegraphics[width=\textwidth]{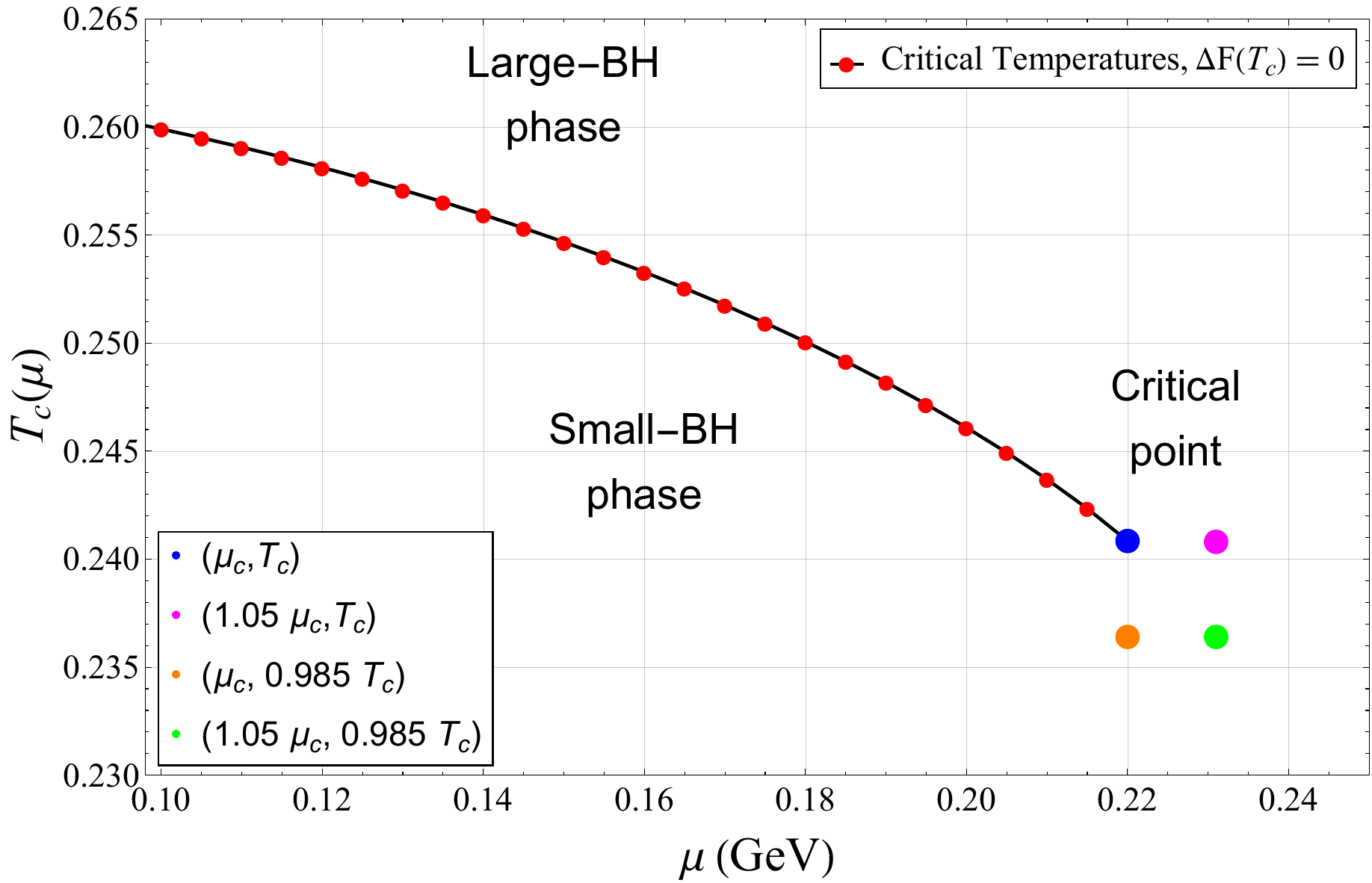}
        \caption{\small Phase structure near the critical end point.}
        \label{fig:squarephase}
    \end{minipage}%
    \hfill
    \begin{minipage}[t]{0.49\textwidth}
        \centering
        \includegraphics[width=\textwidth]{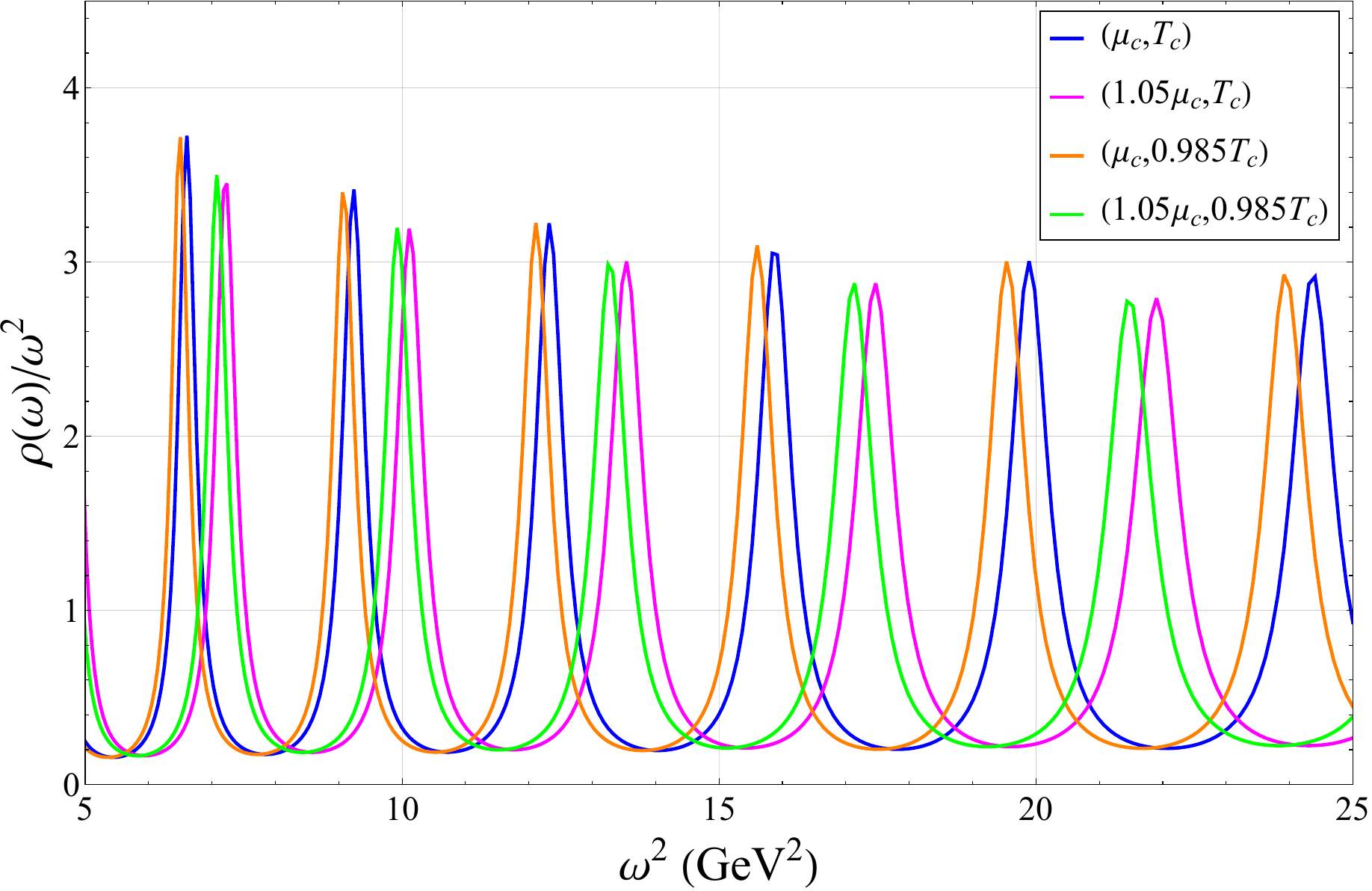}
        \caption{\small Spectral functions around the critical end point.}
        \label{fig:squarespec}
    \end{minipage}
\end{figure}

We then also investigate how the spectral functions change near this critical point $(\mu_c,T_c) \sim (0.22~\text{GeV},0.24~\text{GeV})$. In Fig.~\ref{fig:squarephase} we show, in the $\mu \times T_c$ plane, the chosen points of interest, near the critical end point, and, in Fig~\ref{fig:squarespec}, the corresponding spectral functions at these points.  

Again, we observe a smooth change in the spectral functions near the point where the distinct black hole geometries merge together so the comments made earlier for Fig.~\ref{fig:trans} still persist.

\section{Conclusion} \label{sec8}

In this work, we have developed a self-consistent dynamical holographic QCD model within the Einstein-Maxwell-Dilaton (EMD) framework to investigate the properties of heavy and exotic mesons at finite temperature and baryon density. By solving the coupled EMD equations analytically, we constructed a mathematically consistent holographic dual of QCD that mimics essential features of real QCD such as confinement, deconfinement, and mesonic melting. This approach successfully incorporated a non-quadratic dilaton profile to describe non-linear Regge trajectories that better match the mass spectrum predictions for heavy mesons and models exotic states. The numerically obtained spectra demonstrated good agreement with experimental data, reinforcing the validity of the holographic framework.

At finite temperature, the confinement-deconfinement transition was analyzed through a Hawking-Page phase transition, revealing the sequential melting of mesons as the temperature increased. This was further confirmed by the computation of spectral functions, where the broadening and eventual disappearance of spectral peaks indicated the dissolution of mesonic states. The deformation of the Schrödinger-like effective potential with temperature reinforced the conclusion that bound mesonic states to destabilize in a hot QCD medium.

Significantly, in the final section, we extended the analysis to a finite density plasma by introducing a nonzero chemical potential $\mu$. Here, we observed that increasing the baryon chemical potential had a profound effect on meson stability. The spectral functions revealed a faster broadening and disappearance of peaks compared to the finite-temperature case, indicating an accelerated melting process in a dense medium. This suggests that the presence of a baryonic background leads to more rapid dissociation of mesonic states, a phenomenon that is particularly relevant for heavy-ion collision experiments and the study of the quark-gluon plasma (QGP) in neutron star mergers.

A critical aspect of the finite-density analysis was the distinction between small and large black hole phases in the dual gravitational theory and their impact on mesonic states. The small black hole phase, at sufficiently low temperatures, shows that the spectral function exhibits the presence of bound states, characterized by well-defined sharp peaks. This phase is associated with a confined-like regime where mesonic states persist. The large black hole phase, at sufficiently high temperatures, corresponds to the full deconfinement of quark matter. In this regime, the spectral function loses all well-defined peaks, signaling the complete dissolution of mesons into the QGP. The transition between these two black hole solutions provides a geometric analog of the quarkyonic transition, where matter shifts from a phase dominated by hadronic degrees of freedom to a deconfined quark state. The spectral function evolution clearly illustrates this transition, at least their behavior is quite distinctive when ``deep enough'' in either one of the phases, making the spectral functions a valuable tool for studying QCD at finite temperatures and density. For completeness we note that the spectral functions do not immediately undergo any sudden change when crossing the transition.

A natural next step using this framework is to compute real-time transport coefficients in the hydrodynamic limit. It would be interesting to see how the transport coefficients of heavy mesons are affected by this EMD model that incorporates a non-quadratic Regge trajectory. 

An appropriate perspective includes using gauge/gravity duality, in the context developed heretofore in this work, to also describe pentaquarks \cite{Andreev:2023hmh}. 
Near-threshold photoproduction of charmonium and charmonium-like states was used in the direct production of the hidden-charm 
$P_c(4312)^+$, $P_c(4380)^+$, $P_c(4440)^+$ and $P_c(4457)^+$,  charmonium-pentaquarks observed by LHCb \cite{LHCb:2019kea,LHCb:2015yax}.
In addition, the recent discovery of the  $P_{\psi s}(4338)^0$, whose composition consists of $udsc\bar{c}$ quarks \cite{LHCb:2022ogu}, represents the first confirmed pentaquark containing a strange quark, and it needs to be better understood.  
Our developments may shed some new light on these pentaquarks and other exotic charmonium-like states composing  tetraquarks and hadronic molecules, and their nature.

\section*{Acknowledgments}
B.T.~thanks to the São Paulo Research Foundation -- FAPESP (Grants No. 2023/09097-9 and No. 2022/15325-1) for financial support. R.d.R~thanks to FAPESP
(Grants No. 2021/01089-1 and No. 2024/05676-7), and to  CNPq  (Grants No. 303742/2023-2 and No. 401567/2023-0), for partial financial support. The work of S.S.J.~is supported by Grant No. 09/983(0045)/2019-EMR-I from CSIR-HRDG, India. The work of S.M.~is supported by the core research grant from the Science and Engineering Research Board, a statutory body under the Department of Science and Technology, Government of India, under grant agreement number CRG/2023/007670.

\bibliographystyle{JHEP}
\bibliography{mybib}

\providecommand{\href}[2]{#2}\begingroup\raggedright\begin{thebibliography}{100}

\bibitem{Karch:2006pv}
A.~Karch, E.~Katz, D.T.~Son and M.A.~Stephanov, \emph{{Linear confinement and AdS/QCD}}, \href{https://doi.org/10.1103/PhysRevD.74.015005}{\emph{Phys. Rev. D} {\bfseries 74} (2006) 015005} [\href{https://arxiv.org/abs/hep-ph/0602229}{{\ttfamily hep-ph/0602229}}].

\bibitem{Branz:2010ub}
T.~Branz, T.~Gutsche, V.E.~Lyubovitskij, I.~Schmidt and A.~Vega, \emph{{Light and heavy mesons in a soft-wall holographic approach}}, \href{https://doi.org/10.1103/PhysRevD.82.074022}{\emph{Phys. Rev. D} {\bfseries 82} (2010) 074022} [\href{https://arxiv.org/abs/1008.0268}{{\ttfamily 1008.0268}}].

\bibitem{Colangelo:2008us}
P.~Colangelo, F.~De~Fazio, F.~Giannuzzi, F.~Jugeau and S.~Nicotri, \emph{{Light scalar mesons in the soft-wall model of AdS/QCD}}, \href{https://doi.org/10.1103/PhysRevD.78.055009}{\emph{Phys. Rev. D} {\bfseries 78} (2008) 055009} [\href{https://arxiv.org/abs/0807.1054}{{\ttfamily 0807.1054}}].

\bibitem{Brodsky:2014yha}
S.J.~Brodsky, G.F.~de~Teramond, H.G.~Dosch and J.~Erlich, \emph{{Light-Front Holographic QCD and Emerging Confinement}}, \href{https://doi.org/10.1016/j.physrep.2015.05.001}{\emph{Phys. Rept.} {\bfseries 584} (2015) 1} [\href{https://arxiv.org/abs/1407.8131}{{\ttfamily 1407.8131}}].

\bibitem{Erlich:2005qh}
J.~Erlich, E.~Katz, D.T.~Son and M.A.~Stephanov, \emph{{QCD and a holographic model of hadrons}}, \href{https://doi.org/10.1103/PhysRevLett.95.261602}{\emph{Phys. Rev. Lett.} {\bfseries 95} (2005) 261602} [\href{https://arxiv.org/abs/hep-ph/0501128}{{\ttfamily hep-ph/0501128}}].

\bibitem{Sakai2005Apr}
T.~Sakai and S.~Sugimoto, \emph{{Low Energy Hadron Physics in Holographic QCD}}, \href{https://doi.org/10.1143/PTP.113.843}{\emph{Prog. Theor. Phys.} {\bfseries 113} (2005) 843}.

\bibitem{Sakai2005Nov}
T.~Sakai and S.~Sugimoto, \emph{{More on a Holographic Dual of QCD}}, \href{https://doi.org/10.1143/PTP.114.1083}{\emph{Prog. Theor. Phys.} {\bfseries 114} (2005) 1083}.

\bibitem{Rougemont:2023gfz}
R.~Rougemont, J.~Grefa, M.~Hippert, J.~Noronha, J.~Noronha-Hostler, I.~Portillo et~al., \emph{{Hot QCD phase diagram from holographic Einstein\textendash{}Maxwell\textendash{}Dilaton models}}, \href{https://doi.org/10.1016/j.ppnp.2023.104093}{\emph{Prog. Part. Nucl. Phys.} {\bfseries 135} (2024) 104093} [\href{https://arxiv.org/abs/2307.03885}{{\ttfamily 2307.03885}}].

\bibitem{Maldacena:1997re}
J.M.~Maldacena, \emph{{The Large N limit of superconformal field theories and supergravity}}, \href{https://doi.org/10.4310/ATMP.1998.v2.n2.a1}{\emph{Adv. Theor. Math. Phys.} {\bfseries 2} (1998) 231} [\href{https://arxiv.org/abs/hep-th/9711200}{{\ttfamily hep-th/9711200}}].

\bibitem{Witten:1998qj}
E.~Witten, \emph{{Anti-de Sitter space and holography}}, \href{https://doi.org/10.4310/ATMP.1998.v2.n2.a2}{\emph{Adv. Theor. Math. Phys.} {\bfseries 2} (1998) 253} [\href{https://arxiv.org/abs/hep-th/9802150}{{\ttfamily hep-th/9802150}}].

\bibitem{Gubser:1998bc}
S.S.~Gubser, I.R.~Klebanov and A.M.~Polyakov, \emph{{Gauge theory correlators from noncritical string theory}}, \href{https://doi.org/10.1016/S0370-2693(98)00377-3}{\emph{Phys. Lett. B} {\bfseries 428} (1998) 105} [\href{https://arxiv.org/abs/hep-th/9802109}{{\ttfamily hep-th/9802109}}].

\bibitem{Aharony:1999ti}
O.~Aharony, S.S.~Gubser, J.M.~Maldacena, H.~Ooguri and Y.~Oz, \emph{{Large N field theories, string theory and gravity}}, \href{https://doi.org/10.1016/S0370-1573(99)00083-6}{\emph{Phys. Rept.} {\bfseries 323} (2000) 183} [\href{https://arxiv.org/abs/hep-th/9905111}{{\ttfamily hep-th/9905111}}].

\bibitem{Ammon:2015wua}
M.~Ammon and J.~Erdmenger, \emph{{Gauge/gravity duality}: {Foundations and applications}}, Cambridge University Press, Cambridge (4, 2015), \href{https://doi.org/10.1017/CBO9780511846373}{10.1017/CBO9780511846373}.

\bibitem{Afonin:2022hbb}
S.S.~Afonin, \emph{{Bottom-Up Holographic Approach to Meson Spectroscopy}}, \href{https://doi.org/10.1134/S1063779622020034}{\emph{Phys. Part. Nucl.} {\bfseries 53} (2022) 96}.

\bibitem{dePaula:2009za}
W.~de~Paula and T.~Frederico, \emph{{Scalar mesons within a dynamical holographic QCD model}}, \href{https://doi.org/10.1016/j.physletb.2010.08.045}{\emph{Phys. Lett. B} {\bfseries 693} (2010) 287} [\href{https://arxiv.org/abs/0908.4282}{{\ttfamily 0908.4282}}].

\bibitem{Ballon-Bayona:2017sxa}
A.~Ballon-Bayona, H.~Boschi-Filho, L.A.H.~Mamani, A.S.~Miranda and V.T.~Zanchin, \emph{{Effective holographic models for QCD: glueball spectrum and trace anomaly}}, \href{https://doi.org/10.1103/PhysRevD.97.046001}{\emph{Phys. Rev. D} {\bfseries 97} (2018) 046001} [\href{https://arxiv.org/abs/1708.08968}{{\ttfamily 1708.08968}}].

\bibitem{Ballon-Bayona:2021ibm}
A.~Ballon-Bayona, L.A.H.~Mamani and D.M.~Rodrigues, \emph{{Spontaneous chiral symmetry breaking in holographic soft wall models}}, \href{https://doi.org/10.1103/PhysRevD.104.126029}{\emph{Phys. Rev. D} {\bfseries 104} (2021) 126029} [\href{https://arxiv.org/abs/2107.10983}{{\ttfamily 2107.10983}}].

\bibitem{Karapetyan:2023sfo}
G.~Karapetyan, W.~de~Paula and R.~da~Rocha, \emph{{Configurational information measure of mesonic states in 4-flavor AdS/QCD}}, \href{https://doi.org/10.1016/j.physletb.2023.138174}{\emph{Phys. Lett. B} {\bfseries 846} (2023) 138174} [\href{https://arxiv.org/abs/2304.05122}{{\ttfamily 2304.05122}}].

\bibitem{Bartz:2018nzn}
S.P.~Bartz, A.~Dhumuntarao and J.I.~Kapusta, \emph{{Dynamical AdS/Yang-Mills model}}, \href{https://doi.org/10.1103/PhysRevD.98.026019}{\emph{Phys. Rev. D} {\bfseries 98} (2018) 026019} [\href{https://arxiv.org/abs/1801.06118}{{\ttfamily 1801.06118}}].

\bibitem{Ballon-Bayona:2023zal}
A.~Ballon-Bayona, T.~Frederico, L.A.H.~Mamani and W.~de~Paula, \emph{{Dynamical holographic QCD model for spontaneous chiral symmetry breaking and confinement}}, \href{https://doi.org/10.1103/PhysRevD.108.106016}{\emph{Phys. Rev. D} {\bfseries 108} (2023) 106016} [\href{https://arxiv.org/abs/2308.07503}{{\ttfamily 2308.07503}}].

\bibitem{Aref'eva2018May}
I.~Aref{'}eva and K.~Rannu, \emph{{Holographic anisotropic background with confinement-deconfinement phase transition}}, \href{https://doi.org/10.1007/JHEP05(2018)206}{\emph{J. High Energy Phys.} {\bfseries 2018} (2018) 1}.

\bibitem{Arefeva:2024vom}
I.Y.~Aref'eva, A.~Hajilou, P.~Slepov and M.~Usova, \emph{{Running coupling for holographic QCD with heavy and light quarks: Isotropic case}}, \href{https://doi.org/10.1103/PhysRevD.110.126009}{\emph{Phys. Rev. D} {\bfseries 110} (2024) 126009} [\href{https://arxiv.org/abs/2402.14512}{{\ttfamily 2402.14512}}].

\bibitem{Erdmenger:2020flu}
J.~Erdmenger, N.~Evans, W.~Porod and K.S.~Rigatos, \emph{{Gauge/gravity dual dynamics for the strongly coupled sector of composite Higgs models}}, \href{https://doi.org/10.1007/JHEP02(2021)058}{\emph{JHEP} {\bfseries 02} (2021) 058} [\href{https://arxiv.org/abs/2010.10279}{{\ttfamily 2010.10279}}].

\bibitem{Gursoy:2007cb}
U.~Gursoy and E.~Kiritsis, \emph{{Exploring improved holographic theories for QCD: Part I}}, \href{https://doi.org/10.1088/1126-6708/2008/02/032}{\emph{JHEP} {\bfseries 02} (2008) 032} [\href{https://arxiv.org/abs/0707.1324}{{\ttfamily 0707.1324}}].

\bibitem{Gursoy:2007er}
U.~Gursoy, E.~Kiritsis and F.~Nitti, \emph{{Exploring improved holographic theories for QCD: Part II}}, \href{https://doi.org/10.1088/1126-6708/2008/02/019}{\emph{JHEP} {\bfseries 02} (2008) 019} [\href{https://arxiv.org/abs/0707.1349}{{\ttfamily 0707.1349}}].

\bibitem{Gursoy:2010fj}
U.~Gursoy, E.~Kiritsis, L.~Mazzanti, G.~Michalogiorgakis and F.~Nitti, \emph{{Improved Holographic QCD}}, \href{https://doi.org/10.1007/978-3-642-04864-7_4}{\emph{Lect. Notes Phys.} {\bfseries 828} (2011) 79} [\href{https://arxiv.org/abs/1006.5461}{{\ttfamily 1006.5461}}].

\bibitem{He2013Apr}
S.~He, S.-Y.~Wu, Y.~Yang and P.-H.~Yuan, \emph{{Phase structure in a dynamical soft-wall holographic QCD model}}, \href{https://doi.org/10.1007/JHEP04(2013)093}{\emph{J. High Energy Phys.} {\bfseries 2013} (2013) 1}.

\bibitem{Alho:2013hsa}
T.~Alho, M.~J\"arvinen, K.~Kajantie, E.~Kiritsis, C.~Rosen and K.~Tuominen, \emph{{A holographic model for QCD in the Veneziano limit at finite temperature and density}}, \href{https://doi.org/10.1007/JHEP04(2014)124}{\emph{JHEP} {\bfseries 04} (2014) 124} [\href{https://arxiv.org/abs/1312.5199}{{\ttfamily 1312.5199}}].

\bibitem{Alanen:2009xs}
J.~Alanen, K.~Kajantie and V.~Suur-Uski, \emph{{A gauge/gravity duality model for gauge theory thermodynamics}}, \href{https://doi.org/10.1103/PhysRevD.80.126008}{\emph{Phys. Rev. D} {\bfseries 80} (2009) 126008} [\href{https://arxiv.org/abs/0911.2114}{{\ttfamily 0911.2114}}].

\bibitem{Csaki}
C.~Csaki and M.~Reece, \emph{{Toward a systematic holographic QCD: A Braneless approach}}, \href{https://doi.org/10.1088/1126-6708/2007/05/062}{\emph{JHEP} {\bfseries 05} (2007) 062} [\href{https://arxiv.org/abs/hep-ph/0608266}{{\ttfamily hep-ph/0608266}}].

\bibitem{Gherghetta:2009ac}
T.~Gherghetta, J.I.~Kapusta and T.M.~Kelley, \emph{{Chiral symmetry breaking in the soft-wall AdS/QCD model}}, \href{https://doi.org/10.1103/PhysRevD.79.076003}{\emph{Phys. Rev. D} {\bfseries 79} (2009) 076003} [\href{https://arxiv.org/abs/0902.1998}{{\ttfamily 0902.1998}}].

\bibitem{Bartz:2024dgd}
S.P.~Bartz, R.C.~Meadows and G.~Brock, \emph{{Chiral phase transition in soft-wall AdS/QCD with scalar-dilaton coupling}}, \href{https://doi.org/10.1103/PhysRevD.110.026027}{\emph{Phys. Rev. D} {\bfseries 110} (2024) 026027} [\href{https://arxiv.org/abs/2404.10104}{{\ttfamily 2404.10104}}].

\bibitem{Braga:2017oqw}
N.R.F.~Braga and L.F.~Ferreira, \emph{{Bottomonium dissociation in a finite density plasma}}, \href{https://doi.org/10.1016/j.physletb.2017.08.037}{\emph{Phys. Lett. B} {\bfseries 773} (2017) 313} [\href{https://arxiv.org/abs/1704.05038}{{\ttfamily 1704.05038}}].

\bibitem{Ferreira:2019inu}
L.F.~Ferreira and R.~da~Rocha, \emph{{Pion family in AdS/QCD: the next generation from configurational entropy}}, \href{https://doi.org/10.1103/PhysRevD.99.086001}{\emph{Phys. Rev. D} {\bfseries 99} (2019) 086001} [\href{https://arxiv.org/abs/1902.04534}{{\ttfamily 1902.04534}}].

\bibitem{daRocha:2021imz}
R.~da~Rocha, \emph{{Deploying heavier $\eta$ meson states: Configurational entropy hybridizing AdS/QCD}}, \href{https://doi.org/10.1016/j.physletb.2021.136112}{\emph{Phys. Lett. B} {\bfseries 814} (2021) 136112} [\href{https://arxiv.org/abs/2101.03602}{{\ttfamily 2101.03602}}].

\bibitem{Ferreira:2019nkz}
L.F.~Ferreira and R.~da~Rocha, \emph{{Tensor mesons, AdS/QCD and information}}, \href{https://doi.org/10.1140/epjc/s10052-020-7978-7}{\emph{Eur. Phys. J. C} {\bfseries 80} (2020) 375} [\href{https://arxiv.org/abs/1907.11809}{{\ttfamily 1907.11809}}].

\bibitem{Braga:2018fyc}
N.R.F.~Braga, L.F.~Ferreira and R.~da~Rocha, \emph{{Thermal dissociation of heavy mesons and configurational entropy}}, \href{https://doi.org/10.1016/j.physletb.2018.10.036}{\emph{Phys. Lett. B} {\bfseries 787} (2018) 16} [\href{https://arxiv.org/abs/1808.10499}{{\ttfamily 1808.10499}}].

\bibitem{Braga:2023qee}
N.R.F.~Braga, L.F.~Ferreira and O.C.~Junqueira, \emph{{Configuration entropy of a rotating quark-gluon plasma from holography}}, \href{https://doi.org/10.1016/j.physletb.2023.138265}{\emph{Phys. Lett. B} {\bfseries 847} (2023) 138265} [\href{https://arxiv.org/abs/2301.01322}{{\ttfamily 2301.01322}}].

\bibitem{Colangelo:2018mrt}
P.~Colangelo and F.~Loparco, \emph{{Configurational Entropy can disentangle conventional hadrons from exotica}}, \href{https://doi.org/10.1016/j.physletb.2018.11.053}{\emph{Phys. Lett. B} {\bfseries 788} (2019) 500} [\href{https://arxiv.org/abs/1811.05272}{{\ttfamily 1811.05272}}].

\bibitem{Ferreira:2020iry}
L.F.~Ferreira and R.~da~Rocha, \emph{{Nucleons and higher spin baryon resonances: An AdS/QCD configurational entropic incursion}}, \href{https://doi.org/10.1103/PhysRevD.101.106002}{\emph{Phys. Rev. D} {\bfseries 101} (2020) 106002} [\href{https://arxiv.org/abs/2004.04551}{{\ttfamily 2004.04551}}].

\bibitem{Bernardini:2018uuy}
A.E.~Bernardini and R.~da~Rocha, \emph{{Informational entropic Regge trajectories of meson families in AdS/QCD}}, \href{https://doi.org/10.1103/PhysRevD.98.126011}{\emph{Phys. Rev. D} {\bfseries 98} (2018) 126011} [\href{https://arxiv.org/abs/1809.10055}{{\ttfamily 1809.10055}}].

\bibitem{Barbosa-Cendejas:2018mng}
N.~Barbosa-Cendejas, R.~Cartas-Fuentevilla, A.~Herrera-Aguilar, R.R.~Mora-Luna and R.~da~Rocha, \emph{{Dynamical tachyonic AdS/QCD and information entropy}}, \href{https://doi.org/10.1016/j.physletb.2018.06.007}{\emph{Phys. Lett. B} {\bfseries 782} (2018) 607} [\href{https://arxiv.org/abs/1805.04485}{{\ttfamily 1805.04485}}].

\bibitem{daRocha:2021ntm}
R.~da~Rocha, \emph{{Information entropy in AdS/QCD: Mass spectroscopy of isovector mesons}}, \href{https://doi.org/10.1103/PhysRevD.103.106027}{\emph{Phys. Rev. D} {\bfseries 103} (2021) 106027} [\href{https://arxiv.org/abs/2103.03924}{{\ttfamily 2103.03924}}].

\bibitem{daRocha:2024sjn}
R.~da~Rocha and P.H.O.~Silva, \emph{{Deformed AdS/QCD, mesonic mass spectra, and the differential configurational entropy: Still a margin for heavier resonances}}, \href{https://doi.org/10.1103/PhysRevD.110.126019}{\emph{Phys. Rev. D} {\bfseries 110} (2024) 126019} [\href{https://arxiv.org/abs/2412.02375}{{\ttfamily 2412.02375}}].

\bibitem{MartinContreras:2020cyg}
M.A.~Martin~Contreras and A.~Vega, \emph{{Nonlinear Regge trajectories with AdS/QCD}}, \href{https://doi.org/10.1103/PhysRevD.102.046007}{\emph{Phys. Rev. D} {\bfseries 102} (2020) 046007} [\href{https://arxiv.org/abs/2004.10286}{{\ttfamily 2004.10286}}].

\bibitem{Karapetyan:2021ufz}
G.~Karapetyan and R.~da~Rocha, \emph{{Configurational entropy of heavy-quark QCD exotica}}, \href{https://doi.org/10.1140/epjp/s13360-021-01942-7}{\emph{Eur. Phys. J. Plus} {\bfseries 136} (2021) 993} [\href{https://arxiv.org/abs/2103.10863}{{\ttfamily 2103.10863}}].

\bibitem{MartinContreras:2023oqs}
M.A.~Martin~Contreras and A.~Vega, \emph{{Holographic stability for non-qq\textasciimacron{} candidates}}, \href{https://doi.org/10.1103/PhysRevD.108.126024}{\emph{Phys. Rev. D} {\bfseries 108} (2023) 126024} [\href{https://arxiv.org/abs/2309.02905}{{\ttfamily 2309.02905}}].

\bibitem{DeWolfe:2010he}
O.~DeWolfe, S.S.~Gubser and C.~Rosen, \emph{{A holographic critical point}}, \href{https://doi.org/10.1103/PhysRevD.83.086005}{\emph{Phys. Rev. D} {\bfseries 83} (2011) 086005} [\href{https://arxiv.org/abs/1012.1864}{{\ttfamily 1012.1864}}].

\bibitem{DeWolfe:2011ts}
O.~DeWolfe, S.S.~Gubser and C.~Rosen, \emph{{Dynamic critical phenomena at a holographic critical point}}, \href{https://doi.org/10.1103/PhysRevD.84.126014}{\emph{Phys. Rev. D} {\bfseries 84} (2011) 126014} [\href{https://arxiv.org/abs/1108.2029}{{\ttfamily 1108.2029}}].

\bibitem{Rougemont:2015ona}
R.~Rougemont, J.~Noronha and J.~Noronha-Hostler, \emph{{Suppression of baryon diffusion and transport in a baryon rich strongly coupled quark-gluon plasma}}, \href{https://doi.org/10.1103/PhysRevLett.115.202301}{\emph{Phys. Rev. Lett.} {\bfseries 115} (2015) 202301} [\href{https://arxiv.org/abs/1507.06972}{{\ttfamily 1507.06972}}].

\bibitem{Noronha:2009ud}
J.~Noronha, \emph{{Connecting Polyakov Loops to the Thermodynamics of SU(N(c)) Gauge Theories Using the Gauge-String Duality}}, \href{https://doi.org/10.1103/PhysRevD.81.045011}{\emph{Phys. Rev. D} {\bfseries 81} (2010) 045011} [\href{https://arxiv.org/abs/0910.1261}{{\ttfamily 0910.1261}}].

\bibitem{Noronha:2010hb}
J.~Noronha, \emph{{The Heavy Quark Free Energy in QCD and in Gauge Theories with Gravity Duals}}, \href{https://doi.org/10.1103/PhysRevD.82.065016}{\emph{Phys. Rev. D} {\bfseries 82} (2010) 065016} [\href{https://arxiv.org/abs/1003.0914}{{\ttfamily 1003.0914}}].

\bibitem{He:2013qq}
S.~He, S.-Y.~Wu, Y.~Yang and P.-H.~Yuan, \emph{{Phase Structure in a Dynamical Soft-Wall Holographic QCD Model}}, \href{https://doi.org/10.1007/JHEP04(2013)093}{\emph{JHEP} {\bfseries 04} (2013) 093} [\href{https://arxiv.org/abs/1301.0385}{{\ttfamily 1301.0385}}].

\bibitem{Ballon-Bayona:2020xls}
A.~Ballon-Bayona, H.~Boschi-Filho, E.F.~Capossoli and D.M.~Rodrigues, \emph{{Criticality from Einstein-Maxwell-dilaton holography at finite temperature and density}}, \href{https://doi.org/10.1103/PhysRevD.102.126003}{\emph{Phys. Rev. D} {\bfseries 102} (2020) 126003} [\href{https://arxiv.org/abs/2006.08810}{{\ttfamily 2006.08810}}].

\bibitem{Gubser:2008yx}
S.S.~Gubser, A.~Nellore, S.S.~Pufu and F.D.~Rocha, \emph{{Thermodynamics and bulk viscosity of approximate black hole duals to finite temperature quantum chromodynamics}}, \href{https://doi.org/10.1103/PhysRevLett.101.131601}{\emph{Phys. Rev. Lett.} {\bfseries 101} (2008) 131601} [\href{https://arxiv.org/abs/0804.1950}{{\ttfamily 0804.1950}}].

\bibitem{Bohra:2019ebj}
H.~Bohra, D.~Dudal, A.~Hajilou and S.~Mahapatra, \emph{{Anisotropic string tensions and inversely magnetic catalyzed deconfinement from a dynamical AdS/QCD model}}, \href{https://doi.org/10.1016/j.physletb.2019.135184}{\emph{Phys. Lett. B} {\bfseries 801} (2020) 135184} [\href{https://arxiv.org/abs/1907.01852}{{\ttfamily 1907.01852}}].

\bibitem{Bohra:2020qom}
H.~Bohra, D.~Dudal, A.~Hajilou and S.~Mahapatra, \emph{{Chiral transition in the probe approximation from an Einstein-Maxwell-dilaton gravity model}}, \href{https://doi.org/10.1103/PhysRevD.103.086021}{\emph{Phys. Rev. D} {\bfseries 103} (2021) 086021} [\href{https://arxiv.org/abs/2010.04578}{{\ttfamily 2010.04578}}].

\bibitem{Dudal:2021jav}
D.~Dudal, A.~Hajilou and S.~Mahapatra, \emph{{A quenched 2-flavour Einstein\textendash{}Maxwell\textendash{}Dilaton gauge-gravity model}}, \href{https://doi.org/10.1140/epja/s10050-021-00461-4}{\emph{Eur. Phys. J. A} {\bfseries 57} (2021) 142} [\href{https://arxiv.org/abs/2103.01185}{{\ttfamily 2103.01185}}].

\bibitem{Cai:2012xh}
R.-G.~Cai, S.~He and D.~Li, \emph{{A hQCD model and its phase diagram in Einstein-Maxwell-Dilaton system}}, \href{https://doi.org/10.1007/JHEP03(2012)033}{\emph{JHEP} {\bfseries 03} (2012) 033} [\href{https://arxiv.org/abs/1201.0820}{{\ttfamily 1201.0820}}].

\bibitem{Yang:2015aia}
Y.~Yang and P.-H.~Yuan, \emph{{Confinement-deconfinement phase transition for heavy quarks in a soft wall holographic QCD model}}, \href{https://doi.org/10.1007/JHEP12(2015)161}{\emph{JHEP} {\bfseries 12} (2015) 161} [\href{https://arxiv.org/abs/1506.05930}{{\ttfamily 1506.05930}}].

\bibitem{Ferreira-Martins:2019wym}
A.J.~Ferreira-Martins, P.~Meert and R.~da~Rocha, \emph{{Deformed AdS$_4$ \textendash{}Reissner\textendash{}Nordstr\"om black branes and shear viscosity-to-entropy density ratio}}, \href{https://doi.org/10.1140/epjc/s10052-019-7167-8}{\emph{Eur. Phys. J. C} {\bfseries 79} (2019) 646} [\href{https://arxiv.org/abs/1904.01093}{{\ttfamily 1904.01093}}].

\bibitem{Jena:2022nzw}
S.S.~Jena, B.~Shukla, D.~Dudal and S.~Mahapatra, \emph{{Entropic force and real-time dynamics of holographic quarkonium in a magnetic field}}, \href{https://doi.org/10.1103/PhysRevD.105.086011}{\emph{Phys. Rev. D} {\bfseries 105} (2022) 086011} [\href{https://arxiv.org/abs/2202.01486}{{\ttfamily 2202.01486}}].

\bibitem{Shukla:2023pbp}
B.~Shukla, D.~Dudal and S.~Mahapatra, \emph{{Anisotropic and frame dependent chaos of suspended strings from a dynamical holographic QCD model with magnetic field}}, \href{https://doi.org/10.1007/JHEP06(2023)178}{\emph{JHEP} {\bfseries 06} (2023) 178} [\href{https://arxiv.org/abs/2303.15716}{{\ttfamily 2303.15716}}].

\bibitem{Shukla:2024qlf}
B.~Shukla, J.~Nongmaithem, D.~Dudal and S.~Mahapatra, \emph{{Interplay of magnetic field and chemical potential induced anisotropy and frame dependent chaos of a $Q\bar{Q}$ pair in holographic QCD}},  \href{https://arxiv.org/abs/2411.17279}{{\ttfamily 2411.17279}}.

\bibitem{Dudal:2017max}
D.~Dudal and S.~Mahapatra, \emph{{Thermal entropy of a quark-antiquark pair above and below deconfinement from a dynamical holographic QCD model}}, \href{https://doi.org/10.1103/PhysRevD.96.126010}{\emph{Phys. Rev. D} {\bfseries 96} (2017) 126010} [\href{https://arxiv.org/abs/1708.06995}{{\ttfamily 1708.06995}}].

\bibitem{Song}
S.~He, M.~Huang, Q.-S.~Yan and Y.~Yang, \emph{{Confront Holographic QCD with Regge Trajectories}}, \href{https://doi.org/10.1140/epjc/s10052-010-1239-0}{\emph{Eur. Phys. J. C} {\bfseries 66} (2010) 187} [\href{https://arxiv.org/abs/0710.0988}{{\ttfamily 0710.0988}}].

\bibitem{Rinaldi:2020ybv}
M.~Rinaldi, \emph{{Double parton correlations in mesons within AdS/QCD soft-wall models: a first comparison with lattice data}}, \href{https://doi.org/10.1140/epjc/s10052-020-8241-y}{\emph{Eur. Phys. J. C} {\bfseries 80} (2020) 678} [\href{https://arxiv.org/abs/2003.09400}{{\ttfamily 2003.09400}}].

\bibitem{Afonin:2014nya}
S.S.~Afonin and I.V.~Pusenkov, \emph{{Universal description of radially excited heavy and light vector mesons}}, \href{https://doi.org/10.1103/PhysRevD.90.094020}{\emph{Phys. Rev. D} {\bfseries 90} (2014) 094020} [\href{https://arxiv.org/abs/1411.2390}{{\ttfamily 1411.2390}}].

\bibitem{jk}
J.-K.~Chen, \emph{{Regge trajectories for the mesons consisting of different quarks}}, \href{https://doi.org/10.1140/epjc/s10052-018-6134-0}{\emph{Eur. Phys. J. C} {\bfseries 78} (2018) 648}.

\bibitem{Guo}
F.-K.~Guo, C.~Hanhart, U.-G.~Mei\ss{}ner, Q.~Wang, Q.~Zhao and B.-S.~Zou, \emph{{Hadronic molecules}}, \href{https://doi.org/10.1103/RevModPhys.90.015004}{\emph{Rev. Mod. Phys.} {\bfseries 90} (2018) 015004} [\href{https://arxiv.org/abs/1705.00141}{{\ttfamily 1705.00141}}].

\bibitem{Lebed1}
R.F.~Lebed, R.E.~Mitchell and E.S.~Swanson, \emph{{Heavy-Quark QCD Exotica}}, \href{https://doi.org/10.1016/j.ppnp.2016.11.003}{\emph{Prog. Part. Nucl. Phys.} {\bfseries 93} (2017) 143} [\href{https://arxiv.org/abs/1610.04528}{{\ttfamily 1610.04528}}].

\bibitem{Oncala:2017hop}
R.~Oncala and J.~Soto, \emph{{Heavy Quarkonium Hybrids: Spectrum, Decay and Mixing}}, \href{https://doi.org/10.1103/PhysRevD.96.014004}{\emph{Phys. Rev. D} {\bfseries 96} (2017) 014004} [\href{https://arxiv.org/abs/1702.03900}{{\ttfamily 1702.03900}}].

\bibitem{Jaffe1}
R.L.~Jaffe, \emph{{Exotica}}, \href{https://doi.org/10.1016/j.physrep.2004.11.005}{\emph{Phys. Rept.} {\bfseries 409} (2005) 1} [\href{https://arxiv.org/abs/hep-ph/0409065}{{\ttfamily hep-ph/0409065}}].

\bibitem{ParticleDataGroup:2024cfk}
{\scshape Particle Data Group} collaboration, \emph{{Review of particle physics}}, \href{https://doi.org/10.1103/PhysRevD.110.030001}{\emph{Phys. Rev. D} {\bfseries 110} (2024) 030001}.

\bibitem{Meyer}
C.A.~Meyer and E.S.~Swanson, \emph{{Hybrid Mesons}}, \href{https://doi.org/10.1016/j.ppnp.2015.03.001}{\emph{Prog. Part. Nucl. Phys.} {\bfseries 82} (2015) 21} [\href{https://arxiv.org/abs/1502.07276}{{\ttfamily 1502.07276}}].

\bibitem{Bellantuono:2014lra}
L.~Bellantuono, P.~Colangelo and F.~Giannuzzi, \emph{{Exotic $J^{PC}=1^{-+}$ mesons in a holographic model of QCD}}, \href{https://doi.org/10.1140/epjc/s10052-014-2830-6}{\emph{Eur. Phys. J. C} {\bfseries 74} (2014) 2830} [\href{https://arxiv.org/abs/1402.5308}{{\ttfamily 1402.5308}}].

\bibitem{BESIII:2015cld}
{\scshape BESIII} collaboration, \emph{{Observation of $Z_c(3900)^{0}$ in $e^+e^-\to\pi^0\pi^0 J/\psi$}}, \href{https://doi.org/10.1103/PhysRevLett.115.112003}{\emph{Phys. Rev. Lett.} {\bfseries 115} (2015) 112003} [\href{https://arxiv.org/abs/1506.06018}{{\ttfamily 1506.06018}}].

\bibitem{Liu}
Y.~Liu, M.A.~Nowak and I.~Zahed, \emph{{Heavy Holographic Exotics: Tetraquarks as Efimov States}}, \href{https://doi.org/10.1103/PhysRevD.100.126023}{\emph{Phys. Rev. D} {\bfseries 100} (2019) 126023} [\href{https://arxiv.org/abs/1904.05189}{{\ttfamily 1904.05189}}].

\bibitem{Godfrey:2008nc}
S.~Godfrey and S.L.~Olsen, \emph{{The Exotic XYZ Charmonium-like Mesons}}, \href{https://doi.org/10.1146/annurev.nucl.58.110707.171145}{\emph{Ann. Rev. Nucl. Part. Sci.} {\bfseries 58} (2008) 51} [\href{https://arxiv.org/abs/0801.3867}{{\ttfamily 0801.3867}}].

\bibitem{Lu:2020qmp}
Q.-F.~L\"u, D.-Y.~Chen and Y.-B.~Dong, \emph{{Open charm and bottom tetraquarks in an extended relativized quark model}}, \href{https://doi.org/10.1103/PhysRevD.102.074021}{\emph{Phys. Rev. D} {\bfseries 102} (2020) 074021} [\href{https://arxiv.org/abs/2008.07340}{{\ttfamily 2008.07340}}].

\bibitem{Fang:2022mks}
{\scshape LHCb} collaboration, \emph{{Recent LHCb results on heavy hadron spectroscopy}}, \href{https://doi.org/10.1016/j.nuclphysbps.2022.09.015}{\emph{Nucl. Part. Phys. Proc.} {\bfseries 318-323} (2022) 66}.

\bibitem{LHCb:2021uow}
{\scshape LHCb} collaboration, \emph{{Observation of New Resonances Decaying to $J/\psi K^+$+ and $J/\psi \phi$}}, \href{https://doi.org/10.1103/PhysRevLett.127.082001}{\emph{Phys. Rev. Lett.} {\bfseries 127} (2021) 082001} [\href{https://arxiv.org/abs/2103.01803}{{\ttfamily 2103.01803}}].

\bibitem{Campanella:2018xev}
S.~Campanella, P.~Colangelo and F.~De~Fazio, \emph{{Excited heavy meson decays to light vector mesons: implications for spectroscopy}}, \href{https://doi.org/10.1103/PhysRevD.98.114028}{\emph{Phys. Rev. D} {\bfseries 98} (2018) 114028} [\href{https://arxiv.org/abs/1810.04492}{{\ttfamily 1810.04492}}].

\bibitem{Mahapatra:2020wym}
S.~Mahapatra, S.~Priyadarshinee, G.N.~Reddy and B.~Shukla, \emph{{Exact topological charged hairy black holes in AdS Space in $D$-dimensions}}, \href{https://doi.org/10.1103/PhysRevD.102.024042}{\emph{Phys. Rev. D} {\bfseries 102} (2020) 024042} [\href{https://arxiv.org/abs/2004.00921}{{\ttfamily 2004.00921}}].

\bibitem{Mahapatra:2018gig}
S.~Mahapatra and P.~Roy, \emph{{On the time dependence of holographic complexity in a dynamical Einstein-dilaton model}}, \href{https://doi.org/10.1007/JHEP11(2018)138}{\emph{JHEP} {\bfseries 11} (2018) 138} [\href{https://arxiv.org/abs/1808.09917}{{\ttfamily 1808.09917}}].

\bibitem{Priyadarshinee:2023cmi}
S.~Priyadarshinee and S.~Mahapatra, \emph{{Analytic three-dimensional primary hair charged black holes and thermodynamics}}, \href{https://doi.org/10.1103/PhysRevD.108.044017}{\emph{Phys. Rev. D} {\bfseries 108} (2023) 044017} [\href{https://arxiv.org/abs/2305.09172}{{\ttfamily 2305.09172}}].

\bibitem{Priyadarshinee:2021rch}
S.~Priyadarshinee, S.~Mahapatra and I.~Banerjee, \emph{{Analytic topological hairy dyonic black holes and thermodynamics}}, \href{https://doi.org/10.1103/PhysRevD.104.084023}{\emph{Phys. Rev. D} {\bfseries 104} (2021) 084023} [\href{https://arxiv.org/abs/2108.02514}{{\ttfamily 2108.02514}}].

\bibitem{Daripa:2024ksg}
A.~Daripa and S.~Mahapatra, \emph{{Analytic three-dimensional primary hair charged black holes with Coulomb-like electrodynamics and their thermodynamics}}, \href{https://doi.org/10.1103/PhysRevD.109.124039}{\emph{Phys. Rev. D} {\bfseries 109} (2024) 124039} [\href{https://arxiv.org/abs/2401.04561}{{\ttfamily 2401.04561}}].

\bibitem{MartinContreras:2021bis}
M.A.~Martin~Contreras, S.~Diles and A.~Vega, \emph{{Heavy quarkonia spectroscopy at zero and finite temperature in bottom-up AdS/QCD}}, \href{https://doi.org/10.1103/PhysRevD.103.086008}{\emph{Phys. Rev. D} {\bfseries 103} (2021) 086008} [\href{https://arxiv.org/abs/2101.06212}{{\ttfamily 2101.06212}}].

\bibitem{Mamani:2022qnf}
L.A.H.~Mamani, D.~Hou and N.R.F.~Braga, \emph{{Melting of heavy vector mesons and quasinormal modes in a finite density plasma from holography}}, \href{https://doi.org/10.1103/PhysRevD.105.126020}{\emph{Phys. Rev. D} {\bfseries 105} (2022) 126020} [\href{https://arxiv.org/abs/2204.08068}{{\ttfamily 2204.08068}}].

\bibitem{Herzog:2006ra}
C.P.~Herzog, \emph{{A Holographic Prediction of the Deconfinement Temperature}}, \href{https://doi.org/10.1103/PhysRevLett.98.091601}{\emph{Phys. Rev. Lett.} {\bfseries 98} (2007) 091601} [\href{https://arxiv.org/abs/hep-th/0608151}{{\ttfamily hep-th/0608151}}].

\bibitem{Scadron:2006dy}
M.D.~Scadron, R.~Delbourgo and G.~Rupp, \emph{{Constituent quark masses and the electroweak standard model}}, \href{https://doi.org/10.1088/0954-3899/32/5/009}{\emph{J. Phys. G} {\bfseries 32} (2006) 735} [\href{https://arxiv.org/abs/hep-ph/0603196}{{\ttfamily hep-ph/0603196}}].

\bibitem{Braga:2017bml}
N.R.F.~Braga, L.F.~Ferreira and A.~Vega, \emph{{Holographic model for charmonium dissociation}}, \href{https://doi.org/10.1016/j.physletb.2017.10.013}{\emph{Phys. Lett. B} {\bfseries 774} (2017) 476} [\href{https://arxiv.org/abs/1709.05326}{{\ttfamily 1709.05326}}].

\bibitem{Lucini:2003zr}
B.~Lucini, M.~Teper and U.~Wenger, \emph{{The High temperature phase transition in SU(N) gauge theories}}, \href{https://doi.org/10.1088/1126-6708/2004/01/061}{\emph{JHEP} {\bfseries 01} (2004) 061} [\href{https://arxiv.org/abs/hep-lat/0307017}{{\ttfamily hep-lat/0307017}}].

\bibitem{Hou:2001ig}
W.-S.~Hou, C.-S.~Luo and G.-G.~Wong, \emph{{Glueball states in a constituent gluon model}}, \href{https://doi.org/10.1103/PhysRevD.64.014028}{\emph{Phys. Rev. D} {\bfseries 64} (2001) 014028} [\href{https://arxiv.org/abs/hep-ph/0101146}{{\ttfamily hep-ph/0101146}}].

\bibitem{Dudal:2015kza}
D.~Dudal and T.G.~Mertens, \emph{{Radiation Gauge in AdS/QCD: Inadmissibility and Implications on Spectral Functions in the Deconfined Phase}}, \href{https://doi.org/10.1016/j.physletb.2015.10.074}{\emph{Phys. Lett. B} {\bfseries 751} (2015) 352} [\href{https://arxiv.org/abs/1510.05490}{{\ttfamily 1510.05490}}].

\bibitem{Jena:2024cqs}
S.S.~Jena, J.~Barman, B.~Toniato, D.~Dudal and S.~Mahapatra, \emph{{A dynamical Einstein-Born-Infeld-dilaton model and holographic quarkonium melting in a magnetic field}}, \href{https://doi.org/10.1007/JHEP12(2024)096}{\emph{JHEP} {\bfseries 12} (2024) 096} [\href{https://arxiv.org/abs/2408.14813}{{\ttfamily 2408.14813}}].

\bibitem{Baker_Graves-Morris_1996}
G.A.~Baker and P.~Graves-Morris, \emph{Padé Approximants}, Encyclopedia of Mathematics and its Applications, Cambridge University Press, 2~ed. (1996).

\bibitem{05fc0d9b-f28e-3545-ab7b-94a0385f9451}
A.~Iserles, \emph{On the generalized padé approximations to the exponential function}, {\emph{SIAM Journal on Numerical Analysis} {\bfseries 16} (1979) 631}.

\bibitem{Miranda:2009uw}
A.S.~Miranda, C.A.~Ballon~Bayona, H.~Boschi-Filho and N.R.F.~Braga, \emph{{Black-hole quasinormal modes and scalar glueballs in a finite-temperature AdS/QCD model}}, \href{https://doi.org/10.1088/1126-6708/2009/11/119}{\emph{JHEP} {\bfseries 2009} (2009) 119} [\href{https://arxiv.org/abs/0909.1790}{{\ttfamily 0909.1790}}].

\bibitem{Mamani:2018uxf}
L.A.H.~Mamani, A.S.~Miranda and V.T.~Zanchin, \emph{{Melting of scalar mesons and black-hole quasinormal modes in a holographic QCD model}}, \href{https://doi.org/10.1140/epjc/s10052-019-6902-5}{\emph{Eur. Phys. J. C} {\bfseries 79} (2019) 1} [\href{https://arxiv.org/abs/1809.03508}{{\ttfamily 1809.03508}}].

\bibitem{Fujita:2009wc}
M.~Fujita, K.~Fukushima, T.~Misumi and M.~Murata, \emph{{Finite-temperature spectral function of the vector mesons in an AdS/QCD model}}, \href{https://doi.org/10.1103/PhysRevD.80.035001}{\emph{Phys. Rev. D} {\bfseries 80} (2009) 035001} [\href{https://arxiv.org/abs/0903.2316}{{\ttfamily 0903.2316}}].

\bibitem{Fujita:2009ca}
M.~Fujita, T.~Kikuchi, K.~Fukushima, T.~Misumi and M.~Murata, \emph{{Melting Spectral Functions of the Scalar and Vector Mesons in a Holographic QCD Model}}, \href{https://doi.org/10.1103/PhysRevD.81.065024}{\emph{Phys. Rev. D} {\bfseries 81} (2010) 065024} [\href{https://arxiv.org/abs/0911.2298}{{\ttfamily 0911.2298}}].

\bibitem{deForcrand:2002hgr}
P.~de~Forcrand and O.~Philipsen, \emph{{The QCD phase diagram for small densities from imaginary chemical potential}}, \href{https://doi.org/10.1016/S0550-3213(02)00626-0}{\emph{Nucl. Phys. B} {\bfseries 642} (2002) 290} [\href{https://arxiv.org/abs/hep-lat/0205016}{{\ttfamily hep-lat/0205016}}].

\bibitem{Stephanov:2004wx}
M.A.~Stephanov, \emph{{QCD Phase Diagram and the Critical Point}}, \href{https://doi.org/10.1143/PTPS.153.139}{\emph{Prog. Theor. Phys. Suppl.} {\bfseries 153} (2004) 139} [\href{https://arxiv.org/abs/hep-ph/0402115}{{\ttfamily hep-ph/0402115}}].

\bibitem{Borsanyi:2025dyp}
S.~Borsanyi, Z.~Fodor, J.N.~Guenther, P.~Parotto, A.~Pasztor, C.~Ratti et~al., \emph{{Lattice QCD constraints on the critical point from an improved precision equation of state}},  \href{https://arxiv.org/abs/2502.10267}{{\ttfamily 2502.10267}}.

\bibitem{Andreev:2023hmh}
O.~Andreev, \emph{{QQ\textasciimacron{}qqq quark system, compact pentaquark, and gauge/string duality. II}}, \href{https://doi.org/10.1103/PhysRevD.108.106012}{\emph{Phys. Rev. D} {\bfseries 108} (2023) 106012} [\href{https://arxiv.org/abs/2306.08581}{{\ttfamily 2306.08581}}].

\bibitem{LHCb:2019kea}
{\scshape LHCb} collaboration, \emph{{Observation of a narrow pentaquark state, $P_c(4312)^+$, and of two-peak structure of the $P_c(4450)^+$}}, \href{https://doi.org/10.1103/PhysRevLett.122.222001}{\emph{Phys. Rev. Lett.} {\bfseries 122} (2019) 222001} [\href{https://arxiv.org/abs/1904.03947}{{\ttfamily 1904.03947}}].

\bibitem{LHCb:2015yax}
{\scshape LHCb} collaboration, \emph{{Observation of $J/\psi p$ Resonances Consistent with Pentaquark States in $\Lambda_b^0 \to J/\psi K^- p$ Decays}}, \href{https://doi.org/10.1103/PhysRevLett.115.072001}{\emph{Phys. Rev. Lett.} {\bfseries 115} (2015) 072001} [\href{https://arxiv.org/abs/1507.03414}{{\ttfamily 1507.03414}}].

\bibitem{LHCb:2022ogu}
{\scshape LHCb} collaboration, \emph{{Observation of a J/\ensuremath{\psi}\ensuremath{\Lambda} Resonance Consistent with a Strange Pentaquark Candidate in B-\textrightarrow{}J/\ensuremath{\psi}\ensuremath{\Lambda}p\textasciimacron{} Decays}}, \href{https://doi.org/10.1103/PhysRevLett.131.031901}{\emph{Phys. Rev. Lett.} {\bfseries 131} (2023) 031901} [\href{https://arxiv.org/abs/2210.10346}{{\ttfamily 2210.10346}}].

\end{thebibliography}\endgroup
\end{document}